\documentclass[twoside,11pt]{article}
\usepackage{jmlr2e}
\usepackage{color}
\usepackage{enumerate}
\usepackage{algorithm}
\usepackage{amsmath,bbm}
\usepackage{smile}
\def\T{\mathrm{\scriptstyle T}}
\usepackage{booktabs}
\usepackage{lastpage}

\jmlrheading{23}{2022}{1-\pageref{LastPage}}{10/21; Revised 6/22}{8/22}{21-1269}{Kean Ming Tan, Heather Battey, and Wen-Xin Zhou}

\ShortHeadings{Decentralized Quantile Regression}{Tan, Battey and Zhou}
\firstpageno{1}

\def\SN{\sum_{i=1}^N}

\newcommand*{\var}{\textnormal{var}}
\newcommand{\imd}{{\mathsmaller {\rm imd}}}
\newcommand{\nn}{\nonumber}

\def\##1\#{\begin{align}#1\end{align}}
\def\$#1\${\begin{align*}#1\end{align*}}

\usepackage{relsize}
\newcommand{\HH}{{\mathsmaller {\rm H}}}
\newcommand{\cq}{{\mathsmaller {\rm cq}}}
\newcommand{\dc}{{\mathsmaller {\rm dc}}}
\newcommand{\loc}{{\mathsmaller {\rm loc}}}

\def\sn{\sum_{i=1}^n}

\newcommand{\BB}{\mathbb{B}}

\newcommand{\wt}{\widetilde}

\newcommand{\bfsym}[1]{\ensuremath{\boldsymbol{#1}}}
       \def \bbeta    {\bfsym{\beta}}
       \def \bdelta   {\bfsym{\delta}}

\newcommand{\Rom}[1]{\text{\uppercase\expandafter{\romannumeral #1\relax}}}

\newcommand{\cc}{{\rm c}}

\begin{document}

\title{Communication-Constrained Distributed Quantile Regression with Optimal Statistical Guarantees}

\author{\name Kean Ming Tan \email keanming@umich.edu \\
       \addr Department of Statistics \\
       	     University of Michigan \\
	           Ann Arbor MI, 48109, USA
       \AND
\name Heather Battey \email h.battey@imperial.ac.uk \\
\addr Department of Mathematics \\
Imperial College London \\
London, SW7 2AZ, U.K.
\AND
\name Wen-Xin Zhou \email wez243@ucsd.edu \\
	\addr Department of Mathematical Sciences \\
	University of California, San Diego \\
	La Jolla, CA 92093, USA
}	
\editor{Qiang Liu}
\maketitle

\begin{abstract}%
We address the problem of how to achieve optimal inference in distributed quantile regression without stringent scaling conditions. This is challenging due to the non-smooth nature of the quantile regression (QR) loss function, which invalidates the use of existing methodology. The difficulties are resolved through a double-smoothing approach that is applied to the local (at each data source) and global objective functions. Despite the reliance on a delicate combination of local and global smoothing parameters, the quantile regression model is fully parametric, thereby facilitating interpretation. In the low-dimensional regime, we establish a finite-sample theoretical framework for the sequentially defined distributed QR estimators. This reveals a trade-off between the communication cost and statistical error. 
We further discuss and compare several alternative confidence set constructions, based on inversion of Wald and score-type tests and resampling techniques, detailing an improvement that is effective for more extreme quantile coefficients. 
In high dimensions, a sparse framework is adopted, where the proposed doubly-smoothed objective function is complemented with an $\ell_1$-penalty.
We show that the corresponding distributed penalized QR estimator achieves the global convergence rate after a near-constant number of communication rounds. A thorough simulation study further elucidates our findings.
\end{abstract}

\section{Introduction}
\label{sec:1} 

Quantile regression is indispensable for understanding pathways of dependence irretrievable through a standard conditional mean regression analysis. Since its inception by \cite{KB1978}, appreciable effort has been expended in understanding and operationalizing quantile regression. Statistical aspects have focused on the situation in which all the data are simultaneously available for inference while practical aspects have centered around reformulations of the quantile regression optimization problem for computational efficiency.

Challenges arise when data are distributed, either by the study design or due to storage and privacy concerns. 
The latter have become more prominent, with less centralized systems tending to be preferred both by the individuals whose data are collected and by those responsible for ensuring their security. In such settings, communication costs associated with statistical procedures become a consideration in addition to their theoretical properties. 
Ideally,  inferential tools are sought whose communication costs are as low as possible without sacrificing statistical accuracy, where the latter would be quantified in terms of estimation error or distributional approximation errors for test statistics.

Data may be naturally partitioned because of the way they were collected, or deliberately distributed for other reasons. \cite{LSTS2020} provided examples in which the distributed setting arises: (i) when data are too numerous to be stored in a central location; (ii) when privacy and security are a concern, such as for medical records, necessitating decentralized statistical analyses. We are motivated particularly by situations in which there are separate data-collecting entities such as local governments, research labs, hospitals, or smart phones, and direct data sharing raises concerns over privacy or loss of ownership. Due to privacy concerns over sending raw data, data collected at each location must remain there, which makes communication efficiency critical, especially when the network comprises an enormous number of local data-collecting entities. Communication in the network can be slower than local computation by three orders of magnitude due to limited bandwidth \citep{LLZ2020}. It is therefore desirable to communicate as few rounds as possible, leaving expensive computation to local machines.

Among two general principles that have been proposed for distributed statistical inference, the simple meta-analysis approach of averaging estimates from separate data sources has the advantage of only requiring one round of communication. \cite{JLY2018} highlighted some disadvantages. Notably, in a simpler setting than that posed by quantile regression, a stringent constraint on the number of sources is implicated. To attain the convergence rate hypothetically achievable using the combined sample of size $N$, a meta-analysis must limit the number of sources, $m$, to be far fewer than $\sqrt{N}$. This is due to small sample bias inherent to most nonlinear estimators, which does not diminish upon aggregation. A violation of the scaling condition slows the convergence rate of the estimator. This, while sometimes acceptable for point estimation, is detrimental for statistical inference, as illustrated later in simulations.

By extending the distributed approximate Newton algorithm \citep{SSZ2014}, \cite{WKSZ2017} and \cite{JLY2018}  proposed an alternative principle for distributed inference in parametric models, which requires a controlled amount of further communications to yield statistically optimal estimators without the restriction $m=o(\sqrt{N})$ on the number of machines. Another variant of this principle was considered in \cite{FGW2021}, along with a simultaneous analysis of the optimization and statistical errors. For reasons outlined below, these ideas are not directly applicable to quantile regression without considerable methodological development, guided by the detailed theoretical analysis provided in this paper. 

Quantile regression quantifies dependence of an outcome variable's quantiles on a number of covariates. All quantiles are potentially of interest but to take an important example, quantile regression has bearing on the types of applications for which conditional extreme value analysis might be considered. There are relatively few successful examples of  modeling the extremes. The pioneering work of \cite{EH2020} being a notable exception, suitable when all variables are on an equal footing. When explanation for extreme behavior of a particular variable is sought, as would be the case in many hydrological, sociological, and medical applications, quantile regression provides succinct interpretable conclusions. Subtle graphical structure is deducible from a succession of quantile regression analyses by a result of \cite{Cox2007}, generalizing insights by \cite{Cochran1938}. Besides substantive understanding furnished by a quantile regression model, coefficient estimators enjoy robustness properties in the form of limited sensitivity to anomalous data or leptokurtic tail behavior of the conditional distribution of the outcome.

The associated non-differentiable loss function, otherwise responsible for tortuously slow computation, necessitates linear programming reformulations, solvable by variants of simplex and interior point methods. These algorithms are not compatible with distributed architectures, rendering statistical inference challenging when data are distributed. Even when computation is ignored and the non-differentiable loss function is used directly, both the distributed estimation procedures proposed by \cite{WKSZ2017}, \cite{JLY2018} and  \cite{FGW2021} and the technical devices used therein are unavailable due to their requirements on the loss function. Namely that it be strongly convex and twice differentiable with Lipschitz continuous second derivatives.

Two papers by \cite{VCC2019} and \cite{CLZ2018} are motivated by the challenges of distributed data and the relevance of quantile regression, seeking synthesized estimators of quantile regression coefficients. These papers employed the meta-analysis approach, thereby requiring stringent scaling to achieve the desired theoretical guarantees, although with the advantage of requiring a single round of communication. We discuss these works in greater detail in Section \ref{sec:comparison}. 
In addition to the scaling deficiencies, a generalization of the simple meta-analysis approach to high-dimensional settings has proved elusive for quantile regression. In sparse high-dimensional linear and generalized linear models, the success of meta-analyses hinges on the ability to de-bias suitably penalized estimators \citep{LSLT2017, BFLLZ2018}. Such de-biased estimators are unavailable for penalized quantile regression, except under the stringent assumption that the regression error is independent of the covariates \citep{BK2017}.

 The present paper operationalizes the ideas of \cite{JLY2018} and \cite{WKSZ2017} in the context of quantile regression, enabling distributed estimation and inference in low and high-dimensional regimes. The key idea of our proposal is double-smoothing of the local and global approximate loss functions, which requires different smoothing bandwidths to achieve desirable statistical properties. Specifically, our proposed synthesized estimator achieves the optimal statistical rate of convergence by a delicate combination of local and global smoothing, and number of communication rounds. The latter turns out to be small. 

In the low-dimensional regime, we further detail distributed constructions of confidence sets. Among these is one based on a self-normalized reformulation of a score-type statistic. Modulo estimation of the parameter vector, score constructions rewritten as self-normalized sums enjoy a form of linearity that enables synthesis across data sources without information loss. To our knowledge, this work is the first to provide Berry-Esseen type quantification of distributional approximation errors in a distributed setting, which may be of independent interest.  
In the high-dimensional regime, the proposed doubly-smoothed local and global objective functions are coupled with an $\ell_1$ penalty to encourage sparse solutions, which we solve using a locally adaptive majorize-minimize algorithm.   Theoretically, we show that the resulting estimator is near-optimal under both the $\ell_1$ and $\ell_2$ norms.   The results are presented in Section~\ref{sec:hd}.

\medskip
\noindent
{\sc Notation}: 
For every integer $k\geq 1$, we use $\RR^k$ to denote the the $k$-dimensional Euclidean space. The inner product of any two vectors $\bu=(u_1, \ldots, u_k)^\T, \bv=(v_1, \ldots ,v_k)^\T \in \RR^k$ is defined by $\bu^\T \bv = \langle \bu, \bv \rangle= \sum_{i=1}^k u_i v_i$.
We use $\| \cdot \|_p$ $(1\leq p \leq \infty)$ to denote the $\ell_p$-norm in $\RR^k$: $\| \bu \|_p = ( \sum_{i=1}^k | u_i |^p )^{1/p}$ and $\| \bu \|_\infty = \max_{1\leq i\leq k} |u_i|$. 
Throughout this paper, we use bold capital letters to represent matrices. For $k\geq 2$, $\Ib_k$ represents the identity matrix of size $k$. For any $k\times k$ symmetric matrix $\Ab \in \RR^{k\times k}$, $\| \Ab \|_2$ is the operator norm of $\Ab$.
For a positive semidefinite matrix $\Ab \in \RR^{k\times k}$,  $\| \cdot \|_{\Ab}$ denotes the norm linked to $\Ab$ given by $\| \bu \|_{\Ab} = \| \Ab^{1/2} \bu \|_2$, $\bu \in \RR^k$.
Moreover, given $r \geq 0$, define the Euclidean ball and sphere in $\RR^k$ as $\BB^k(r) = \{ \bu \in \RR^k : \| \bu \|_2 \leq r\}$ and $\mathbb{S}^{k-1}(r) = \partial \mathbb B^k(r) = \{ \bu \in \RR^k: \| \bu \|_2 =r \}$, respectively. In particular, $\mathbb{S}^{k-1} \equiv \mathbb{S}^{k-1}(1)$ denotes the unit sphere.
For two sequences of non-negative numbers $\{ a_n \}_{n\geq 1}$ and $\{ b_n \}_{n\geq 1}$, $a_n \lesssim b_n$ indicates that there exists a constant $C>0$ independent of $n$ such that $a_n \leq Cb_n$; $a_n \gtrsim b_n$ is equivalent to $b_n \lesssim a_n$; $a_n \asymp b_n$ is equivalent to $a_n \lesssim b_n$ and $b_n \lesssim a_n$.

\section{Distributed Inference for Quantile Regression}
\label{sec2}

\subsection{Conquer: convolution-smoothed quantile regression}
\label{sec:smoothing}

For a quantile level $\tau \in (0,1)$, we consider a linear conditional quantile model for the data vector $(y, \bx) \in \RR \times \RR^p$:
\begin{equation}
	 Q_\tau( y  | \bx  ) = \bx^\T \bbeta^* = \sum_{j=1}^p x_j \beta^*_j   , \label{qr.model}
\end{equation}
where $Q_\tau(y  | \bx  )$ denotes the conditional $\tau$-quantile of $y$ given $\bx  = (x_{1}, \ldots, x_{p})^\T$ with $x_{1} \equiv 1$. Here, $\bbeta^* = \bbeta^*(\tau) \in \RR^p$  is the regression coefficient vector that minimizes the criterion function
\begin{equation}
	\cQ(\bbeta )  : = \EE  \bigl\{ \rho_\tau( y - \bx^\T \bbeta) \bigr\} , \label{population.QR}
\end{equation}
where $\rho_\tau(u) = u \{\tau - \mathbbm{1}(u<0) \}$ is the asymmetric absolute deviation function, also known as the {\it check function} or {\it pinball loss}. Given a random sample  $\{ (y_i ,\bx_i) \}_{i=1}^N$ of size $N >p$ from $(y, \bx)$, the linear quantile regression estimator of $\bbeta^*$ is defined as a minimizer of the empirical analog of $\cQ(\cdot)$:
\begin{equation}
	\hat \bbeta \in \argmin_{\bbeta \in \RR^p }  \hat \cQ(\bbeta) , ~\mbox{ where }~ \hat \cQ(\bbeta) := \frac{1}{N} \sum_{i=1}^N \rho_\tau ( y_i - \bx_i^\T \bbeta) .  \label{empirical.QR}
\end{equation}
Since the seminal work of \cite{KB1978}, quantile regression (QR) has been extensively studied from both statistical and computational perspectives. We refer to \cite{K2005} and \cite{KCHP2017} for a systematic introduction of quantile regression under various settings.

By the convexity of the check function, the population loss function $\cQ(\cdot)$ in \eqref{population.QR} is also convex. Moreover, under mild conditions, $\cQ(\cdot)$ is twice differentiable and strongly convex in a neighborhood of $\bbeta^*$ with Hessian matrix $\Hb := \nabla^2 \cQ(\bbeta^*) = \EE \{ f_{\varepsilon | \bx} (0) \bx \bx^\T \}$, where $f_{\varepsilon | \bx}(\cdot)$ denotes the conditional density of $\varepsilon$ given $\bx$. In contrast, the empirical loss $\hat \cQ(\cdot)$ is not differentiable at $\bbeta^*$, and its ``curvature energy'' is concentrated at a single point. 
This is substantially different from other widely used loss functions that are at least locally strongly convex, such as the squared or logistic loss.  
To deal with the non-smoothness issue, \cite{H1998} proposed to smooth the objective function, or equivalently the check function $\rho_\tau(\cdot)$, to obtain $\rho_\tau^{\HH} (u) =  u\{ \tau - G(-u/h) \}$, where $G(\cdot)$ is a smooth function and $h>0$ is the smoothing parameter or bandwidth. See also \cite{WSZ2012}, \cite{WMY2015}, \cite{GK2016} and \cite{CLZ2019} for extensions of such a smoothed objective function approach with more complex data.
However, Horowitz's smoothing gains smoothness at the cost of convexity, which inevitably raises optimization issues especially when $p$ is large.
On the other hand, by the first-order condition, the population parameter $\bbeta^*$ satisfies the moment condition $\nabla \cQ(\bbeta^*) = \EE  [   \{   \mathbbm{1}( y < \bx^\T \bbeta) - \tau   \} \bx   ] |_{\bbeta = \bbeta^*}  = \textbf{0}$. 
This property motivates a smoothed estimating equation (SEE) estimator \citep{W2006, KS2017}, defined as the solution to the smoothed moment condition
\begin{equation}
  \frac{1}{N}  \sum_{i=1}^N \bigl\{    G\bigl(  ( \bx_i^\T \bbeta - y_i )/h \bigr)  -\tau \bigr\} \bx_i = \textbf{0} . \label{see}
\end{equation}
From an $M$-estimation viewpoint, the aforementioned SEE estimator can be equivalently defined as a minimizer of the empirical smoothed loss function
\begin{equation}
\hat \cQ_{h}(\bbeta) = \frac{1}{N} \sum_{i=1}^N \ell_{h}(y_i -   \bx_i^\T  \bbeta  ) ~~\mbox{ with }~~\ell_{h}(u) =  (\rho_\tau * K_h )(u) = \int_{-\infty}^{\infty} \rho_\tau(v) K_h(v- u ) \, {\rm d} v , \label{conquer}
\end{equation}
where $K(\cdot)$ is a kernel function, $K_h(u) = (1/h) K(u/h)$, and $*$ is the convolution operator.  This approach will be referred to as {\it conquer}, which stands for  convolution-type smoothed quantile regression. The ensuing estimator is then denoted by $\hat \bbeta^{\cq}  = \hat \bbeta^{\cq}_h \in \argmin_{\bbeta \in \RR^p}  \hat \cQ_{h}(\bbeta)$.

To see the connection between SEE and conquer methods, define $\overbar K(u) =\int_{-\infty}^u K(v) \, {\rm d}v$, and note that the empirical loss $\hat  \cQ_{h}(\cdot)$ in \eqref{conquer} is twice continuously differentiable with gradient and Hessian given by $	\nabla  \hat \cQ_{  h}(\bbeta)  =  (1/N) \sum_{i=1}^N  \{ \overbar{K}  (  (  \bx_i^\T \bbeta  - y_i )/h  ) - \tau  \} \bx_i $ and $\nabla^2  \hat \cQ_{  h}(\bbeta)  =  (1/N) \sum_{i=1}^N K_h  (   y_i - \bx_i^\T \bbeta  )  \cdot\bx_i \bx_i^\T$, respectively. When a non-negative kernel is used, $\hat \cQ_{h}(\cdot)$ is convex so that any minimizer of $\bbeta \mapsto \hat \cQ_h(\bbeta)$ satisfies the first-order moment condition \eqref{see} with $G = \overbar K$.

When the dimension $p$ is fixed, asymptotic properties of the SEE or conquer estimator have been studied by \cite{KS2017} and \cite{FGH2019}, although the former focused on a more challenging instrumental variables quantile regression problem.
 In the finite sample setup, \cite{HPTZ2020} established exponential-type concentration inequalities and nonasymptotic Bahadur representation for the conquer estimator, while allowing the dimension $p$ to grow with the sample size $n$. 
 Their results reveal a key feature of the smoothing parameter: the bandwidth should adapt to both the sample size $n$  and dimensionality $p$, so as to achieve a trade-off between statistical accuracy and computational stability. For statistical inference, \cite{HPTZ2020} suggested and proved the validity of the {\it multiplier bootstrap} for conquer, which has desirable finite sample performance under various settings, including those at extreme quantile levels. We refer to Section~5 of \cite{HPTZ2020} for further details on the computational aspects of conquer.

\subsection{Distributed quantile regression with conquer}
\label{sec:distributed.conquer}
Before detailing an approach for distributed inference for QR coefficients, motivated primarily by situations in which the data are distributed, we start with some remarks on computation. 
The optimization problem in \eqref{empirical.QR} can be recast as a convex linear program, solvable by the simplex or interior point methods. The latter has a computational complexity of order $\cO(N^{1+a} p^3 \log N)$ for some $a \in (0,1/2)$. An efficient algorithm, the Frisch-Newton algorithm with preprocessing, has an improved complexity of $\cO\{ (N p )^{2(1+a)/3 } p^3 \log N + N p \}$ \citep{PK1997}. While not inordinate relative to the $\cO(p^2 N)$ complexity of least squares, to achieve the same quality of distributional approximation, quantile regression requires a considerably larger sample size. Thus for formal inference there are sometimes computational advantages to parallelized inference even when data are available in their totality.

For ease of exposition, assume that the $m$ data sources are of equal sample size $n$, so that $N=m \cdot n$. The combined data set is $\{ (y_i, \bx_i) \}_{i=1}^N$, where $\bx_i$ is a $p$-dimensional vector. For $j =1,\ldots, m$, the $j$th location stores a subsample of $n$ observations, denoted by $\cD_j = \{ (y_i, \bx_i) \}_{i\in \cI_j}$, and $\{\cI_j\}_{j=1}^{m}$ are disjoint index sets satisfying $\cup_{j=1}^m \cI_j = \{ 1,\ldots, N\}$ and $| \cI_j | = n$, where $|\cI_j|$ is the cardinality of $\cI_j$.

Under a conditional quantile regression model, the observations $(y_1,\bx_1), \ldots, (y_N, \bx_N)$ are i.i.d.~sampled from $(y, \bx) \sim P$ satisfying $Q_{\tau}(y|\bx) = \bx^\T \bbeta^*$, and the model parameter $\bbeta^*$ is equivalently defined as
\#
	\bbeta^* = \argmin_{\bbeta \in \RR^p }  \cQ(\bbeta) , \quad   \cQ(\bbeta) :=   \EE_{(y,\bx) \sim P} \bigl\{   \rho_\tau \bigl( y -  \bx^\T \bbeta ) \bigr\} ,  \label{model.parameter}
\#
where $\rho_\tau(\cdot)$ is the check function. Unlike the model setting considered by \cite{WKSZ2017} and \cite{JLY2018} in which the target loss function is twice differentiable and has Lipschitz continuous second derivative, the non-smooth check function is not everywhere differentiable, which prevents gradient-based optimization methods from being efficient.

Given two bandwidths $h , b >0$, we define the global and local smoothed quantile loss functions as
\#
	 \hat \cQ_h (\bbeta)   =  \frac{1}{N} \sum_{i=1}^N \ell_h(y_i - \bx_i^\T \bbeta ) ~~\mbox{ and }~~
	 \hat \cQ_{j,b} (\bbeta) = \frac{1}{n} \sum_{i \in \cI_j}  \ell_b( y_i - \bx_i^\T \bbeta )  , \ \ j = 1,\ldots , m, \label{local.loss.functions}
\#
where the loss function $\ell_h(\cdot)$ is as defined in~\eqref{conquer}. 
Hereafter, $h$ and $b$ will be referred to as the {\it global bandwidth} and {\it local bandwidth}, respectively, and we assume $b \geq h>0$.
In the context of quantile regression,  we extend the approximate Newton-type method proposed by \cite{SSZ2014} through convolution smoothing; see also \cite{WKSZ2017} and \cite{JLY2018}. Notably, the ideas behind these Newton-type methods coincide, to some extent, with the classical one-step construction \citep{B1975}, which focused on improving an initial estimator that is already consistent but not efficient.

Starting with an initial estimator $\wt \bbeta^{(0)}$ of $\bbeta^*$, we define the shifted conquer loss function
\#
	\wt \cQ(\bbeta) =  \hat  \cQ_{1,b} (\bbeta)  - \big\langle \nabla  \hat  \cQ_{1,b} ( \wt \bbeta^{(0)} ) - \nabla  \hat \cQ_h( \wt \bbeta^{(0)} ),  \bbeta    \big\rangle  , \label{surrogate.loss}
\#
which leverages local higher-order information and global first-order information, and therefore depends on both local and global bandwidths $b$ and $h$.
The resulting communication-efficient estimator is given by
\#
	\wt \bbeta^{(1)} = \wt \bbeta^{(1)}_{b,h} \in \argmin_{\bbeta \in \RR^p}  	\wt \cQ(\bbeta) .  \label{one.step.conquer}
\#

Informal motivation for the aforementioned approach is provided by a Taylor series expansion of the global loss function around the initial estimator $\tilde{\bbeta}^{(0)}$. For a suitable choice of $\tilde{\bbeta}^{(0)}$, the approximation error is well controlled in view of the heuristic argument outlined by \cite{JLY2018}. Furthermore, on writing  $\tilde{Q}(\bbeta)$ more explicitly as $\tilde{Q}(\bbeta;\tilde{\bbeta}^{(0)})$, it can be arranged, through bandwidths $b$ and $h$, that $\argmin_{\bbeta\in \RR^p} \tilde{Q}(\bbeta;\tilde{\bbeta}^{(0)})$ is a contraction mapping in a suitable neighborhood of $\bbeta^*$, to be defined. Intuitively, in view of Banach's fixed point theorem, the sequence of minimizers obtained through iteration of this procedure converges to the global conquer estimator, itself converging to $\bbeta^*$. In the limit of increasing iterations, there is no information loss over the oracle procedure with access to all data simultaneously, in spite of the data being distributed. The theoretical results of this section establish the delicate choices of $h$, $b$, and the number of iterations in order for the \emph{synthesis error} to match the statistical error of the global conquer estimator.

Under the conditional quantile model \eqref{qr.model}, the generic data vector $(y,\bx)$ can be written in a linear form $ y =   \bx^\T \bbeta^* + \varepsilon$, where the model error $\varepsilon$ satisfies $Q_\tau(\varepsilon | \bx) = 0$.
Let $f_{\varepsilon | \bx}(\cdot)$ be the conditional density function of $\varepsilon$ given $\bx$.
Given i.i.d.~observations $\{ (y_i ,\bx_i) \}_{i=1}^N$, we write $\varepsilon_i = y_i - \bx_i^\T \bbeta^*$, satisfying $\PP(\varepsilon_i\leq 0 | \bx_i )=\tau$. 

To investigate the statistical properties of $\wt \bbeta^{(1)}$, we impose some regularity conditions.

\medskip
\noindent
(C1). There exist $\bar f  \geq \underbar{$f$} >0$ such that $ \underbar{$f$}  \leq f_{\varepsilon |\bx}(0 ) \leq \bar f $ almost surely (over $\bx$).
Moreover, there exists some $l_0 >0$ such that $|f_{\varepsilon |\bx}(u)- f_{\varepsilon |\bx}(v)| \leq l_0  |u-v|$ for all $u , v \in \RR$ almost surely.
 
\noindent
(C2). $K(\cdot)$ is a symmetric and non-negative kernel that satisfies $\kappa_2 := \int_{-\infty}^\infty u^2 K(u) \, {\rm d} u <\infty$, $\kappa_u := \sup_{u\in \RR} K(u) <\infty$ and $\kappa_l := \min_{|u|\leq 1 }K(u) >0$. 

\noindent
(C3).
The predictor $\bx \in \RR^p$ is {\it sub-Gaussian}: there exists $ \upsilon_1>0$ such that $\PP( | \bz^\T \bu |    \geq \upsilon_1    t  ) \leq  2e^{-t ^2/2}$ for every unit vector $\bu \in \mathbb{S}^{p-1}$ and $t \geq 0$, where $\bz = \Sigma^{-1/2} \bx$ and $\Sigma = \EE(\bx \bx^\T)$ is positive definite. 

\medskip

Condition~(C1) imposes regularity conditions on the conditional density function. These are standard in quantile regression.
In (C2), the requirement $\min_{|u|\leq 1} K(u) >0$ is for technical simplicity and can be relaxed to  $ \min_{|u|\leq c} K(u) >0$ for some $c\in (0,1)$, which will only change the constant terms in all of our theoretical results.
In particular, for kernels that are compactly supported on $[-1,1]$, we may choose $c=1/2$ and assume  $\kappa_l  = \min_{|u|\leq 1/2 }K(u) >0$ instead. Distributions with heavier tails than Gaussian on $\bx$ are excluded by Condition (C3) in order to guarantee exponential-type concentration bounds for estimators of quantile regression coefficients.


For some radii $r, r_* >0$, define the events
\#
\cE_0 (r) = \big\{ \wt \bbeta^{(0)} \in \Theta(r) \big\} ~~\mbox{ and }~~  \cE_*(r_*)  =\big\{  \|   \nabla \hat  \cQ_h(\bbeta^* )  \|_{\Omega} \leq r_* \big\}  , \label{event0}
\#
where 
$$
	\Theta(r)  :=   \{  \bbeta\in \RR^p: \| \bbeta - \bbeta^* \|_{\Sigma} \leq r   \} ~~\mbox{ and }~~\Omega := \Sigma^{-1}.
$$
In particular, $\cE_0(r)$ is a ``good" event on which the initial estimator $\wt \bbeta^{(0)}$ falls into a local neighborhood $\Theta(r)$ around $\bbeta^*$. Recall that $\Hb = \EE \{ f_{\varepsilon | \bx}(0) \bx \bx^\T \}$ is the Hessian of  the population quantile loss $\bbeta \mapsto  \EE  \{ \rho_\tau( y - \bx^\T \bbeta) \}$ at $\bbeta^*$.
The following theorem provides the statistical properties of $\tilde{\bbeta}^{(1)}$.

\begin{theorem} \label{thm:one-step}
Assume Conditions~(C1)--(C3) hold, and let $\cE_0(r_0) $ and $\cE_*(r_*)$ be the events defined in \eqref{event0} for some $r_0 \gtrsim r_* >0$.
Let $x>0$, and suppose the bandwidths $b \geq h>0$ satisfy $\max\{  r_0 , \sqrt{(p+x)/n}  \, \}  \lesssim b \lesssim   1 $ and $\sqrt{(p+x)/N} \lesssim h$. Conditioned on the event $\cE_0(r_0) \cap  \cE_*(r_*)$, the one-step estimator $\wt \bbeta^{(1)}$ defined in \eqref{one.step.conquer} satisfies 
\#
	   \| \wt \bbeta^{(1)} - \bbeta^*  \|_{\Sigma} \lesssim \Bigg(    \sqrt{\frac{p+x}{n b}} +   \sqrt{\frac{p+x}{N h}}    + b \Bigg) \cdot r_0     +   r_* \label{one-step.error} 
\#
and
\#
 \| \,   \Hb  (  \wt \bbeta^{(1)} - \bbeta^* ) +  \nabla \hat \cQ_h(\bbeta^* )  \,\|_{\Omega} \lesssim  \Bigg(    \sqrt{\frac{p+x}{n b}} +   \sqrt{\frac{p+x}{N h}}    + b \Bigg) \cdot r_0   \label{one-step.bahadur}
\#
with probability at least $1-3e^{-x}$.
\end{theorem}

Equation \eqref{one-step.error} is the prediction error for the estimator obtained from running a single iteration of our proposed method, while equation \eqref{one-step.bahadur} provides bounds on a linear Bahadur representation of the estimator, used later for detailed statistical inference on $\bbeta^*$ or functionals thereof.

Before proceeding, we first discuss some implications of Theorem~\ref{thm:one-step}. The parameter $r_0$ captures the  convergence rate of the initial estimator $\wt \bbeta^{(0)}$. 
It can be constructed either on a single local machine that has access to $n$ observations or via averaging all the local estimators. The former is communication-free, while the latter usually improves the statistical accuracy at the cost of one round of communication.  
Therefore, we may expect a conservative convergence rate of the initial estimator, which is of order $\sqrt{p/n}$. In this case, the rate $r_0 \asymp \sqrt{p/n}$ is sub-optimal compared to that of the global QR estimator $\hat \bbeta$ in \eqref{empirical.QR} or the conquer estimator $\hat \bbeta^{\cq}$ in \eqref{conquer}.
Large sample properties of $\hat \bbeta^{\cq}$ have been examined by \cite{FGH2019} when $p$ is fixed, and by \cite{HPTZ2020} under the increasing-$p$ regime. According to the latter, the expected prediction error of $\hat \bbeta^{\cq}$, namely $\| \hat \bbeta^{\cq} - \bbeta^* \|_{\Sigma}$, is primarily determined by $\|  \nabla \hat \cQ_h(\bbeta^* ) \|_{\Omega}$ which is of order $\sqrt{p/N} + h^2$; see Lemma~\ref{lem:global.score} in the Appendix. Therefore, the second term on the right-hand side of \eqref{one-step.error} corresponds to the optimal   statistical rate, provided that $(p/N)^{1/2} \lesssim h\lesssim (p/N)^{1/4}$ when all the data are used. 
Turning to the first term, we see that with properly chosen bandwidths $b$ and $h$, say $b \asymp (p/n)^{1/3}$ and $h\asymp  (p/N)^{1/3}$, the one-step estimator $\wt \bbeta^{(1)}$ refines the statistical accuracy of $\wt \bbeta^{(0)}$ by a factor of order $(p/n)^{1/3}$.

We can repeat the one-step procedure in~\eqref{one.step.conquer} using $\wt \bbeta^{(1)}$ as an initial estimator, thereby obtaining $\wt \bbeta^{(2)}$. After $T$ iterations, we denote the resulting distributed QR estimator by $\wt \bbeta^{(T)}$. Since the statistical error is reduced by a factor of $(p/n)^{1/3}$, with high probability, at each iteration, we expect that after $\Omega\big(  \lceil \log(m)/\log(n/p) \rceil \big)$ iterations, the communication-efficient distributed estimator $\wt \bbeta^{(T)}$ will achieve the same convergence rate as the global estimator $\hat \bbeta$ or $\hat \bbeta^{\cq}$.

We formally describe the above iterative procedure as follows, starting at iteration 0 with an initial estimate $\wt \bbeta^{(0)}$.
At iteration $t = 1, 2, \ldots$, construct the shifted conquer loss function 
\#
	\wt  \cQ^{(t)} (\bbeta ) = \hat  \cQ_{1,b} ( \bbeta )  - \big\langle \nabla \hat  \cQ_{1,b} ( \wt \bbeta^{(t-1)} ) - \nabla \hat \cQ_h(\wt \bbeta^{(t-1)} ) , \bbeta \big\rangle , \label{surrogate.loss.l}
\#
yielding $\wt \bbeta^{(t)}$ that minimizes $\wt \cQ^{(t)} (\cdot)$, that is, 
\#
\wt \bbeta^{(t)} \in \argmin_{\bbeta \in \RR^p} \wt \cQ^{(t)} (\bbeta ).  \label{iterate.l}
\#
As before, $b \geq h>0$ are the local and global bandwidths, respectively. The details are described in Algorithm~\ref{algo:dqr}.
Notably, the shifted loss $\wt \cQ^{(t)}(\cdot)$ ($t\geq 1$) is twice-differentiable, convex and (provably) locally strongly convex.
To solve the shifted conquer loss minimization problem in \eqref{iterate.l}, in Section~\ref{sec:algorithm} of the Appendix, we describe a gradient descent (GD) algorithm modified by the application of a Barzilai-Borwein step \citep{BB1988}. Such a first-order algorithm is computationally scalable to large dimensions.

In reminiscence of the classical one-step estimator of \cite{B1975}, we may instead seek an approximate solution to the minimization problem \eqref{iterate.l} at each iteration by performing one step of Newton's method. At iteration $t$, $\nabla \wt \cQ^{(t)}(\wt \bbeta^{(t-1)}) =   \nabla \hat \cQ_h(\wt \bbeta^{(t-1)})$ and $\nabla^2 \wt \cQ^{(t)}(\wt \bbeta^{(t-1)}) = \nabla^2 \hat \cQ_{1,b}(\wt \bbeta^{(t-1)})$. Thus, starting with an initialization $\overbar{\bbeta}^{(0)}$, the Newton step computes the update $\overbar{\bbeta}^{(t)} =   \overbar \bbeta^{(t-1)} -  \{  \nabla^2 \hat \cQ_{1,b}(\overbar \bbeta^{(t-1)})   \}^{-1}   \nabla \hat \cQ_h(\overbar \bbeta^{(t-1)})$ for $t=1,2, \ldots$.
At each iteration, the above one-step update essentially performs a Newton-type step based on $\overbar \bbeta^{(t-1)}$. While computationally advantageous, the desirable statistical properties of this one-step estimator rely on uniform convergence of the sample Hessian, which typically requires stronger scaling with the sample size.

Theorem~\ref{thm:multi-step} below provides the statistical properties of the distributed conquer estimator $\wt \bbeta^{(T)}$, including high probability bounds on both estimation error and Bahadur linearization error. The latter serves as an intermediate step for establishing the asymptotic distribution of $\wt \bbeta^{(T)}$. Similar results can be obtained for $ \overbar{\bbeta}^{(T)}$. In fact, the analysis in this case is much simpler due to the closed-form expression, and is therefore omitted.

\begin{algorithm}[!t]
    \caption{ {\small Distributed Quantile Regression via Convolution Smoothing.}}
    \label{algo:dqr}
    \textbf{Input}: data batches $\{(y_i, \bx_i)\}_{i\in \cI_j}$, $j=1,\ldots, m$, stored on $m$ local machines, quantile level $\tau\in (0,1)$, bandwidths $b , h>0$, initialization $\wt{\bbeta}^{(0)}$, maximum number of iterations $T$, $g_0 = 1$.
    
    \begin{algorithmic}[1]
      \FOR{$t = 1, 2 \ldots, T$}
       \STATE  Broadcast $\wt \bbeta^{(t-1)}$ to all local machines.
			\FOR{$j=1,\ldots, m$}         
          \STATE  Compute $\nabla \hat \cQ_{j,h}(\wt \bbeta^{(t-1)})$ on the $j$th local machine, and send it to the master (first) machine.
          \ENDFOR
          \STATE Compute the global gradient $\nabla \hat \cQ_h(\wt \bbeta^{(t-1)}) = (1/m) \sum_{j=1}^m  \nabla \hat \cQ_{j,h}(\wt \bbeta^{(t-1)})$ and its $\ell_\infty$-norm $g_t = \|  \nabla \hat \cQ_h(\wt \bbeta^{(t-1)}) \|_\infty$ on the master.
       \STATE {\bf if} $g_t > g_{t-1}$ or $g_t < 10^{-5}$ {\bf break} 
       \STATE  {\bf otherwise} Compute $\nabla \hat \cQ_{1,b}(\wt \bbeta^{(t-1)})$, and solve $\wt \bbeta^{(t)} \in \argmin_{\bbeta \in \RR^p} \wt \cQ^{(t)}(\bbeta)$ on the master.
      \ENDFOR 
    \end{algorithmic}
      \textbf{Output}: $\wt \bbeta^{(T)}$.
\end{algorithm}

\begin{theorem} \label{thm:multi-step}
Assume the same set of conditions in Theorem~\ref{thm:one-step}.
Then, conditioned on $\cE_0(r_0) \cap \cE_*(r_*)$, the distributed estimator $\wt \bbeta^{(T)}$ with $T \gtrsim \log (r_0 / r_* ) / \log(1/b)$ satisfies
\begin{equation}
\begin{split}
 \| \wt \bbeta^{(T)} - \bbeta^*  \|_{\Sigma} &\lesssim r_*,      \vspace{0.2cm}\\
\| \,\Hb (  \wt \bbeta^{(T)}- \bbeta^*  )  +  \nabla \hat \cQ_h(\bbeta^*) \, \|_{\Omega}  &\lesssim   \big\{    \sqrt{(p+x)/(nb)} +   \sqrt{(p+x)/(Nh)}   + b \big\} \cdot  r_*, 
  \label{multi-step.bound}
\end{split}
\end{equation}
with probability at least $1- (2T+1) e^{-x}$.
\end{theorem}

\begin{remark} \label{rmk:tradeoff}
From the proof of Theorem~\ref{thm:multi-step}, we see that the multi-round estimate $\wt \bbeta^{(t)}$ after $t$ iterations satisfies with high probability that
$$
		 \| \wt \bbeta^{(t)} - \bbeta^* \|_2 \lesssim  \delta ^{t} \cdot r_0 + r_*  ~~\mbox{ with }~~ \delta =  \sqrt{\frac{p }{n b}} +   \sqrt{\frac{p }{N h}}    + b ,
$$ 
where $r_0$ and $r_*$ represent, respectively, the initial convergence rate and the global rate (attainable by the centralized estimator). This result also characterizes the trade-off between communication cost and estimation accuracy. After running the algorithm for $t$ rounds, the communication cost for each local machine/node is $ \cO(p t)$. On the other hand, since the statistical limit of distributed estimation is determined by $r_*$, we need as many as $\cO(\log(r_*/r_0)/\log(1/\delta))$ communication rounds for the proposed distributed estimator to achieve the optimal rate, resulting in a total communication cost  $\cO(p \log(r_*/r_0)/\log(1/\delta))$ for each local machine. Ignoring logarithmic factors, the above parameters $(r_0, r_*, \delta)$  will be taken as
$$
 r_0 \asymp \sqrt{\frac{p}{n}}, \quad r_* \asymp \sqrt{\frac{p}{N}}, ~\mbox{ and }~ \delta \asymp \bigg( \frac{p}{n} \bigg)^{1/3}.
$$
\end{remark}

Now let us discuss the construction of the initial estimator $\wt \bbeta^{(0)}$.
Using a local sample from a single source, we can take $\wt \bbeta^{(0)}$ to be either  the standard QR estimator \citep{KB1978} or the conquer estimator described in Section~\ref{sec:smoothing}. That is, 
\#
\wt \bbeta^{(0)} \in \argmin_{\bbeta \in \RR^p} \frac{1}{n} \sum_{i\in\cI_1} \rho_\tau(y_i - \bx_i^\T \bbeta)   ~~\mbox{ or }~~ 	 \wt \bbeta^{(0)} \in \argmin_{\bbeta \in \RR^p} \hat  \cQ_{1, b} ( \bbeta)  , \label{initial.estimator}
\#
where $b>0$ is the local bandwidth. 
In the diverging-$p$ regime, explicit high probability error bounds for the QR and conquer estimators can be found in \cite{PZ2020} and \cite{HPTZ2020}, respectively.

\begin{theorem} \label{thm:final.rate}
Assume that Conditions~(C1)--(C3) hold, and choose the bandwidths $b , h>0$ as $ b \asymp \{(p+   \log(n\log m) ) /n \}^{1/3}$ and $h \asymp \{ (p+  \log(n\log m))/N \}^{\gamma}$ for any $\gamma \in [1/3, 1/2]$. Moreover, suppose the sample size per source satisfies $n\gtrsim p+\log (n\log m)$. Then, starting at iteration 0 with an initial estimate $\wt \bbeta^{(0)}$ given in \eqref{initial.estimator}, the multi-round distributed estimator $\wt \bbeta = \wt \bbeta^{(T)}$ with $ T \asymp \lceil  \frac{\log(m)}{\log(n /p)}  \rceil$ satisfies 
\#
	\| \wt \bbeta  - \bbeta^*  \|_{\Sigma} \lesssim \sqrt{\frac{p+  \log(n \log m)}{N}}  \label{final.rate}
\#
and
\# \label{final.bahadur}
  \bigg\| \Hb( \wt \bbeta - \bbeta^*  )+  \frac{1}{N}\sum_{i=1}^N \bigl\{ \overbar K(-\varepsilon_i/ h ) - \tau \bigr\} \bx_i  \bigg\|_{\Omega}  \lesssim  \frac{  ( p+   \log( n \log m) )^{5/6}}{n^{1/3} N^{1/2} }   + \frac{p + \log(n \log m) }{N h^{1/2}}
\#
with probability at least $1- C n^{-1}$, where $m= N/n$ is the number of sources.
\end{theorem}

\begin{remark} \label{rmk:bandwidth}
Guided by the theoretically ``optimal" choice of the local and global bandwidths stated in Theorem~\ref{thm:final.rate}, in practice we suggest to choose
\#
	 b = c \cdot  \bigg(\frac{p+ \log n}{n} \bigg)^{1/3} ~~\mbox{ and }~~ h = c \cdot  \bigg( \frac{ p + \log N}{N} \bigg)^{1/3},  \label{bwd.choice}
\#
for some positive constant $c$. For preprocessed data that has constant-level scales, we may choose $c$ from $\{0.5, 1, 2.5, 5 \}$ using a validation set. More generally, we consider a heuristic, dynamic method for choosing $c$. To solve the optimization problem in \eqref{iterate.l}, Section~\ref{sec:algorithm} describes a quasi-Newton-type algorithm, namely the gradient descent with step size automatically determined by the Barzilai-Borwein method. At each iteration, we set $c$ in \eqref{bwd.choice} to be the minimum between the sample standard deviation and the median absolute deviation (multiplied by 1.4826) of the residuals from the previous iterate. The resulting estimate is then scale-invariant.
\end{remark}

\begin{figure}[!t]
	\begin{center}
		\subfigure[]{\includegraphics[scale=0.8]{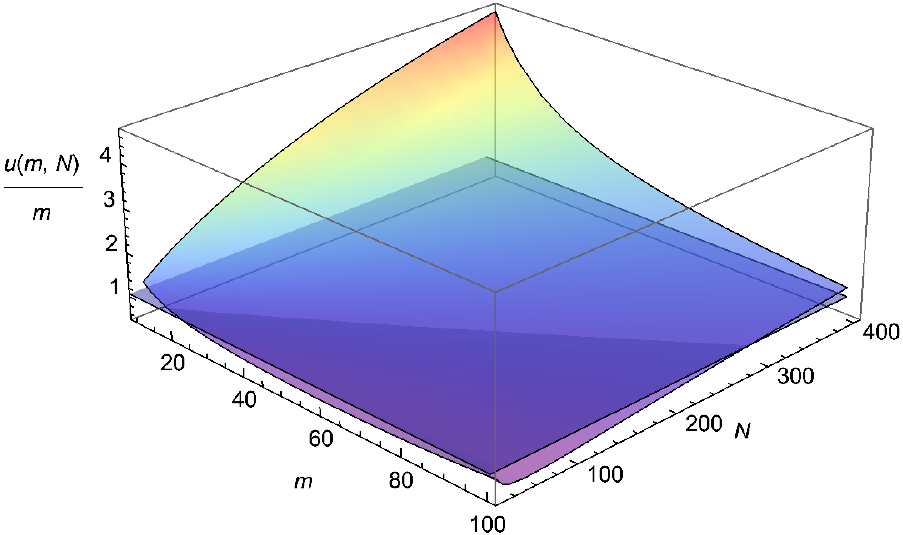}}\quad 
		\subfigure[]{\includegraphics[scale=0.8]{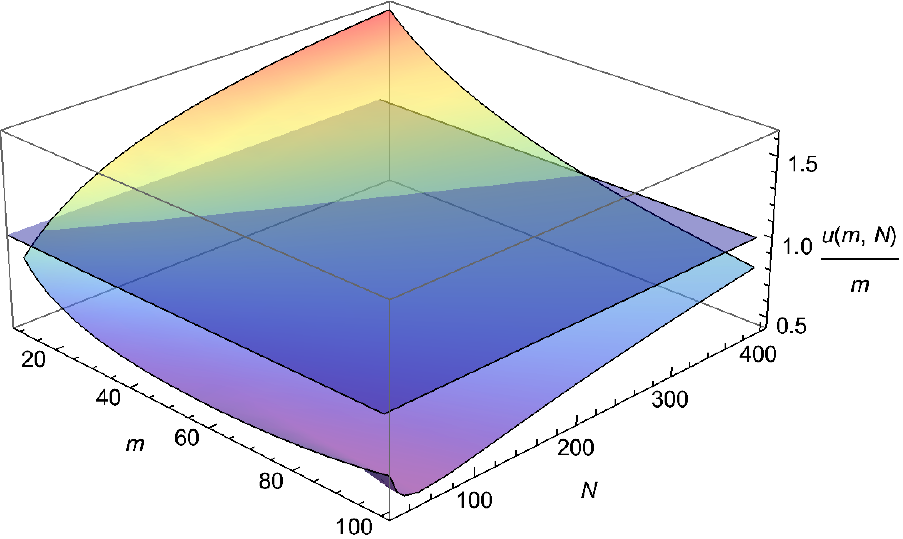}}\quad
	\end{center}
	\vspace{-0.5cm}
	\caption{\label{fig:fixedpoint} Plot of $u(m,N)/m$ against $m$ and $N$ overlaid with the constant function to indicate the fixed point of $u(m,N)$ for (a) $p=n/10$ and (b) $p=n/2$.}
\end{figure}

The scaling condition $n\gtrsim p+  \log(n \log m)$, while not as easy to parse as the $m \lesssim \sqrt{N}$ condition implicated by simple meta analyses, is appreciably less stringent. To visualize this constraint, we introduce the function
\[
u(m,N):=\frac{N}{p+\log(N/m)+\log(\log m)}
\]
so that the knife-edge permissible value of $m$ arises when $m=u(m,N)$. This fixed point equation is thus solved when $u(m,N)/m=1$. Figure \ref{fig:fixedpoint} plots $u(m,N)/m$ against $m$ and $N$ for $p=n/10$ and $p=n/2$. The permissible scaling of $m$ with $N$ is the curve traced out by the intersection of $u(m,N)/m$ with the constant function, taking value 1 for all values of the argument.
From Figure~\ref{fig:fixedpoint}, the permissible scaling of $m$ with $N$ is visibly faster than $\sqrt{N}$. By comparing Figures~\ref{fig:fixedpoint}(a) and (b), we see that this scaling is made more severe by proportional increases in $p$.

Using a local estimator as the initialization is most efficient in terms of storage, communication, and computational complexity.
Alternatively, one can use the so called divide-and-conquer (meta analysis) estimator based on simply averaging the local QR estimators \citep{VCC2019} as the initialization. This improves the statistical stability at the cost of one round of communication. For $j=1,\ldots, m$, define the local empirical loss functions $\hat \cQ_j (\bbeta)= (1/n) \sum_{i\in \cI_j} \rho_\tau(y_i - \bx_i^\T \bbeta)$, and the corresponding local QR estimators $\hat \bbeta^{\loc}_j \in \argmin_{\bbeta \in \RR^p} \hat \cQ_j(\bbeta)$. Estimators obtained from the separate sources are combined after one round of communication to construct a global estimator, namely the divide-and-conquer quantile regression (DC-QR) estimator
\#
	 \hat \bbeta^{\dc} = \frac{1}{m} \sum_{j=1}^m \hat \bbeta^{\loc}_j .  \label{dc.qr}
\#
For quantile regression, \cite{VCC2019} derived  the estimation error of $\hat \bbeta^{\dc}$ when the (random) covariates have fixed dimension $p$ and are uniformly bounded, that is, $\max_{1\leq i\leq n} \| \bx_i \|_2 \leq c_p$ for some $c_p>0$. Under regularity conditions that are similar to Condition~(C1), Theorem~3.1 therein implies
\#
	  \|   \hat \bbeta^{\dc} - \hat \bbeta   \|_2  = \cO_{\PP}  \Biggl( \frac{  \log N}{n}  + \frac{  (\log N)^{7/4}}{n^{1/4} N^{1/2} }  \Biggr) + o_{\PP}(N^{-1/2} ) \nn
\#  
as long as  $m = o( N / \log N)$. If communication constraints allow, we recommend using the DC-QR estimator $ \hat \bbeta^{\dc}$ as the initial estimator, and setting $T=  \max\{ \lceil \log m \rceil, 2\}$ in Algorithm~\ref{algo:dqr}. The whole procedure hence requires at most $T+1$ communication rounds.

In Section~\ref{subsec:inferencesim}, we demonstrate via numerical studies that the bias of the DC-QR estimator is visibly larger than the bias of the proposed distributed conquer estimator under extreme quantile regression models with heteroscedastic errors. As a result, confidence sets based on a normal approximation to the DC-QR Wald statistic are susceptible to severe undercoverage in linear heteroscedastic models.

\begin{remark}
Just as the statistical aspects of extreme value theory are challenged by the limitation of data beyond extreme thresholds, QR coefficients at extreme quantiles are notoriously hard to estimate. Section \ref{sec:extreme} of the Appendix details a minor adaptation of our procedure which improves its performance at extreme quantile levels.
\end{remark}

\subsection{Distributed inference}
 \label{subsec:inference}
 
 \subsubsection{Wald-type confidence sets}
\label{subsec:wald} 
With a view to more detailed statistical inference beyond point estimation, we first establish a distributional approximation in the form of a Berry-Esseen bound. This forms the basis for a Wald test, which can be inverted to give confidence sets for $\bbeta^*$ and linear functionals thereof. Construction of the pivotal test statistic relies on a consistent estimator of the asymptotic variance, which is typically obtained using a nonparametric estimate of the conditional density function of the response given the covariates.

\begin{theorem} \label{corollary:clt}
Under the same set of conditions in Theorem~\ref{thm:final.rate}, the distributed conquer estimator $\wt \bbeta = \wt \bbeta^{(T)}$ satisfies
 \#
 \sup_{x\in \RR, \, \ba \in \RR^p}   & \big| \PP\big\{   N^{1/2}   \ba^\T(\wt \bbeta  - \bbeta^*)   /\sigma_{\tau, h}  \leq x\big\} - \Phi(x) \big| \nn \\
 & \lesssim \frac{p+  \log(n \log m)}{(N h)^{1/2}}  + N^{1/2} h^2 +  \frac{  ( p+  \log(n \log m) )^{5/6}}{n^{1/3}  }  ,
 \#
 where $\sigma_{\tau, h}^2 =  \ba^\T \Hb^{-1}  \EE[ \{ \overbar K(-\varepsilon/h) - \tau\}^2 \bx \bx^\T ]  \Hb^{-1}\ba$ and $\Phi(\cdot)$ is the standard normal distribution function.
 In particular, under the scaling $p + \log(\log m) = o(\min \{  n^{2/5} , N^{3/8} \} )$, the distributed estimator $\wt \bbeta $ with bandwidths $b \asymp \{(p  + \log(n \log m) ) /n \}^{1/3}$ and $h \asymp \{(p+   \log(n \log m) ) / N \}^{2/5}$ satisfies
 \#
N^{1/2}  \sigma_{\tau, h}^{-1} \, \ba^\T(\wt \bbeta  - \bbeta^*) \xrightarrow {{\rm d}}  \cN(0,1) ~~\mbox{ and }~~  \frac{ N^{1/2} \ba^\T(\wt \bbeta  - \bbeta^*) }{   ( \ba^\T \Hb^{-1} \Sigma \Hb^{-1} \ba )^{1/2}  }  \xrightarrow {{\rm d}}  \cN\bigl( 0,\tau(1-\tau ) \bigr)    \nn
 \#
uniformly over $\ba \in \RR^p$ as $n\to \infty$, where $\xrightarrow {{\rm d}}$ is a shorthand for convergence in distribution.
\end{theorem}

The accuracy of the normal approximation hinges on both the global and local bandwidths, and on the scaling of $m$ with $N$ and $p$. The role of $b$ is via \eqref{multi-step.bound}, in view of which, the upper bound in Theorem \ref{corollary:clt} is of order 
$$ (p+x)^{1/2}  \Bigg(  \sqrt{\frac{p+x}{nb}} + b + \sqrt{\frac{p+x}{N h}} \,\Bigg) + N^{1/2} h^2,$$
where $x =  \log(n\log m)$. Minimizing as a function of $(h,b)$ delivers the rate in Theorem \ref{corollary:clt} by taking
$$
 b \asymp \bigg( \frac{p+x}{n} \bigg)^{1/3} ~~\mbox{ and }~~  h \asymp \bigg( \frac{p+x}{N} \bigg)^{2/5}.
$$
To our knowledge, Theorem \ref{corollary:clt} is the first Berry-Esseen inequality with explicit error bounds depending on both $n$ and $p$ in a distributed setting.
 
 We first describe methods that use the normal distribution with estimated variance for calibration. Let $\wt \bbeta = \wt \bbeta^{(T)}$ be the communication-efficient estimator discussed in the previous subsection. Under mild conditions, Theorem~\ref{corollary:clt} establishes the asymptotic normality that for every $1\leq j\leq p$,
\#
  \frac{  N^{1/2} (\wt \beta_j - \beta^*_j )  }{   (\Hb^{-1} \Sigma \Hb^{-1})_{jj}^{1/2} }  \xrightarrow {{\rm d}} \cN\bigl( 0,\tau(1-\tau ) \bigr)  ,  \nn
\#
where $\Hb = \EE \{ f_{\varepsilon |\bx} (0) \bx \bx^\T \}$ and $\Sigma= \EE(\bx \bx^\T)$. The problem is then reduced to estimating the pointwise variance $ (\Hb^{-1} \Sigma \Hb^{-1})_{jj}$, i.e., the $j^{{\rm th}}$ diagonal entry of $\Hb^{-1} \Sigma \Hb^{-1}$.

To this end, define the residual function and fitted residuals as
\#
   \varepsilon_i(\bbeta) = y_i - \bx_i^\T \bbeta ~\mbox{ for }~ \bbeta \in \RR^p ~\mbox{ and }~ \hat  \varepsilon_i =  \varepsilon_i(\wt \bbeta) = y_i - \bx_i^\T \wt \bbeta , \ \ i=1,\ldots, N. \label{fitted.residuals}
\#
In a nondistributed setting, the $p \times p$ Hessian matrix $\Hb$ can be estimated by the following variant of Powell's kernel-type estimator \citep{P1991}
\#
\hat \Hb_b = \frac{1}{ m } \sum_{j=1}^m  \hat \Hb_{j,b}  ~\mbox{ with }~   \hat \Hb_{j,b}  = \frac{1}{n b} \sum_{i\in \cI_j} \phi\bigl( \hat \varepsilon_i  / b\bigr) \bx_i \bx_i^\T ,  ~ j =1,\ldots, m,  \label{kernel.matrix.est1}
\#
where $\phi(\cdot)$ is the standard normal density function and $b  >0$ is a bandwidth that may differ from the previous one. 
Moreover, define $\hat \Sigma = (1/m) \sum_{j=1}^m \hat \Sigma_j$ and $\hat \Sigma_b(\tau) = (1/m) \sum_{j=1}^m \hat \Sigma_{j,b}(\tau)$, where
\#
	\hat \Sigma_j = \frac{1}{n} \sum_{i\in \cI_j}  \bx_i \bx_i^\T~~\mbox{ and }~~  \hat   \Sigma_{j ,b}(\tau)   = \frac{1}{n} \sum_{i\in \cI_j}  \bigl\{  \overbar K(-\hat \varepsilon_i / b  )  -\tau  \bigr\}^2 \bx_i \bx_i^\T . \label{kernel.matrix.est2}
\#
Computing the full matrix estimators $\hat \Hb_b$ and $\hat \Sigma$ or $\hat \Sigma_b(\tau)$ requires each machine  to communicate  $p\times p$ local estimators $\hat \Hb_{j,b} $ and $\hat \Sigma_j$ or $\hat \Sigma_{j,b}(\tau)$ to the master machine. This incurs excessive communication cost.  

To achieve a trade-off between communication efficiency and statistical accuracy, we instead use a local pointwise variance estimator 
\#
   \tau (1-\tau )  \, \bigl( \hat \Hb_{1,b} ^{-1} \hat \Sigma_1 \hat \Hb_{1,b} ^{-1}  \bigr)_{jj}   ~~\mbox{ or }~~
 \bigl(  \hat \Hb_{1,b}^{-1} \hat \Sigma_{1,b} (\tau)  \hat \Hb_{1,b}^{-1}  \bigr)_{jj}   . \label{var.est}
\#
For the latter, note that $\hat \Sigma_{1,b} (\tau)$ can be viewed as a sample analog of $\EE \{ \overbar K(-\varepsilon / b ) - \tau\}^2 \bx \bx^\T$, which is closely related to the asymptotic variance of $\wt \bbeta$ as revealed by Theorem~\ref{corollary:clt}. Moreover, as discussed in \cite{FGH2019}, the width of a confidence interval based on $\hat \Hb_{1,b}^{-1} \hat \Sigma_{1,b} (\tau)  \hat \Hb_{1,b}^{-1} $ for any element of $\bbeta^*$ is asymptotically narrower than that  based on the na{\"i}ve variance estimator $ \tau (1-\tau )  \,  \hat \Hb_{1,b} ^{-1} \hat \Sigma_1 \hat \Hb_{1,b} ^{-1} $.

For every $\bbeta \in \RR^p$, define the matrix-valued function 
$$
	\hat \Hb_{1,b}(\bbeta) =  \frac{1}{nb} \sum_{i\in \cI_1} \phi\bigl( \varepsilon_i(\bbeta) / b \bigr) \bx_i \bx_i^\T , 
$$
where $\varepsilon_i(\bbeta) = y_i - \bx_i^\T \bbeta$ are as in \eqref{fitted.residuals}.
Under this notation, $\hat \Hb_{1,b} = \hat \Hb_{1,b}(\wt \bbeta )$. The next result provides a uniform convergence result for $ \hat \Hb_{1,b}(\bbeta) $ over $\bbeta$ in a local neighborhood of $\bbeta^*$.
For any symmetric matrix $\Ab \in \RR^{p\times p}$, we use $\| \cdot \|_{\Omega}$ ($\Omega=\Sigma^{-1}$) to denote the relative operator norm, that is, $\| \Ab \|_{\Omega} = \| \Sigma^{-1/2} \Ab \Sigma^{-1/2}\|_2$. With this notation, we have $\| \Sigma\|_{\Omega} = 1$. 

\begin{proposition} \label{prop:covariance.estimation}
Conditions~(C1)--(C3) ensure that, for any $r, x>0$,
\#
	\sup_{\bbeta \in \Theta(r)}  \|   \hat \Hb_{1,b}(\bbeta) - \Hb  \|_{\Omega}
	\lesssim \sqrt{\frac{p\log n + x}{nb }}  +  b + r 
\#
with probability at least $1-3 e^{-x}$ as long as $n\gtrsim p+x$ and $b\gtrsim (p\log n+ x)/n$. In addition, if $f_{\varepsilon | \bx}'(\cdot)$ is Lipschitz continuous, that is, $|f_{\varepsilon | \bx}'(u)- f_{\varepsilon | \bx}'(v) | \leq l_1 |u-v|$ for all $u,v\in \RR$ almost surely (over $\bx$), then $\sup_{\bbeta \in \Theta(r)}   \|  \hat \Hb_{1,b}(\bbeta) - \Hb    \|_{\Omega} \lesssim \sqrt{(p\log n + x)/(nb)} + b^2 + r$ with high probability.
\end{proposition}

\begin{theorem} \label{thm:variance.consistency}
Under the same set of assumptions in Theorem~\ref{thm:final.rate}, the local estimators $\hat \Sigma_1$ and $\hat \Hb_{1,b}$ with $b\asymp \{  p \log(n)/n \}^{1/3}$ satisfy the bounds
\#
  \|   \hat \Sigma_1 -  \Sigma   \|_{\Omega}  & \lesssim    (p+\log n)^{1/2} n^{-1/2} ~~\mbox{ and }~~ 	  \|   \hat \Hb_{1,b} - \Hb      \|_{\Omega}    \lesssim  ( p \log n)^{1/3}   n^{-1/3} \nn
\#
with probability at least $1-C n^{-1}$ as long as $n\gtrsim p \log n$.
\end{theorem}

 In a simpler case where $f_{\varepsilon | \bx} (0)= f_\varepsilon(0)$ is independent of $\bx$, we have  $\Hb^{-1} \Sigma \Hb^{-1} = \{  f_\varepsilon(0) \}^{-2} {\Omega}$  so that it suffices to estimate the univariate density function $f_{\varepsilon | \bx}(\cdot)$ at $0$.
Arguably the most commonly used method is the following kernel density estimator: 
\#
 	\hat f_\varepsilon (0)  = \frac{1}{N b} \sum_{i=1}^N   K\bigl(  \hat \varepsilon_i / b\bigr) = \frac{1}{m} \sum_{j=1}^m \hat f_{  \varepsilon, j} (0)  , \label{KDE}
\#
where $\hat f_{\varepsilon , j} (0) =  (nb)^{-1} \sum_{i \in \cI_j} K ( \hat \varepsilon_i / b )$ for $j=1,\ldots, m$. Therefore, $\hat f_\varepsilon(0)$ can be easily computed in a distributed manner. For convenience, we use the standard normal density as kernel function and the rule-of-thumb bandwidth by \cite{HS1988}, that is,
$$
	b_N^{{\rm rot}} = N^{-1/3}\cdot  \Phi^{-1}(1-\alpha/2)^{2/3} \Biggl\{  \frac{1.5 \cdot  \phi(\Phi^{-1}(\tau))^2 }{2 \Phi^{-1}(\tau)^2 + 1}\Biggr\}^{1/3} ,
$$
where $\alpha$ is a prepecified probability of miscoverage.  For the kernel matrix estimators $\hat \Hb_{1,b}$ and $\hat \Sigma_{1,b}(\tau)$ defined in \eqref{kernel.matrix.est1} and \eqref{kernel.matrix.est2}, we use the same local bandwidth $b$ as in Algorithm~\ref{algo:dqr} for efficient distributed quantile regression. 
The corresponding normal-based confidence intervals for $\beta^*_j$ ($j=1,\ldots,p$) are given by
\#
  \left[ \wt \beta_j - \Phi^{-1}(1-\alpha/2) \cdot \hat \sigma_j \cdot  N^{-1/2}    , ~~\wt \beta_j + \Phi^{-1}(1-\alpha/2) \cdot  \hat \sigma_j \cdot  N^{-1/2}  \, \right] ,  \label{normal.CI}
\#
where $\hat \sigma_j  = (  \hat \Hb_{1,b}^{-1} \hat \Sigma_{1,b}(\tau) \hat \Hb_{1,b }^{-1} )^{1/2}_{jj}$, $ \sqrt{\tau (1-\tau )}  \, ( \hat \Hb_{1,b} ^{-1} \hat \Sigma_1 \hat \Hb_{1,b} ^{-1}    )^{1/2}_{jj}$ or $ \hat f_\varepsilon(0)^{-1}  ( \hat \Sigma_1 )_{jj}^{1/2}\sqrt{ \tau (1-\tau) } $. The first two variance estimates are preferred under general heteroscedastic  models in which $\Hb= \EE\{ f_{\varepsilon | \bx}(0) \bx \bx^\T \}$ no longer takes the form $f_\varepsilon(0) \cdot  \Sigma$.

\begin{remark}
The construction of normal-based confidence intervals as in \eqref{normal.CI} depends crucially on the asymptotic variance estimation. The validity of $\hat \sigma_j = \hat f_\varepsilon(0)^{-1}  ( \hat \Sigma_1 )_{jj}^{1/2}\sqrt{ \tau (1-\tau) }$  relies on the assumption that  $\Hb= \EE\{ f_{\varepsilon | \bx}(0) \bx \bx^\T \}$ takes the form $f_\varepsilon(0) \cdot  \Sigma$.
This holds trivially when the model error $\varepsilon$ and covariates $\bx$ are independent, which is arguably too restrictive in the context of quantile regression. More generally, let us consider a standard location-scale model  $y = \bx^\T \bbeta^* + \sigma(\bx) \cdot e$, where $e \sim f_e(\cdot)$ is independent of $\bx$ and $\sigma(\cdot)$ is a non-negative function. In this case, we have $\varepsilon = \sigma(\bx) \cdot e$, whose conditional and unconditional densities at 0 are $f_{\varepsilon | \bx}(0) =  f_e(0)  / \sigma(\bx)$ and $f_{\varepsilon }(0) =   f_e(0) \cdot \EE  \{ 1/ \sigma(\bx)\}$. This reveals that $\Hb = f_e(0) \cdot   \EE \{  \bx \bx^\T /\sigma(\bx)\}$ and $f_\varepsilon(0) \cdot \Sigma$ are generally unequal, and therefore the use of $\hat \sigma_j = (  \hat \Hb_{1,b}^{-1} \hat \Sigma_{1,b}(\tau) \hat \Hb_{1,b}^{-1} )_{jj}^{1/2}$ or $  \hat \sigma_j =  \sqrt{\tau (1-\tau )}  \, ( \hat \Hb_{1,b} ^{-1} \hat \Sigma_1 \hat \Hb_{1,b} ^{-1}  )^{1/2}_{jj}$. is more robust and preferable under heteroscedastic  models.
\end{remark}

 \subsubsection{Score-type confidence sets}\label{sec:score}
 
 While the Wald test inverts to give explicit confidence intervals as in equation \eqref{normal.CI}, confidence sets based on other types of test acknowledge that the set of parameter values consistent with the data need not form an interval.

 
 For some $k=1,\ldots,p$, consider the hypothesis
 \#
 H^k_0 : \beta^*_k =  c_k  ~~\mbox{ versus }~~ H^k_1  : \beta^*_k \neq c_k,
 \# 
 where $c_k$ is a predetermined constant. Let $\wt{\bbeta}_{H^k} =(\wt \beta_{H^k,1},\ldots, \wt \beta_{H^k,p} )^\T \in \RR^p$ denote the distributed quantile regression estimator with its $k$th coordinate constrained at the hypothesized value, i.e., $\wt \beta_{H^k,k}=c_k$. To construct a score test, define the   gradient 
 \#
 \hat \bS = (\hat S_1, \ldots, \hat S_p)^\T =  N \cdot  \nabla \hat \cQ_h(\wt \bbeta_{H^k})  =  \sum_{j=1}^m \sum_{i\in \cI_j}    \hat \xi_i  \bx_i, 
 \#
 where $\hat \xi_i  = \overbar K   \{ ( \bx_i^\T \wt \bbeta_{H^k} - y_i )/h  \} -\tau $.  Under the null hypothesis $H^k_0$, it is reasonable to expect the $t$-statistic $\hat T_k$, which is defined as $N^{1/2}\hat S_k$ divided by the estimated standard deviation, to be asymptotically normally distributed. We can write the $t$-statistic in terms of the self-normalized sum $\hat T_k$ as \citep{Efron1969}:
 \#
 \hat T_k = \frac{ \hat S_k /\hat V_k }{\sqrt{ \{N-(\hat S_k /\hat V_k )^2\}/(N-1) }},
 \#
 where  $ \hat S_k  = 	\sum_{j=1}^m \sum_{i\in \cI_j}  \hat \xi_i x_{ik}$ and $\hat V_k^2 =  	\sum_{j=1}^m \sum_{i\in \cI_j}  ( \hat \xi_i  x_{ik} )^2$. This representation has the advantage that the quantities $ \hat S_k$ and $\hat V_k$ can be calculated in a distributed manner without information loss.

 Write $\xi_i = \overbar K(-\varepsilon_i/h) - \tau$ and $\mu_k = \EE(\xi_i x_{ik})$, where $\varepsilon_i = y_i-\bx_i^\T \bbeta^*$. Denote the ``oracle" version of $\hat T_k$ by $T_k$:
 \$
 T_k =  \frac{  N^{-1/2}\sum_{i=1}^N  ( \xi_i x_{ik} - \mu_k)}{\sqrt{(N-1)^{-1} \sum_{i=1}^N   ( \xi_i x_{ik}  - N^{-1}  \sum_{\ell=1}^N  \xi_\ell x_{\ell k}  )^2   }} = \frac{    S_k / V_k }{\sqrt{ \{N-(  S_k /  V_k )^2\}/(N-1)  }},
 \$
 where $S_k =\sum_{i=1}^N ( \xi_i x_{ik} -\mu_k)$ and $V_k^2 =\sum_{i=1}^N ( \xi_i x_{ik} -\mu_k)^2$. Note that $S_k$ is a sum of independent zero-mean random variables. Asymptotic properties of the self-normalized sum $S_k/V_k$ have been well established in the literature \citep{DLS2009}. On writing $\hat{T}_k$ more explicitly as $\hat{T}_k(c_k)$, we define the $\alpha$-level confidence set associated with the score test as
 \#
 \bigl\{ \, c_k: \Phi^{-1}(\alpha/2) \leq \hat{T}_k(c_k) \leq \Phi^{-1}(1-\alpha/2)\, \bigr\}. \label{eq:scoreset}
 \#
 This will often, but need not always, deliver intervals. The  possibility of non-interval confidence sets should be viewed as an advantage, as exemplified by Fieller's problem \citep{Fieller1954}. The disadvantage of using the score statistic for constructing confidence sets is that $\hat{T}_{k}(c_k)$ has to be evaluated for a multitude of $c_k$ values, in practice over a fine grid of points. The computational burden of this is considerable relative to the Wald construction in Section~\ref{subsec:wald}.

 \subsubsection{Resampling-based confidence sets}\label{sec:bootstrap}
 
 An alternative widely used approach treats an interval as the primary mode of inference rather than the significance test and constructs the former directly by resampling methods such as the bootstrap. Resampling approaches typically provide tighter confidence limits than the Wald-based interval due to their implicit higher-order accuracy over limiting distributional approximations. However, the computational burden is high in the present context.

Recall from Theorem~\ref{thm:final.rate} that the multi-round distributed estimator $\wt \bbeta = (\wt \beta_1, \ldots, \wt \beta_p)^\T$ admits the following asymptotic linear (Bahadur) representation:
\#
     N^{1/2} ( \wt \bbeta - \bbeta^* )  =  - \Hb^{-1} \frac{1}{\sqrt{N}} \sum_{i=1}^N  \bigl\{  \overbar K(-\varepsilon_i / h ) -  \tau   \bigr\} \bx_i  + o_{\PP}(1)   . \nn
\#
Motivated by this asymptotic representation, \cite{BCFH2017} suggested and proved the validity of the {\it multiplier score bootstrap}, which is based on randomly perturbing the asymptotic linear forms of the nonlinear quantile regression estimators. Intuitively, the distribution of $ N^{1/2} ( \wt \bbeta - \bbeta^* )$ can be approximately estimated by the bootstrap draw of 
\#
N^{1/2} ( \wt \bbeta^\flat -  \wt \bbeta  ) :=  -  \hat \Hb^{-1} \frac{1}{\sqrt{N}} \sum_{i=1}^N  e_i  \bigl\{  \overbar K(- \hat \varepsilon_i / h ) -  \tau   \bigr\} \bx_i   ,    
\#
where $e_1,\ldots, e_N$ are i.i.d.~standard normal random variables, $\hat \Hb$ denotes a generic (consistent) estimator of $\Hb$, and $\hat \varepsilon_i$ are fitted residuals. In the distributed framework, each bootstrap draw requires one round of communication. The  composite communication cost can be exorbitant when the number of bootstrap replications is large, say 1000 or 2000.

Recently, \cite{YCC2020} proposed two bootstrap methods for constructing simultaneous confidence intervals with distributed data. 
To operationalize their proposals in the present context, define $\hat \bxi_i =  \{  \overbar K(- \hat \varepsilon_i / h ) -  \tau   \} \bx_i$ for  $i=1,\ldots, N$, and let $\hat \Hb_1$ be a local estimator of $\Hb$ using the $n$ samples on the first machine. For example, $\hat \Hb_1$ can be taken as either $ \hat \Hb_{1,b}$ given in \eqref{kernel.matrix.est1} or $\hat f_\varepsilon(0) \cdot \hat \Sigma_1$. 
Then, consider the following two multiplier bootstrap statistics 
\#
\bw^\sharp = (w^\sharp_1, \ldots, w^\sharp_p )^\T  =  - \hat \Hb_{1 }^{-1}  \frac{1}{\sqrt{m}} \sum_{j=1}^m e_{j} \cdot n^{1/2} \nabla \hat \cQ_{j,h}(\wt \bbeta)    \label{boot1}
\#
and
\#
	\bw^\flat = (w^\flat_1, \ldots, w^\flat_p)^\T  =  - \hat \Hb_{1 }^{-1} \frac{1}{\sqrt{n+m-1}}  \Bigg\{  \sn e_i  \cdot \hat \bxi_i + \sum_{j=2}^m e_{n+j-1}  \cdot n^{1/2}   \nabla \hat \cQ_{j,h}( \wt \bbeta)  \Bigg\} , \label{boot2}
\#
both of which only require one additional round of communication, and therefore are communication-efficient.
As before, $e_1, \ldots, e_{n+m-1}$ are i.i.d.~standard normal variables. 

For  any $q \in (0,1)$ and $1\leq j\leq p$, let $\cc_j^\sharp(q)$ and $\cc_j^\flat(q)$ be the (conditional) $q$-quantiles of $w^\sharp_j$ and $w_j^\flat$, respectively, defined as $ \cc_j^\sharp(q)= \inf \{ t\in \RR: \PP^*( w_j^\sharp \leq t ) \geq q   \}$ and $\cc^\flat(q)= \inf \{ t\in \RR: \PP^*( w_j^\flat \leq t ) \geq q   \}$, where $\PP^*(\cdot) = \PP(\cdot \, |\, y_1, \bx_1, \ldots, y_N,  \bx_N)$ denotes the conditional probability given the observed samples.
The ensuing bootstrap confidence intervals for $\beta^*_j$ ($j=1,\ldots, p$) are given by 
\#
	\Bigg[ \wt \beta_j -   \frac{\cc_j^\sharp(1-\alpha/2)}{\sqrt{N}} , \, \wt \beta_j -  \frac{\cc_j^\sharp( \alpha/2) }{\sqrt{N}}  \Bigg]  ~\mbox{ and }~ 	\Bigg[ \wt \beta_j -  \frac{ \cc_j^\flat(1-\alpha/2)}{\sqrt{N}} , \, \wt \beta_j -  \frac{ \cc_j^\flat( \alpha/2)}{\sqrt{N}} \Bigg] . \label{boot.CIs}
\#
Our simulations in Section~\ref{subsec:inferencesim} show that the two bootstrap methods have nearly identical performance when $m$ is large, while the latter is more stable and thus preferable when $m$ is relatively small. We leave the theoretical analysis of these distributed bootstrap methods in the future as a significant amount of additional work is still needed.

\subsection{Comparison with prior work}
\label{sec:comparison}
The problem of distributed quantile regression has been considered in two earlier papers. \cite{VCC2019} established the statistical properties of the estimator obtained by averaging $m$ local estimators, each constructed according to equation \eqref{empirical.QR}. The single round of communication and direct use of the check function means that there are no tuning parameters. However, as indicated in Section \ref{sec:1}, the permissible scaling of $m$ with $N$ required to ensure the optimal statistical properties is restrictive, and violation of this constraint leads to under-coverage of resulting confidence sets. Under-coverage is particularly severe under the highly plausible scenario in which the quantile regression error depends on the covariates. See Section \ref{sec:simulation} for an empirical demonstration.

For $M$-estimation with a convex loss, \cite{CLZ2018} proposed a general multi-round distributed procedure paired with stochastic gradient descent. When applied to quantile regression, their approach is a variant of stochastic subgradient descent. For minimizing a convex but non-differentiable function, subgradient methods typically exhibit very slow (sublinear) convergence and hence are not computationally stable. This explains the unpopularity of subgradient approaches among other computational methods for quantile regression. Theoretically, their distributed QR estimator needs a sufficiently large local sample size---namely $n\gtrsim (N p)^{1/2} \log(N)$, to achieve the optimal rate $\cO_{\PP}(\sqrt{p/N})$; see Theorem~4.7 therein. In addition to the suboptimal scaling, \cite{CLZ2018} only derived the convergence rate for point estimation without the uncertainty quantification sought in the present work.

The same authors \citep{CLZ2019} proposed a procedure specifically for distributed quantile regression. Their smoothed loss function is closer to that of \cite{H1998} and incompatible with the ideas of \cite{JLY2018} and \cite{WKSZ2017} due to violation of the uniform Lipschitz continuity condition of the second derivative. They instead exploit a representation of the estimator in terms of estimator-dependent ``sufficient statistics''. Since the representation is not of closed form, an iterative approach is required, using the estimate at iteration $t$ to update the sufficient statistics at each component source. While the limited communication improves the permissible scaling of $m$ with $N$ over the approach of \cite{VCC2019}, the construction is such that $m$ $p\times p$ matrices (and other quantities), are communicated at each iteration. Communication of hessian matrices  is generally viewed as too communication intensive, particularly when $p$ is large.

We note that none of these approaches is generalizable to the sparse high-dimensional setting. The penalization required to enforce sparsity in high dimensions exacerbates bias so that meta-analysis hinges of the ability to de-bias such estimators prior to aggregation. Attempts to construct de-biased estimators for quantile regression have, so far, relied on an unrealistic assumption that the quantile regression error is independent of the covariates. The key representation used by \cite{CLZ2019} is violated upon penalization of their smooth quantile regression estimator, and suffers from singularity when $p>n$ if penalization is not applied.

An anonymous reviewer pointed out concurrent work by \cite{JY2021} (the first version of our manuscript dates back to late September, 2020) who also proposed communication-efficient algorithms for distributed quantile regression  by means of convolution smoothing. The main difference between this work and ours concerns the theoretical aspects. Under similar regularity and moment conditions, we provide explicit non-asymptotic concentration bounds as well as Berry-Esseen-type bounds for normal approximation. These results complement the conventional $\cO_{\PP}$ statements in \cite{JY2021}.
To achieve the global convergence rate in low-dimensions, Theorem~3.2 in \cite{JY2021} requires $(p, n , N)$ to satisfy $n = N^r$ and $p \asymp N^c$ for some $0< r\leq 1$ and $0< c < \min(3/8, r)$, while our result (Theorem~\ref{thm:final.rate}) only requires $n\gtrsim p + \log\log(N/n)$. In high-dimensions, from Theorem~4.1 in \cite{JY2021} and its proof we see that the dimension $p$ cannot exceed the sample size $N$ in the sense that $p\asymp N^c$ for some $c\in (0, 1)$. Our results, detailed in Section~\ref{sec:hd}, show that the penalized distributed QR estimator achieves the global rate under the sample size requirements $n\gtrsim  s^2 \log p$ and $N\gtrsim s^3 \log p$, which considerably relax those in \cite{JY2021}.

\section{Distributed Penalized Quantile Regression in High Dimensions}
\label{sec:hd}

In this section, we consider quantile regression in high-dimensional sparse models with distributed data. 
In such models, the total number of predictors $p$ can be very large, while the number of important predictors is significantly smaller. 
As before, assume that the data set $\{ (y_i, \bx_i)\}_{i=1}^N$ with $N=n\cdot m$ is distributed across $m$ sources, so that each source $j$ contributes $n$ i.i.d. observations $\cD_j = \{ (y_i, x_i) \}_{i\in \cI_j}$ indexed by $\cI_j$.
Assume further that the sparsity $\| \bbeta^* \|_0 := \sum_{j=1}^p \mathbbm{1}(\beta^*_j \neq 0)$ is at most $s$, which is much smaller than the local sample size, that is, $s=o(n)$.

\subsection{Penalized conquer with distributed data}

To fit sparse models in high dimensions, the use of $\ell_1$ penalization has become a common practice since the seminal work of \cite{Tib1996}. The $\ell_1$-penalized quantile regression ($\ell_1$-QR) estimator is defined as
\#
	\hat \bbeta \in \argmin_{\bbeta \in \RR^p }  \, \frac{1}{N} \sum_{i=1}^N \rho_\tau ( y_i - \bx_i^\T \bbeta) + \lambda \cdot \| \bbeta \|_1  =  \argmin_{\bbeta \in \RR^p }   \,\hat \cQ(\bbeta) + \lambda \cdot \| \bbeta \|_1 , \label{lasso-qr}
\#
where $\lambda>0$ is a regularization parameter.
Statistical  properties and computational methods for $\ell_1$-QR have been  well studied in the past decade; see, for example, \cite{WLJ2007}, \cite{WL2008}, \cite{LZ2008}, \cite{BC2011}, \cite{WWL2012}, \cite{YH2016} and \cite{Gu2018}.
Recently, \cite{TWZ2020} studied the $\ell_1$-penalized conquer ($\ell_1$-conquer) estimator, which is a solution to the following optimization problem
\#
 	\min_{\bbeta \in \RR^p }  \,  \underbrace{ \frac{1}{N} \sum_{i=1}^N (\rho_\tau * K_h) ( y_i - \bx_i^\T \bbeta) }_{\hat \cQ_h(\bbeta) }+ \, \lambda \cdot \| \bbeta \|_1 ,  \label{l1.conquer}
\#	
where $K(\cdot)$ is a non-negative kernel and $h>0$ is the bandwidth. 

Notably, the smoothed loss function $\hat \cQ_h(\cdot)$ is (provably) strongly convex in a local neighborhood of $\bbeta^*$ with high probability. With a proper initialization, the corresponding optimization problem with $\ell_1$-penalization can be efficiently solved via first-order algorithms.
In a distributed setting, we extend the iterative algorithm in Section~\ref{sec2} as follows.
Let $\wt \bbeta^{(0)} \in \RR^p$ be an initial regularized estimator. Denote by $\wt \cQ(\bbeta) = \hat  \cQ_{1,b}(\bbeta)  - \langle \nabla \hat  \cQ_{1,b}(\wt \bbeta^{(0)}) - \nabla \hat  \cQ_h(\wt \bbeta^{(0)}), \bbeta \rangle$ the same shifted conquer loss as in \eqref{surrogate.loss}, where $b$ and $h$ are the local and global bandwidths. Analogously to \eqref{one.step.conquer}, the communication-efficient penalized conquer estimator is defined as
\begin{equation}
\label{one.step.l1.conquer}
	 \wt \bbeta^{(1)}    \in \argmin_{\bbeta \in \RR^p} \, \wt \cQ(\bbeta) + \lambda \cdot  \| \bbeta \|_1  , 
\end{equation}
where $\lambda>0$ is a regularization parameter.  Optimization problem~\eqref{one.step.l1.conquer} is convex, which we solve using a local adaptive majorize-minimize algorithm detailed in Section~\ref{sec:algorithmhighd} of the Appendix.

Let $\cS\subseteq \{1,\ldots, p\}$ be the support of $\bbeta^*$, and assume that data are generated from a sparse conditional quantile model \eqref{qr.model} with $| \cS | \leq s$. Define the $\ell_1$-cone 
\#
	\Lambda = \Lambda(s,p) = \bigl\{ \bbeta \in \RR^p: \|  \bbeta - \bbeta^* \|_1 \leq 4 s^{1/2} \|   \bbeta - \bbeta^*  \|_{\Sigma} \bigr\}. \label{l1.cone}
\#
Given $r>0$ and $\lambda_* >0$, we define the ``good" events
\#
	\cE_0(r ) = \bigl\{ \wt \bbeta^{(0)} \in \Theta(r ) \cap \Lambda  \bigr\}    ~~\mbox{ and }~~ \cE_*(\lambda_*) = \bigl\{   \| \nabla \hat \cQ_h(\bbeta^* )  - \nabla \cQ_h(\bbeta^*)   \|_\infty \leq \lambda_* \bigr\}  ,
\# 
which, with slight abuse of notation, extend those given in \eqref{event0} to the high-dimensional setting.

In the following, we first establish upper bounds for the $\ell_1$- and $\ell_2$-errors of the one-step penalized estimator $\wt \bbeta^{(1)}$, provided that the initial estimator $\wt \bbeta^{(0)}$ falls in a local neighborhood of $\bbeta^*$. In parallel to Condition~(C3), we impose the following moment condition on the  high-dimensional random vector $\bx \in \RR^p$ of covariates.

\noindent
(C4).
The predictor $\bx = (x_1,\ldots, x_p)^\T \in \RR^p$ (with $x_1\equiv 1$) has bounded components and uniformly bounded kurtosis. That is, there exists $B\geq 1$ such that $\max_{1\leq j\leq p} |x_j| \leq B$ almost surely, and $\mu_4 := \sup_{\bu \in \mathbb{S}^{p-1}} \EE ( \bz^\T \bu)^4 <\infty$, where $\bz = \Sigma^{-1/2} \bx$ and $\Sigma= (\sigma_{jk})_{1\leq j, k\leq p} = \EE(\bx \bx^\T)$ is positive definite. 
Write $\sigma_u = \max_{1\leq j\leq p} \sigma_{jj}^{1/2}$ and $\lambda_l = \lambda_{\min} ( \Sigma  ) \in (0,1]$. For convenience, we assume $\lambda_l=1$.

For technical reasons, the bounded covariates assumption is also imposed in \cite{WKSZ2017} and \cite{JLY2018} for sparse linear regression and generalized linear models.

\begin{theorem} \label{thm:hd.one-step}
 Assume Conditions~(C1), (C2), and (C4) hold. For $\delta\in (0,1)$ and $r_0, \lambda_*>0$, let $b\geq h>0$ and $\lambda =2 .5( \lambda_*  +  \varrho )>0$ satisfy 
\#
	\varrho \asymp \max  \Bigg[ \Bigg\{ \frac{1}{b} \sqrt{ \frac{\log(p/\delta)}{n}}  +  \frac{1}{h}\sqrt{ \frac{\log(p/\delta)}{N}}   \Bigg\} s^{1/2} r_0 , \,  s^{-1/2}\big( b r_0 + h^2\big) \Bigg]  ~\mbox{ and }~ s^{1/2} \lambda  \lesssim b \lesssim 1 . \nn 
\#  
Conditioned on the event $\cE_0(r_0 ) \cap \cE_*(\lambda_*)$, the one-step estimator $\wt \bbeta^{(1)}$ defined in \eqref{one.step.l1.conquer} satisfies  $\wt \bbeta^{(1)}  \in \Lambda$ and
\#
   \| \wt \bbeta^{(1)} - \bbeta^* \|_{\Sigma} \lesssim  \left\{  \frac{s}{b} \sqrt{ \frac{\log(p/\delta)}{n}}  +  b + \frac{s}{h}\sqrt{ \frac{\log(p/\delta)}{N} }    \right\}     r_0   +      s^{1/2}  \lambda_*  + h^2    \label{hd.one-step.error}
\# 
with probability at least $1-\delta$.
\end{theorem}

In Theorem~\ref{thm:hd.one-step}, the prespecified parameter $r_0>0$ quantifies the accuracy of the initial regularized QR estimator $\wt \bbeta^{(0)}$ under $\ell_2$-norm. Using a subsample of size $n$ to construct such an estimator, the nearly minimax-optimal rate is $\sqrt{s \log(p)/n}$ \citep{BC2011,W2019}. With a suitable choice for the regularization weight $\lambda$, we ensure that $\wt \bbeta^{(1)}$ must lie in the restricted set $\Lambda$, and bandwidths $b, h>0$,  the estimation error of $\wt \bbeta^{(1)}$ is of the order
\$
  \underbrace{ \left\{  \frac{s}{b} \sqrt{ \frac{\log(p/\delta)}{n}}  +  b + \frac{s}{h}\sqrt{ \frac{\log(p/\delta)}{N}}    \right\}   }_{{\rm contraction~ factor}}   r_0   +    \underbrace{   s^{1/2}  \lambda_*  + h^2   }_{{\rm near-optimal~rate}}.
\$
As we shall see, the second term is related to the near-optimal rate when the entire dataset is used.  The first term involves a contraction factor that is of the order $\frac{s}{b}\sqrt{\log(p/\delta)/n} + b + \frac{s}{h}\sqrt{\log(p/\delta)/N}$.
With sufficiently many samples per source---namely, $n\gtrsim s^2 \log(p/\delta)$,  the above one-step estimation procedure, which uses one round of communication, improves the statistical accuracy of $\wt \bbeta^{(0)}$ as long as 
$s\sqrt{\log(p/\delta)/n} \lesssim b \lesssim 1$ and $s\sqrt{\log(p/\delta)/N} \lesssim h \lesssim 1$.

Next, we describe an iterative, multi-round procedure for estimating a sparse $\bbeta^*\in \RR^p$ in a distributed setting. Let $\hat Q_{j,b}(\cdot)$ and $\hat Q_{j,h}(\cdot)$, $j=1,\ldots, m$, be the local empirical loss functions given in \eqref{local.loss.functions}.
At iteration 0,  the first (master) machine computes an initial estimator $\wt \bbeta^{(0)}$ as well as $\nabla \hat \cQ_{1,b}(\wt \bbeta^0)$, and broadcast $\wt \bbeta^{(0)}$ to all local machines.    For $j=1,\ldots, m$, the $j$th local machine then computes gradients $\nabla \hat \cQ_{j,h}(\wt \bbeta^{(0)})$, which are then transmitted back to the first. At iteration $t =1,2,\ldots, T$, the first machine solves the $\ell_1$-penalized shifted conquer loss minimization 
\#
	\wt \bbeta^{(t)} \in \argmin_{\bbeta \in \RR^p} \, \underbrace{   \hat \cQ_{1,b}(\bbeta)  - \bigl\langle  \nabla \hat \cQ_{1,b}(\wt \bbeta^{(t-1)} ) - \nabla  \hat \cQ_h(\wt \bbeta^{(t-1)} )  , \bbeta \bigr\rangle  }_{ = : \,\wt \cQ^{(t)}(\bbeta) } + \,\lambda_t \cdot \| \bbeta \|_1 ,
\#
where $\nabla  \hat \cQ_h(\wt \bbeta^{(t-1)} ) = (1/m)\sum_{j=1}^m \nabla \hat \cQ_{j,h}(\wt \bbeta^{(t-1)})$, and $\lambda_t >0$ are regularization parameters.

\begin{theorem} \label{hd.final.rate}
 Assume Conditions~(C1), (C2), and (C4) hold. Given $\delta\in (0,1)$, choose the local and global bandwidths as 
 \#
  b \asymp s^{1/2} \bigl\{   \log(p/\delta) / n \bigr\}^{1/4} ~~\mbox{ and }~~   h \asymp\bigl\{  s  \log(p/\delta) / N \bigr\}^{1/4} . \label{bandwidths.constraint}
 \#
 For $r_0, \lambda_* >0$, write $r_* = s^{1/2}\lambda_*$ and set $\lambda_t = 2.5( \lambda_* + \varrho_t )>0$ ($t\geq 1$)  with 
\$
 \varrho_t  \asymp  \max \left\{  \gamma^{t } s^{-1/2} r_0 + \gamma s^{-1/2} ( r_*  +  h^2 )\mathbbm{1}(t\geq 2)  , \,  \sqrt{\log(p/\delta)/N} \right\}   ,
\$
where $\gamma = \gamma (s,p,n,N,\delta) \asymp s^{1/2}   \max  \{ \log(p/\delta) / n, s  \log(p/\delta) /N   \}^{1/4}$.
Let the sample size per source and total sample size satisfy $n\gtrsim s^2 \log(p/\delta)$ and $N\gtrsim s^3 \log(p/\delta)$, so that $\gamma <1$. Moreover, assume $r_0\lesssim \min\{ 1, (m/s)^{1/4} \}$ and $r_* \lesssim b$.
Then, conditioned on the event $\cE_*(\lambda_*) \cap \cE_0(r_0)$,  the $T^{{\rm th}}$ iterate $\wt \bbeta^{(T)}$ with $T \gtrsim \log(r_0 / r_*)/\log(1/\gamma)$ satisfies
\#
	  \| \wt \bbeta^{(T) } - \bbeta^*   \|_{\Sigma} \lesssim s^{1/2} \lambda_*  + h^2 ~~\mbox{ and }~~ 
 \| \wt \bbeta^{(T) } - \bbeta^*   \|_1 \lesssim s \lambda_* + s^{1/2} h^2   \label{hd.est.error}
\#
with probability at least $1-T \delta$.
\end{theorem}

According to Theorem~\ref{hd.final.rate}, the success of the iterative  procedure described above relies on a sufficiently accurate initial estimator $\wt \bbeta^{(0)}$. For example, we may choose $\wt \bbeta^{(0)}$ to be a local $\ell_1$-conquer  estimator 
\#
	\wt \bbeta^{(0)} \in \argmin_{\bbeta \in \RR^p}  \, \hat \cQ_{1,b_0}(\bbeta) + \lambda_0 \cdot \| \bbeta \|_1 , \label{first.l1.conquer}
\#
or a local $\ell_1$-QR estimator which is a minimizer of the  program
\#
	\argmin_{\bbeta \in \RR^p} \,  \frac{1}{n} \sum_{i\in \cI_1}  \rho_\tau ( y_i - \bx_i^\T \bbeta)  + \lambda_0 \cdot \| \bbeta \|_1 . \label{first.l1.qr}
\#
High probability  estimation error bounds for $\ell_1$-QR were derived by \cite{BC2011}, \cite{W2013}, and more recently by \cite{W2019} under weaker assumptions. The estimation error for $\ell_1$-conquer is provided by the following result, which is a variant of Theorem~4.1 in \cite{TWZ2020}.

\begin{proposition} \label{prop:hd.initial}
Assume Conditions~(C1), (C2) and (C4) hold.   For $\delta\in (0,1)$, set the regularization parameter $\lambda_0 \asymp \sqrt{\tau(1-\tau) \log(p/\delta)/n}$.
Provided that  $\sqrt{s\log(p/\delta)/n} \lesssim b_0 \lesssim 1$, the local $\ell_1$-conquer estimator $\wt \bbeta^{(0)}$ given in \eqref{first.l1.conquer} satisfies 
\#
	  \| \wt \bbeta^{(0)} - \bbeta^*   \|_{\Sigma} \lesssim  s^{1/2} \lambda_0 + b_0^2  \label{initial.hd.rate}
\#
with probability at least $1-\delta$. If in addition $b_0 \lesssim \{ s \log(p/\delta)/n \}^{1/4}$, then $\wt \bbeta^{(0)} \in \Lambda$.
\end{proposition}

With the above preparations, we are now ready to state the estimator error bound for the distributed regularized conquer  estimator $\wt \bbeta^{(T)}$ in high dimensions.

\begin{theorem} \label{hd.ceqr}
Assume Conditions~(C1), (C2) and (C4) hold, and that the data are generated from a sparse conditional quantile model \eqref{qr.model} with $\| \bbeta^* \|_0 \leq s$.  Suppose the sample size per source and total sample size satisfy $n\gtrsim s^2 \log  (p)$ and $N\gtrsim s^3 \log (p)$. 
Choose the bandwidths $b, h>0$ and regularization parameters $\lambda_t$ ($t\geq 1$) as $b \asymp s^{1/2} \{ \log(p) /n \}^{1/4}$, $h \asymp \{ s \log(p)/ N \}^{1/4}$ and 
\#
 \lambda_t \asymp \sqrt{\frac{\log  (p)}{N}}  + \max\left\{ \frac{s^2 \log(p)}{n} , \frac{s^3 \log(p)}{N} \right\}^{t/4} \sqrt{\frac{\log  (p)}{n}} . \nn
\#
Starting at iteration 0 with an initial estimate $\wt \bbeta^{(0)}$ as described in Proposition~\ref{prop:hd.initial}, the distributed estimator $\wt \bbeta = \wt \bbeta^{(T)}$ with $T\asymp \lceil \log(m) \rceil$ communication rounds satisfies  the error bounds
\#
 \| \wt \bbeta - \bbeta^*   \|_2 \lesssim \sqrt{\frac{s \log(p)}{N}}~~\mbox{ and }~~  \| \wt \bbeta - \bbeta^*   \|_1 \lesssim s \sqrt{\frac{  \log(p)}{N}}
\#
with probability at least $1-C \log(m)/N$.
\end{theorem}

Theorems~\ref{thm:hd.one-step}--\ref{hd.ceqr} are non-trivial extensions of Theorem~3 in \cite{WKSZ2017} to the context of quantile regression. The latter can be applied to the squared loss for linear regression and logistic loss for classification. 
Let $\ell(\cdot)$ be the loss function of interest, and it is assumed therein that
$$
	|\ell'(u) - \ell'(v)| \leq L |u-v|  ~\mbox{ for any }~ u, v\in \RR ~~\mbox{ and }~~  \sup_{u\in \RR} |\ell'''(u)| \leq M .
$$
The key of the proof is to control the difference between the gradient vectors $\nabla \wt \cQ^{(t)} ( \wt \bbeta^{(t-1)} )$ and $\nabla \wt \cQ^{(t)} ( \bbeta^*)$ at each iteration. For this purpose, the proof of Theorem~3 in \cite{WKSZ2017} is based on the second-order Taylor's series expansion, so that the above parameters $L$ and $M$ arise and are treated as constants. In particular, $M=0$ for the quadratic loss. In our context, if we take $\ell(\cdot)$ to be the local conquer loss $(\rho_\tau * K_b)(\cdot)$, then it is easy to see that $L \asymp b^{-1}$ and $M \asymp b^{-2}$. Since the bandwidth $b$ decays as a function of $(n,p)$, neither the result nor proof argument in \cite{WKSZ2017} apply to quantile regression even with smoothing. In the Appendix, we provide a self-contained proof of Theorems~\ref{thm:hd.one-step} and \ref{hd.final.rate}, which relies on a uniform control of the fluctuations of gradient processes and a restricted strong convexity property for the empirical conquer loss.

\subsection{Distributed quantile regression via ADMM}

In this subsection, we describe an alternative algorithm based on the alternating direction method of multiplier (ADMM) for penalized quantile regression with distributed data.
ADMM, which was first introduced by \cite{DR1956} and \cite{GM1976}, has a number of successful applications in modern statistical machine learning. We refer to \cite{Boyd2011} for a comprehensive review on ADMM. In the context of quantile regression, \cite{Yu2017} and \cite{Gu2018} respectively proposed ADMM-based algorithms for fitting penalized QR with both convex and folded-concave penalties.

As argued in \cite{Boyd2011}, ADMM is  well suited  for  distributed convex optimization problems under minimum structural assumption.  For solving penalized QR, in the following we revisit the parallel implementation of the ADMM-based algorithm proposed in \cite{Yu2017}. Recall that the total dataset $\{ (y_i, \bx_i)\}_{i=1}^N$ with $N=n\cdot m$ is distributed across $m$ sources, each containing a data batch indexed by $\cI_j$ ($j=1,\ldots, m$). Write 
\$
	\by = ( y_1, \ldots, y_N)^\T = (\by_{1}^{\T} , \ldots, \by_{m}^{\T} )^\T  ~\mbox{ and }~ \bX = (\bx_1, \ldots, \bx_N )^\T  = (\bX_1^\T, \ldots, \bX_m^\T )^\T \in \RR^{N\times p} , 
\$
where $ \by_j = \by_{\cI_j} \in \RR^n$ and $\bX_j \in \RR^{n\times p}$. Under this set of notation, the $\ell_1$-QR problem \eqref{lasso-qr} can be recast into an equivalent problem 
$$
	\underset{\br_j, \bbeta_j, \bbeta }{\mathrm{minimize}} ~\bigg\{ \sum_{j=1}^m \rho_\tau( \br_j) + \lambda_N \| \bbeta \|_1 \bigg\}~~{\rm such~that ~}~ \by_j - \bX_j \bbeta_j = \br_j, \ \ \bbeta_j = \bbeta, \ \ j=1,\ldots, m ,
$$
where $\lambda_N = N \lambda$.
Here we write $\rho_\tau(\br) = \sn \rho_\tau(r_i)$ for $\br = (r_1,\ldots, r_n)^\T$. To solve this linearly constrained optimization problem, the ADMM updates at iteration $k=0, 1,\ldots$ are
\#
	\bbeta^{k+1} &= \argmin_{\bbeta} \bigg\{ \frac{m \gamma }{2}  \big\| \bbeta - \bar \bbeta^k  - \bar \bdelta^k / \gamma \big\|_2^2 + \lambda_N \| \bbeta \|_1 \bigg\} ,  \label{beta-step}\\
	 \br_j^{k+1} & = \argmin_{\br_j}  \bigg\{ \rho_\tau(\br_j) + \frac{\gamma}{2} \big\| \by_j - \bX_j \bbeta^k_j + \bu^k_j/ \gamma - \br_j \big\|_2^2 \bigg\} ,   \label{r-step}  \\
	 \bbeta_j^{k+1} & =  ( \bX_j^\T \bX_j + \Ib_p )^{-1} \big\{ \bX_j^\T \big( \by_j - \br_j^{k+1} + \bu_j^k / \gamma \big) - \bdelta_j^k / \gamma + \bbeta^{k+1} \big\} , \nn \\
	 \bu_j^{k+1} & = \bu_j^k + \gamma \big( \by_j - \bX_j \bbeta_j^{k+1} - \br_j^{k+1} \big) ,  \nn \\
	 \bdelta_j^{k+1} & =  \bdelta_j^{k } + \gamma \big( \bbeta^{k+1}_j - \bbeta^{k+1} \big) ,   \nn
\#
where $\bar \bbeta^k = (1/m) \sum_{j=1}^m \bbeta^k_j$, $\bar \bdelta^k = (1/m) \sum_{j=1}^m \bdelta_j^k$, and $\gamma>0$ is the augmentation parameter. In particular, the $\bbeta$-update in \eqref{beta-step} and the $\br$-update in \eqref{r-step} have explicit expressions, which are
\$
\bbeta^{k+1} &= \big( \bar \bbeta^k + \bar \bdelta^k/\gamma - \lambda_N/(m\gamma) {\bf 1}_p \big)_+ -  \big( - \bar \bbeta^k - \bar \bdelta^k/\gamma - \lambda_N/(m\gamma) {\bf 1}_p \big)_-  ~~\mbox{ and }  \\
 \br^{k+1} & = \big( \by_j - \bX_j \bbeta^{k}_j + \bu^k_j /\gamma - \tau \gamma^{-1}  {\bf 1}_n \big)_+ -  \big(- \by_j + \bX_j \bbeta^{k}_j - \bu^k_j /\gamma + (\tau-1) \gamma^{-1}  {\bf 1}_n  \big)_+  , 
\$
respectively, where ${\bf 1}_q := (1, \ldots, 1)^\T \in \RR^q$ for each integer $q\geq 1$.

The above parallel version of the ADMM to solve \eqref{lasso-qr}  involves primal variables $\bbeta \in \RR^p$, $(\br_1^{\T}, \ldots , \br_m^{\T})^\T \in \RR^N$ and the dual variable $(\bu_1^{\T}, \ldots, \bu_m^{\T})^\T \in \RR^N$. As a general-purpose algorithm, its convergence can be quite slow when applied to large-scale datasets. For example, under a numerical setting with $p=100$, $N=30,000$ and $m \in \{1, 10, 100\}$ considered in \cite{Yu2017}, it takes more than 100 iterations for the parallel implementation of the ADMM to converge. In a distributed framework, this amounts to (at least) 100 communication rounds in order to achieve the desired level of statistical accuracy.
At even larger data scales, our numerical results (see Figure~\ref{fig:simhighd} below) show evidence that the proposed multi-round, distributed estimator can perform as well as the global estimator within $T=10$ communication rounds.

\section{Numerical Studies}
\label{sec:simulation}


\subsection{Distributed quantile regression}\label{subsec:estimation}

Starting with the low-dimensional setting, we compare the proposed multi-round procedure with the following methods: (i) global QR estimator using all of the available $N=m n$ observations; 
(ii) the averaging-based estimator based on local QR estimators;  (iii) the proposed method with $T\in\{1,4,10\}$ communication rounds; and (iv) a non-smooth version of the proposed method, which uses the subgradient of the QR loss as the global gradient, with $T\in\{1,4,10\}$ communication rounds. 
We employ the \texttt{R} packages \texttt{conquer} and \texttt{quantreg} to compute the conquer and standard QR estimators, respectively.  

As shown in \citet{HPTZ2020}, the performance of conquer is insensitive to the choice of kernel functions, and thus we use the Gaussian kernel wherever smoothing is required. 
Our proposed method involves an initial estimator $\wt{\bbeta}^{(0)}$ and two smoothing parameters $h$ and $b$.
There are multiple ways to obtain an adequate initialization.  For instance, as suggested by \citet{JLY2018}, one can use the simple averaging estimator as the initialization, i.e., the average of  local QR estimators across $m$ sources.    
For simplicity, we take $\wt{\bbeta}^{(0)}$ to be a conquer estimator computed based on $n$ independent data points from one source. 
For the bandwidths, we set $h = 2.5\cdot \{(p+\log N)/N\}^{1/3}$ and $b = 2.5\cdot \{(p+\log n)/n\}^{1/3}$ according to the theoretical analysis in Section~\ref{sec:distributed.conquer}.

To generate the data, we consider two types of heteroscedastic models:
\begin{enumerate}
\item Linear heteroscedasticity: $y_i = \bx_i^\T \bbeta^* + (0.2 x_{ip}+1)\{\varepsilon_i- F^{-1}_{\varepsilon_i}(\tau)\}$;
\item Quadratic heteroscedasticity: $y_i = \bx_i^\T \bbeta^* + 0.5 \{1+ (0.25x_{ip}-1)^2\}\{\varepsilon_i- F^{-1}_{\varepsilon_i}(\tau)\}$,
\end{enumerate}
where $\bx_i$ is generated from a multivariate uniform distribution on the cube $3^{1/2} \cdot[-1,1]^{p+1}$ with covariance matrix $\bSigma=(0.5^{|j-k|})_{1\le,j,k\le p+1}$, and $\bbeta^*= \mathbf{1}_p$ is a $p$-vector of ones.  The random noise is generated from a  $t$-distribution with 2 degrees of freedom, denoted by $t_{2}$.
To evaluate the performance across different methods, we report the estimation error under the $\ell_2$-norm, i.e., $\|\hat{\bbeta}-\bbeta^*\|_2$.   
Table~\ref{sim:table1} presents the results when $n=300$, $p=10$, $m\in\{50,100,200,400,600,1000\}$, and $\tau=0.8$, averaged over 100 trials. 
With the same $p$ and $\tau$, we report the results with a fixed total sample size $N=150,000$, $m = N/n$, and varying local sample size  $n \in \{300,500,1000,1500,3000,6000\}$ in Table~\ref{sim:table2}.

 The global QR estimator, which always has the smallest error as expected, serves as a benchmark for communication-efficient methods.
From Table~\ref{sim:table1}, we see that the proposed multi-round distributed estimator yields the best performance among the communication-efficient estimators, and as the number of communication rounds grows, it becomes almost as good as the global QR estimator even though the one-step estimator ($T=1$) performs rather poorly.  
The performance of the averaging-based QR is comparable to that of the proposed method when the number of machines $m$ is smaller than the local sample size $n$. 
As suggested by the theoretical analyses, when $m$ is larger than $n$, the proposed method outperforms the averaging-based QR.
To highlight the importance of smoothing for distributed quantile regression, we also implement the multi-round procedure using the subgradient of the QR loss, namely, $\nabla \hat \cQ(\bbeta) = (nm)^{-1} \sum_{j=1}^m \sum_{i\in \cI_j} \{ \mathbbm{1}(y_i< \bx_i^\T \bbeta) - \tau\} \bx_i$, instead of $\nabla \hat{\cQ}_h(\cdot)$.
Note that the estimation error of this subgradient-based method is barely improvable as the number of machines increases.  
When the number of total samples $N$ is fixed, from Table~\ref{sim:table2}, we find that the subgradient-based method only performs well if the local sample size is extremely large, which makes all the methods desirable. This demonstrates the importance of smoothing in the context  of distributed learning with non-smooth loss functions. 

Next, we perform a sensitivity analysis to assess the effect of the initial estimator on the final solution of the proposed method with $T=10$. 
To this end, we conduct additional numerical studies where we consider different initial estimators, computed using different sample sizes $n_{\mathrm{init}}= \{150, 300,500,1000,5000\}$.  
Specifically, we consider the aforementioned linear and quadratic heteroscedastic models with $n=300$, $p=10$, $m=400$, and $\tau=0.8$.  
The average estimation error for the proposed method with $T=\{1,4,10\}$ and that of global QR estimator are summarized in Table~\ref{sim:initial}.
From Table~\ref{sim:initial}, we see that the estimation error for the proposed method with $T=1$ decreases as we increase the sample size used to calculate the initial estimator.  Moreover, we see that implementing the proposed method with $T=\{4,10\}$ improves the estimation error significantly, and that the estimation error is no longer sensitive to the initial sample size.   The proposed method with $T=10$ yields an estimator that performs as well as the global QR (implemented using dataset from all sources) even when $n_{\mathrm{init}}  = 150$.  
The results suggest that the proposed method is not sensitive to the sample size used to calculate the initial estimator after some rounds of  communication.
 
\begin{table}[!htp]
	\fontsize{8}{9}\selectfont
	\centering
	\caption{Estimation error under linear and quadratic heteroscedastic models with $t_2$ noise, averaged over 100 trials.  Results for $\tau = 0.8$, $n=300$, and $p=10$, across $m=\{50,100,200,400,600,1000\}$ are reported.}
	\begin{tabular}{  l |   c  c c c  c c c}
	\hline
	\multicolumn{7}{c}{Linear Heteroscedastic Model with $\tau = 0.8$, $n = 300$, and $p=10$}\\ \hline
	 Methods & $m=50$ & $m=100$ & $m=200$ & $m=400$ & $m=600$ & $m=1000$ \\
		\hline

\texttt{averaging-based QR} & 0.077 & 0.060& 0.047& 0.041& 0.037 & 0.035\\		
\texttt{distributed  QR} ($T=1$)& 0.197&0.216&0.223& 0.213& 0.192 & 0.198\\
\texttt{distributed  QR} ($T=4$)& 0.173& 0.223& 0.256& 0.202& 0.187& 0.175\\
\texttt{distributed  QR} ($T=10$)& 0.259& 0.341& 0.427& 0.313& 0.271& 0.305\\
\texttt{distributed smoothed QR} ($T=1$)&0.159& 0.163& 0.163& 0.151& 0.138& 0.143\\
\texttt{distributed smoothed QR} ($T=4$)& 0.076& 0.066& 0.051& 0.032& 0.027& 0.029\\
\texttt{distributed smoothed QR} ($T=10$)&0.075& 0.071& 0.039& 0.027& 0.021& 0.020\\
\texttt{global QR} & 0.069 & 0.050 & 0.035& 0.025& 0.019& 0.016\\		

	 \hline \hline 
	\multicolumn{7}{c}{Quadratic Heteroscedastic Model with $\tau = 0.8$, $n=300$, and $p=10$}\\ \hline	 
	
\texttt{averaging-based QR} & 0.079 & 0.063& 0.050& 0.043& 0.038 & 0.036\\			
\texttt{distributed  QR} ($T=1$)& 0.206&0.210&0.233& 0.204& 0.198 & 0.205\\
\texttt{distributed  QR} ($T=4$)& 0.198& 0.232& 0.281& 0.211& 0.235& 0.184\\
\texttt{distributed  QR} ($T=10$)& 0.263& 0.401& 0.423& 0.330& 0.396& 0.332\\
\texttt{distributed smoothed QR} ($T=1$)&0.162& 0.160& 0.163& 0.151& 0.148& 0.148\\
\texttt{distributed smoothed QR} ($T=4$)& 0.079& 0.062& 0.051& 0.033& 0.034& 0.033\\
\texttt{distributed smoothed QR} ($T=10$)&0.077& 0.057& 0.041& 0.027& 0.024& 0.021\\
\texttt{global QR} & 0.071 & 0.050 & 0.036& 0.025& 0.020& 0.017\\		
	
	\hline
\label{sim:table1}	
		\end{tabular}
\end{table}

\begin{table}[!htp]

	\fontsize{8}{9}\selectfont
	\centering
	\caption{Estimation error under linear and quadratic heteroscedastic models with $t_2$ noise, averaged over 100 trials.  Results for $\tau = 0.8$, $N= nm = 150,000$ and $p=10$, across $n=\{300,500,1000,1500,3000,6000\}$ are reported.}
	\begin{tabular}{  l |   c  c c c  c c c}
	\hline
	\multicolumn{7}{c}{Linear Heteroscedastic Model with $\tau = 0.8$, $N =  150,000$, $m = N/n$, and $p=10$}\\ \hline
	 Methods & $n=300$ & $n=500$ & $n=1000$ & $n=1500$ & $n=3000$ & $n=6000$  \\
		\hline

\texttt{averaging-based QR} & 0.037 & 0.029& 0.025& 0.023& 0.022 & 0.022\\		
\texttt{distributed  QR} ($T=1$)& 0.206&0.140&0.071& 0.053& 0.034 & 0.026\\
\texttt{distributed  QR} ($T=4$)& 0.252& 0.100& 0.026& 0.023& 0.022& 0.022\\
\texttt{distributed  QR} ($T=10$)& 0.389& 0.124& 0.024& 0.023& 0.022& 0.022\\
\texttt{distributed smoothed QR} ($T=1$)&0.149& 0.092& 0.049& 0.038& 0.028& 0.024\\
\texttt{distributed smoothed QR} ($T=4$)& 0.042& 0.024& 0.023& 0.023& 0.023& 0.023\\
\texttt{distributed smoothed QR} ($T=10$)&0.029& 0.023& 0.023& 0.023& 0.023& 0.023\\
\texttt{global QR} & 0.023 & 0.022 & 0.022& 0.022& 0.022& 0.022\\		

	 \hline \hline 
	\multicolumn{7}{c}{Quadratic Heteroscedastic Model  with $\tau = 0.8$, $N =  150,000$, $m = N/n$, and $p=10$}\\ \hline	 
	
\texttt{averaging-based QR} & 0.039 & 0.030& 0.026& 0.024& 0.023 & 0.022\\		
\texttt{distributed  QR} ($T=1$)& 0.215&0.143&0.075& 0.053& 0.034 & 0.027\\
\texttt{distributed  QR} ($T=4$)& 0.236& 0.101& 0.028& 0.023& 0.022& 0.022\\
\texttt{distributed  QR} ($T=10$)& 0.371& 0.123& 0.026& 0.023& 0.022& 0.022\\
\texttt{distributed smoothed QR} ($T=1$)&0.151& 0.095& 0.051& 0.040& 0.029& 0.025\\
\texttt{distributed smoothed QR} ($T=4$)& 0.037& 0.025& 0.024& 0.024& 0.023& 0.023\\
\texttt{distributed smoothed QR} ($T=10$)&0.025& 0.024& 0.024& 0.024& 0.023& 0.023\\
\texttt{global QR} & 0.023 & 0.023 & 0.023& 0.023& 0.022& 0.022\\		
	
	\hline
\label{sim:table2}	
		\end{tabular}
\end{table}

\begin{table}[!htp]
	\fontsize{8}{9}\selectfont
	\centering
	\caption{Estimation error under linear and quadratic heteroscedastic models with $t_2$ noise, averaged over 100 trials.  Results for $\tau = 0.8$, $n=300$, $p=10$, $m=400$, with initial estimator computed using different sample size, $n_{\mathrm{init}}= \{150, 300,500,1000,5000\}$,  are reported.}
	\begin{tabular}{  l |   c  c c c   c c}
	\hline
	\multicolumn{6}{c}{Linear Heteroscedastic Model with $\tau = 0.8$, $n = 300$, $p=10$, and $m=400$ }\\ \hline
	 Methods & $n_{\mathrm{init}}=150$ & $n_{\mathrm{init}}=300$ & $n_{\mathrm{init}}=500$ & $n_{\mathrm{init}}=1000$ & $n_{\mathrm{init}}=5000$   \\
		\hline

\texttt{distributed smoothed QR} ($T=1$)&0.211& 0.181& 0.151& 0.108& 0.085\\
\texttt{distributed smoothed QR} ($T=4$)& 0.041& 0.042& 0.032& 0.033& 0.031\\
\texttt{distributed smoothed QR} ($T=10$)&0.027& 0.038& 0.027& 0.026& 0.027\\
\texttt{global QR} & 0.025 & 0.025 & 0.025& 0.024& 0.025\\		

	 \hline \hline 
	\multicolumn{6}{c}{Quadratic Heteroscedastic Model with $\tau = 0.8$, $n=300$, $p=10$, and $m=400$}\\ \hline	 
\texttt{distributed smoothed QR} ($T=1$)&0.214& 0.151& 0.112& 0.087& 0.047\\
\texttt{distributed smoothed QR} ($T=4$)& 0.039& 0.033& 0.032& 0.032& 0.028\\
\texttt{distributed smoothed QR} ($T=10$)&0.027& 0.027& 0.029& 0.027& 0.027\\
\texttt{global QR} & 0.025 & 0.025 & 0.025& 0.025&0.025\\		
	
	\hline
\label{sim:initial}	
		\end{tabular}
\end{table}

\subsection{Distributed confidence construction}
\label{subsec:inferencesim}

In terms of  uncertainty quantification, we assess the performance of the proposed method for constructing confidence intervals by calculating the coverage probability and width of the confidence interval for each regression coefficient. For point estimation, we implement Algorithm~\ref{algo:dqr} with $T=10$ and employ the averaging-based QR estimator as the initialization.
The bandwidths are set to be $h = 1.5\cdot \{(p+\log N)/N\}^{1/3}$ and $b = 1.5\cdot \{(p+\log n)/n\}^{1/3}$.
For confidence construction, we first consider four methods:  the asymptotic normal-based interval  \eqref{normal.CI} for the proposed communication-efficient estimator (CE-Normal),  the normal-based  interval \eqref{normal.CI} with $\tilde{\bbeta}$ replaced by $\hat{\bbeta}^{\text{dc}}$ of equation \eqref{dc.qr} (DC-Normal), and the two communication-efficient bootstrap constructions as in \eqref{boot1} (CE-Boot (a)) and in \eqref{boot2} (CE-Boot (b)).

\begin{figure}[!htp]
\begin{center}
   \vspace{-2mm}
         \subfigure[]{\includegraphics[scale=0.41]{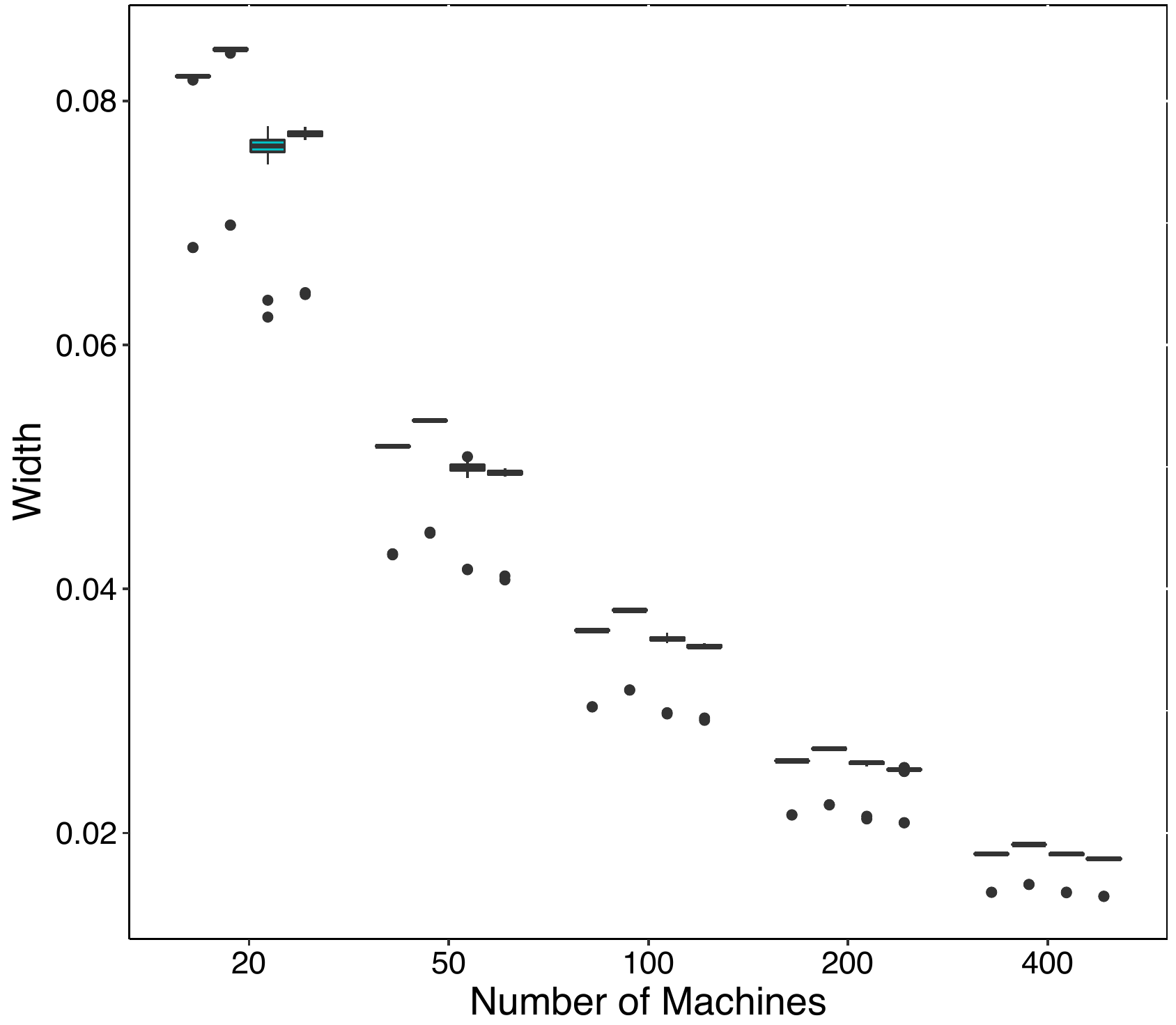}}
                  \subfigure[]{\includegraphics[scale=0.403]{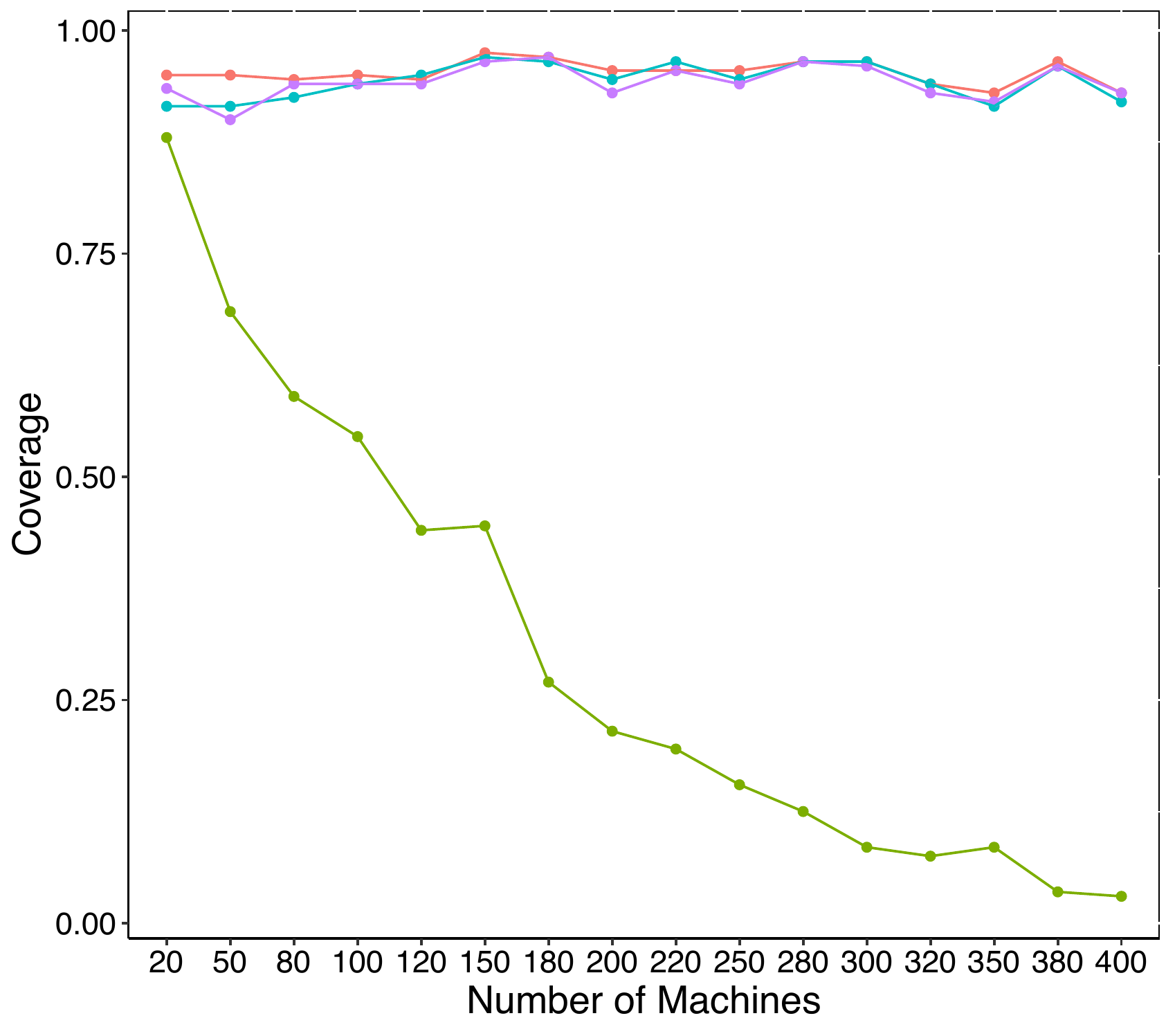}}
   \subfigure[]{\includegraphics[scale=0.41]{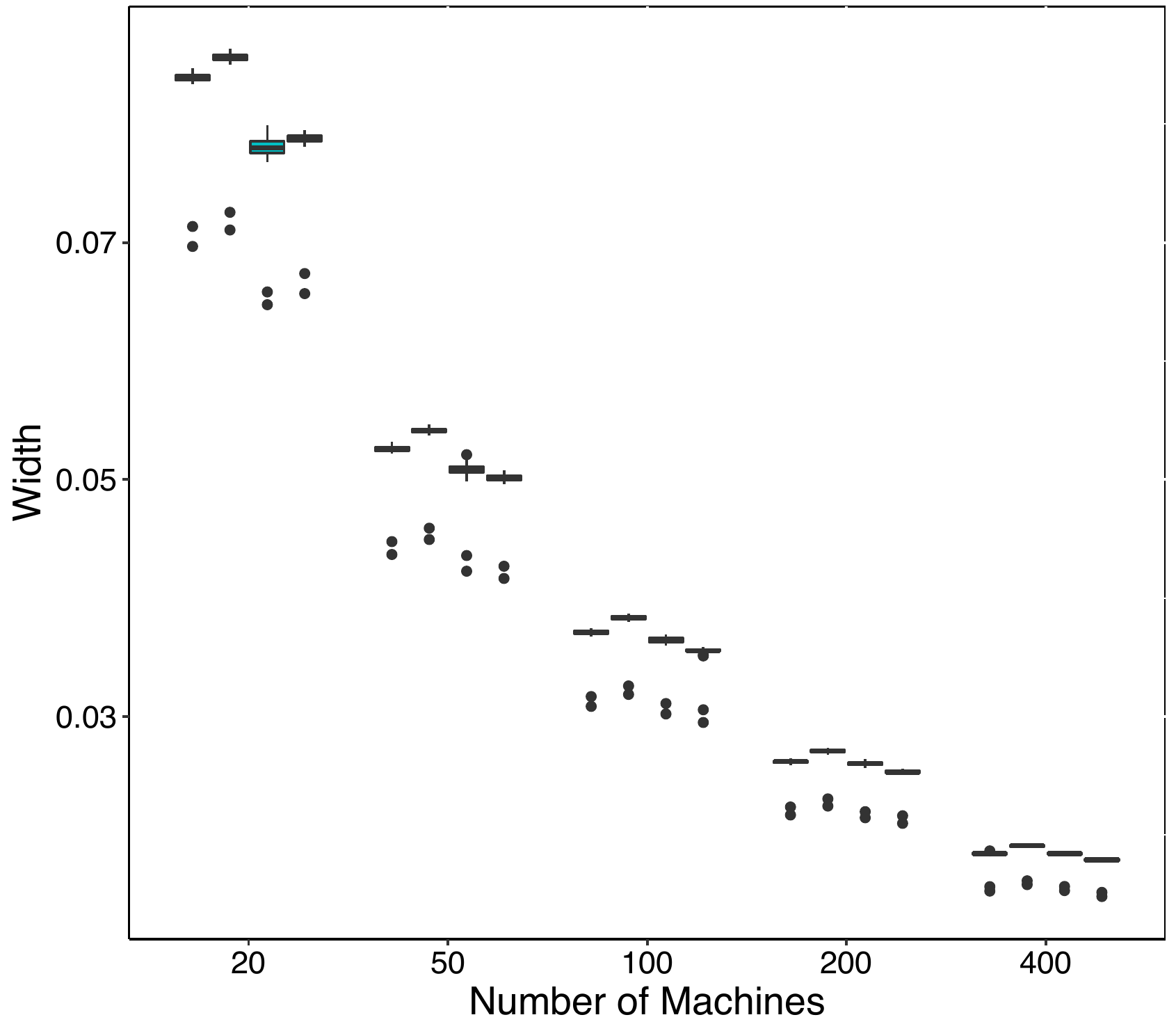}}
   \subfigure[]{\includegraphics[scale=0.403]{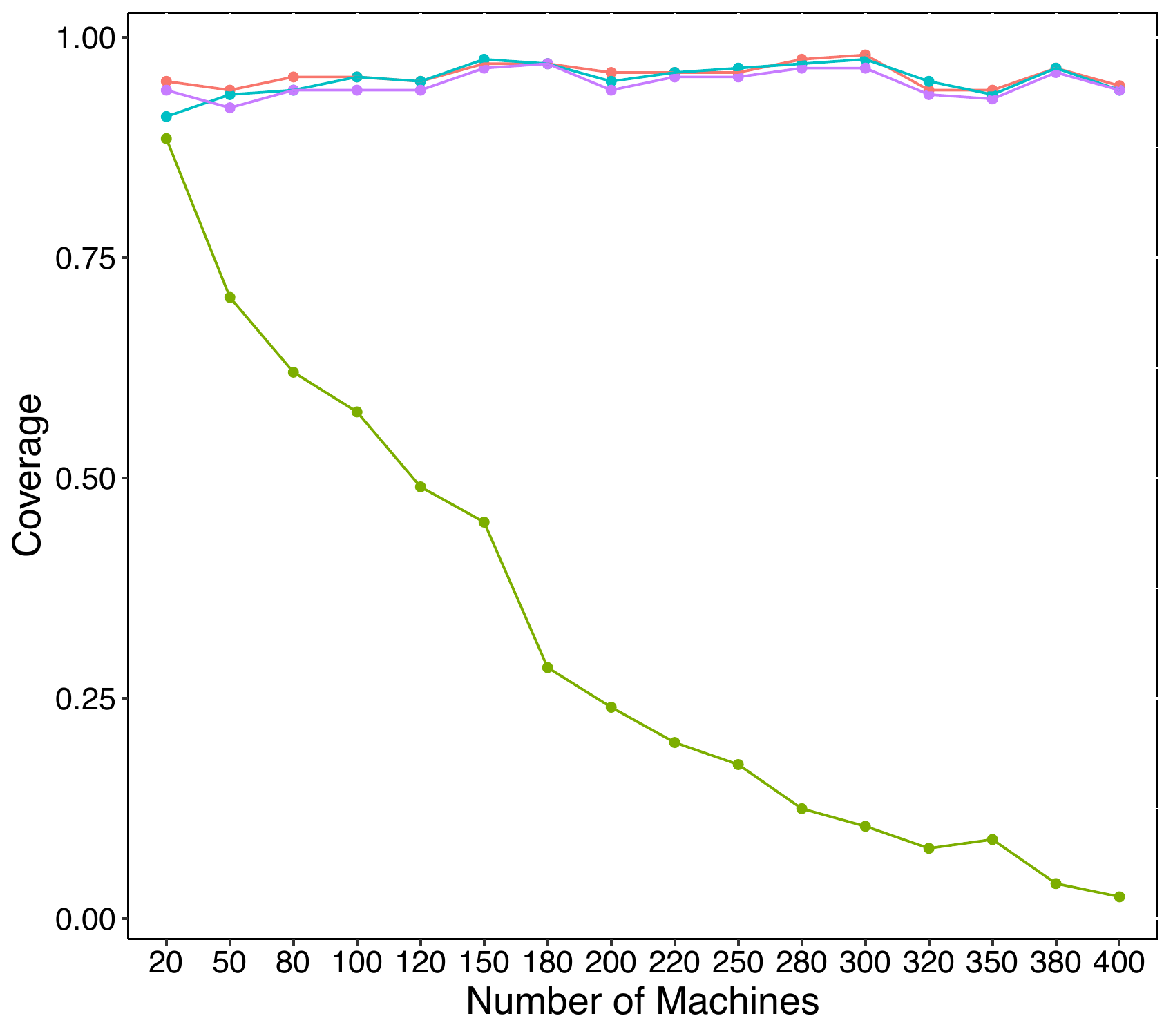}}
      \vspace{-1mm}
\end{center}
\caption{\label{fig:inf1} Properties of confidence intervals for the regression coefficients under  model \eqref{eq:lowhet} with $t_{1.5}$ noise when $\tau=0.8$ and $(n,p)=(2000,50)$ using type I variance  (first row) and type II variance estimators (second row). 
Panels (a) and (c) depict the widths of the confidence intervals for the last regression coefficient over Monte Carlo replicates using: CE-Normal \protect\includegraphics[height=1em]{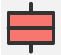};  DC-Normal \eqref{normal.CI} \protect\includegraphics[height=1em]{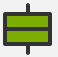}; CE-Boot (a) \protect\includegraphics[height=1em]{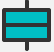};
CE-Boot (b) \protect\includegraphics[height=1em]{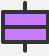}. Empirical coverage probabilities for the last coefficient are shown in panels (b) and (d).
}
\end{figure}

\begin{figure}[!htp]
\begin{center}
         \subfigure[]{\includegraphics[scale=0.41]{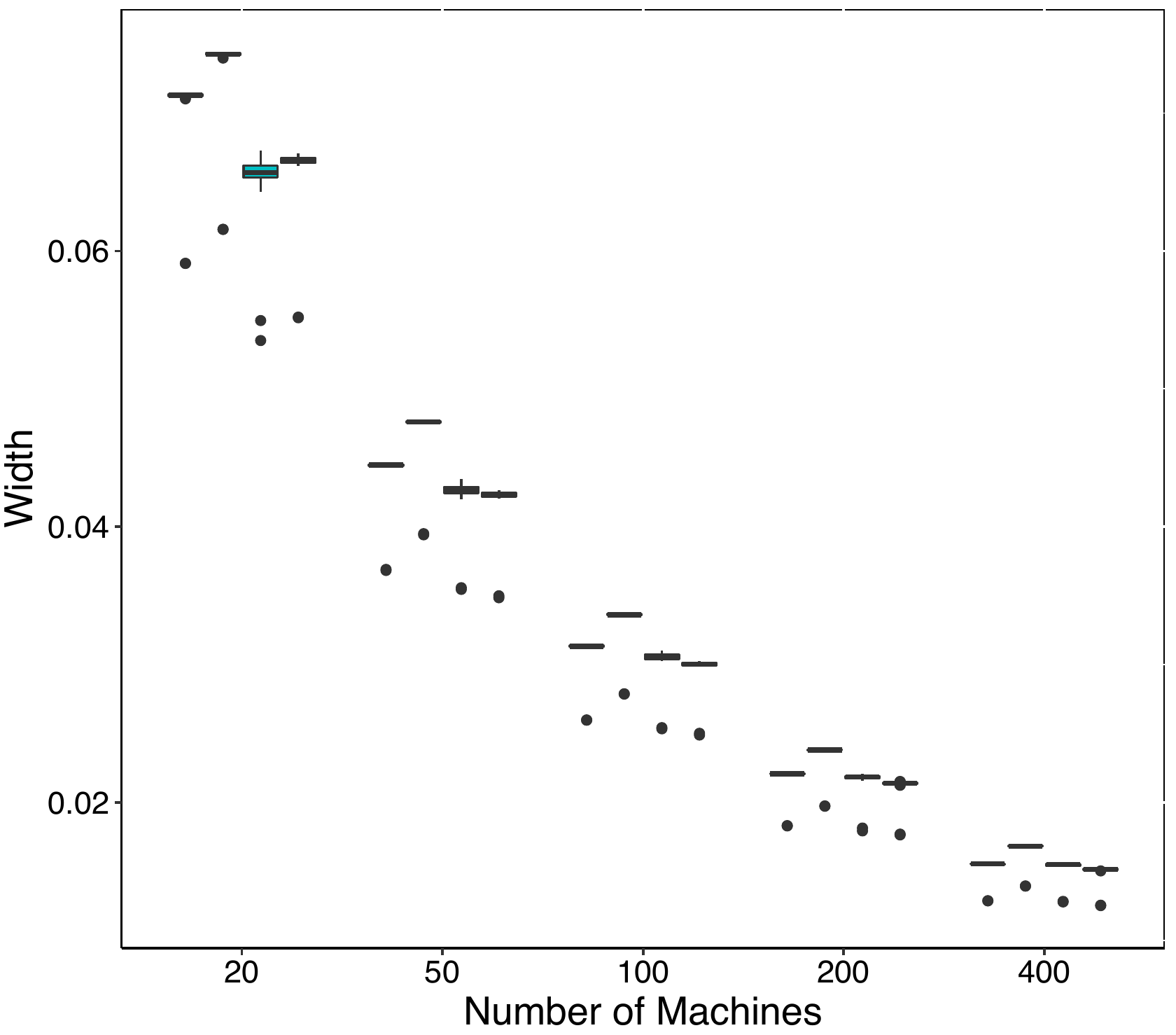}}
                  \subfigure[]{\includegraphics[scale=0.403]{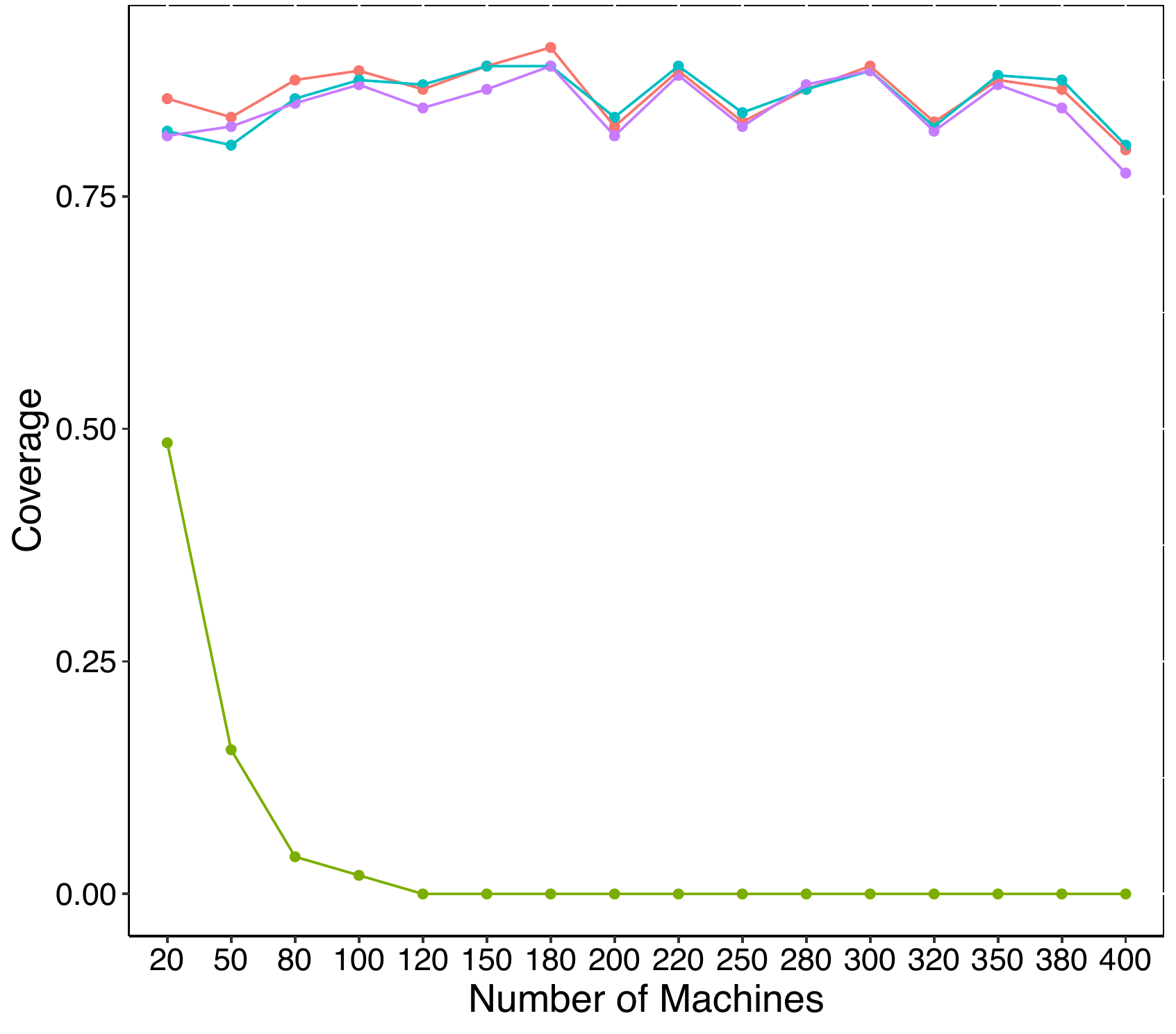}}
   \subfigure[]{\includegraphics[scale=0.41]{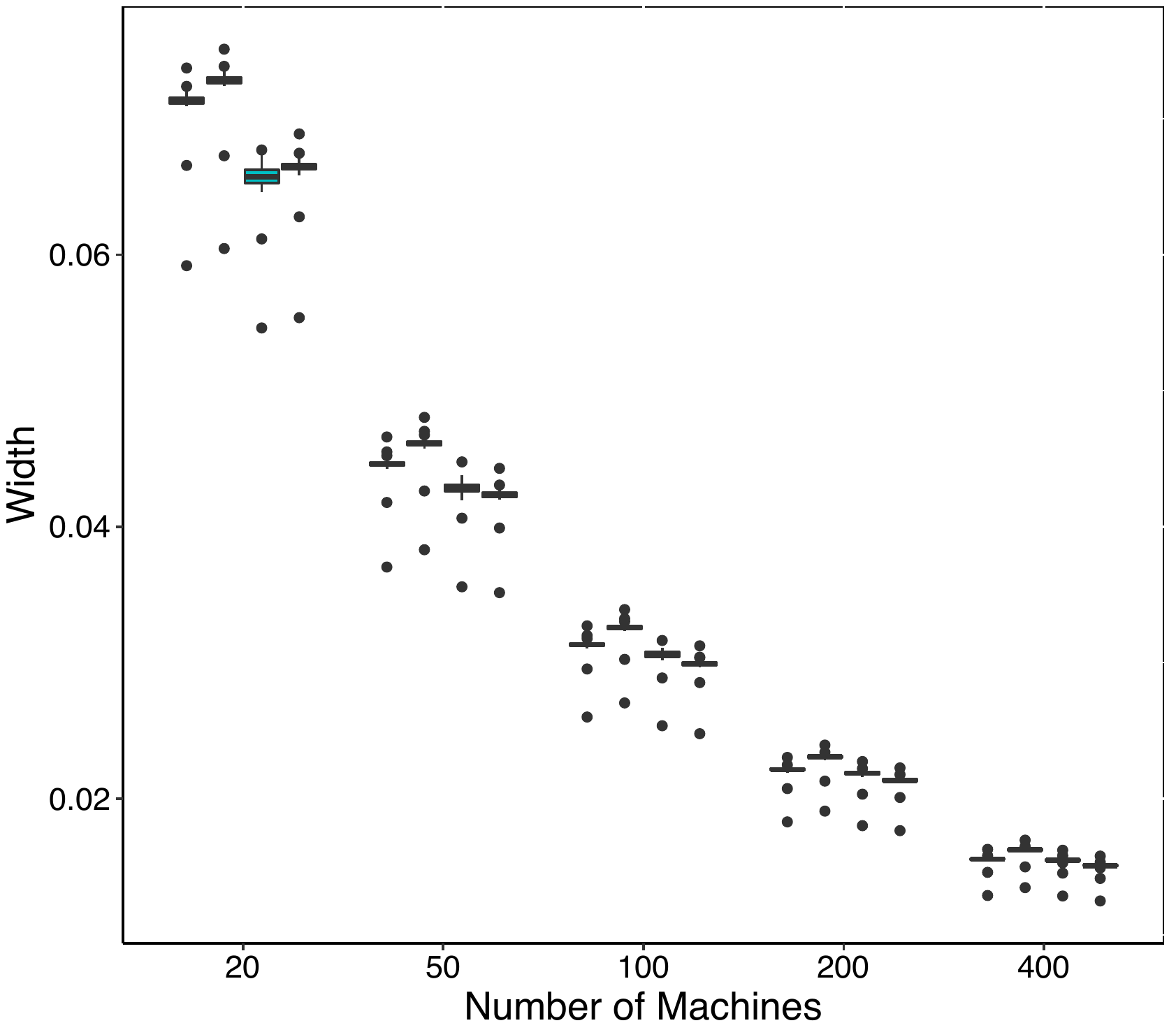}}
   \subfigure[]{\includegraphics[scale=0.403]{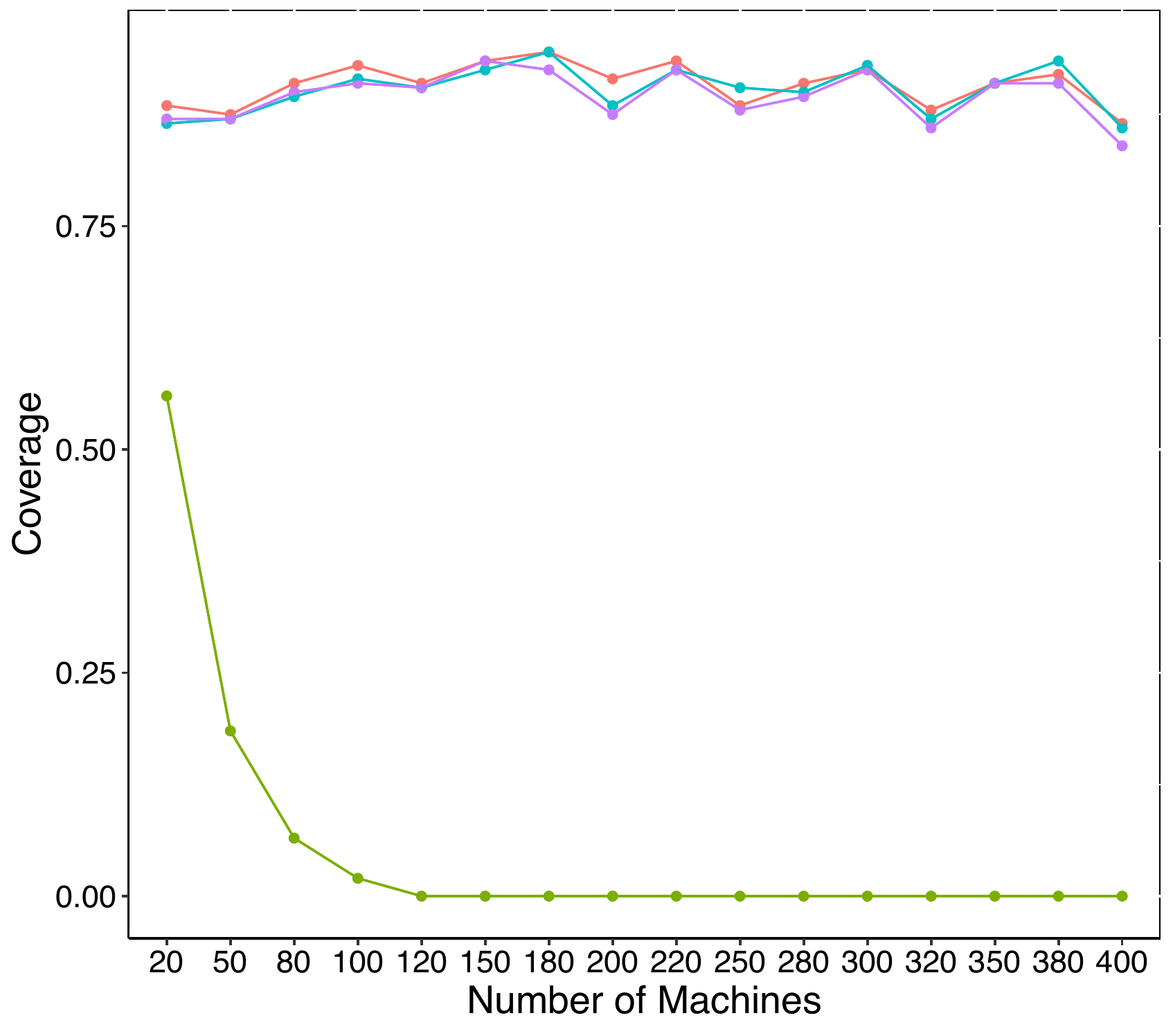}}
\end{center}
\caption{\label{fig:inf2} Confidence intervals for the regression coefficients under model \eqref{eq:highhet}. Other details are the same as Figure~\ref{fig:inf1}. }
\end{figure}

Recall that the normal-based method requires estimating the asymptotic variances $\sigma_j^2$, and the bootstrap methods depend on $\Hb^{-1}$.
 We consider two types of variance estimators.  The first one is easier to implement but relies on the assumption that $\Hb = f_{\varepsilon}(0) \cdot  \Sigma$, which holds when $\varepsilon$ is independent of $\bx$.  In this case, $\sigma_j^2 = \tau(1-\tau)\{ f_\varepsilon(0) \}^{-2} (\Sigma^{-1})_{jj}$. We compute the  global density estimator $\hat{f}_{\varepsilon}(0)$ as in \eqref{KDE} with a rule-of-thumb bandwidth $b_N^{\mathrm{rot}}=N^{-1/3}\cdot  \Phi^{-1}(1-\alpha/2)^{2/3} [  \{1.5 \cdot  \phi(\Phi^{-1}(\tau))^2 \}/\{2 \Phi^{-1}(\tau)^2 + 1\}]^{1/3} $, and a local covariance matrix estimator $\hat \Sigma_1$.
The second estimator is more general and takes the form $\hat \sigma_j  =    \{\tau (1-\tau )\}^{1/2}  \, \bigl( \hat \Hb_{1,b} ^{-1} \hat \Sigma_1 \hat \Hb_{1,b} ^{-1}  \bigr)^{1/2}_{jj}$, where $ \hat \Hb_{1,b}$ is given in \eqref{kernel.matrix.est1}. 
We generate the design matrix  the same way as in Section~\ref{subsec:estimation}, and focus on the following linear heteroscedastic models with different levels of heterogeneity:
\begin{equation}
\label{eq:lowhet}
y_i = \bx_i^\T \bbeta^* + (0.2 x_{ip}+1)\{\varepsilon_i- F^{-1}_{\varepsilon_i}(\tau)\};
\end{equation}
\begin{equation}
\label{eq:highhet}
y_i = \bx_i^\T \bbeta^* + (0.4 x_{ip}+1)\{\varepsilon_i- F^{-1}_{\varepsilon_i}(\tau)\}.
\end{equation}
We set $p=50$, $n=2000$, $\tau=0.4$ and let $m$ vary from 20 to 400. The results for 95\% confidence intervals  are reported in Figures~\ref{fig:inf1} and \ref{fig:inf2}.

For all of our numerical results, we found that the coverage probabilities and widths of the 95\% confidence intervals for the first $p-1$ regression coefficients (independent of the random noise) are similar across all methods.  
Specifically, the proposed CE-Normal and CE-Boot (a) \& (b) methods perform very well across various model settings.  Since the results across all methods are similar, they are omitted due to limited space. 

We focus on reporting the empirical coverage probabilities and widths of the 95\% confidence intervals   for the last regression coefficient in Figures~\ref{fig:inf1} and \ref{fig:inf2}.  
We use the first type of variance estimators in the top panels and the second type in the bottom panels.
From panels (b) and (d) in Figures~\ref{fig:inf1} and \ref{fig:inf2}, we see that the normal-based method for the simple averaging estimator suffers from severe undercoverage when the heterogeneous covariate effect is strong, which in our case, comes from the last covariate.

Score-based confidence sets, while computationally more intensive due to inversion of the test, are extremely efficient due to the linearity of the self-normalized representation exploited in our construction. We illustrate the improvements in a smaller scale simulation study in Section~\ref{sec:extraSim} of the Appendix.

\subsection{Distributed penalized quantile regression}
\label{subsec:estimationhighd}
The following numerical study illustrates the performance of the procedure proposed in Section \ref{sec:hd}, when the dimension $p$ is larger than $n$ for each of the $m$ sources. For comparison purposes, we also consider the $\ell_1$-penalized conquer ($\ell_1$-conquer) fitted to all $N=nm$ observations, which is practically infeasible for the problems that motivated our work, and the simple averaging estimator--the average of $m$ local $\ell_1$-conquer estimates. The performance of the proposed procedure is shown for $T=1$ and for $T$ chosen adaptively using the stopping criterion in Algorithm~\ref{algo:dqr}. As previously discussed, we use the Gaussian kernel for smoothing and the simple averaging estimator as the initialization.

We consider the following heteroscedastic models with $s=5$ significant variables:
\begin{enumerate}
\item Linear heteroscedasticity: $y_i = 3 + \sum_{j=1}^5 x_{ij} + (0.2 x_{i1}+1)\{\varepsilon_i- F^{-1}_{\varepsilon_i}(\tau)\}$;
\item Quadratic heteroscedasticity: $y_i =3 + \sum_{j=1}^5 x_{ij} + 0.5 \{1+ (0.25x_{ip}-1)^2\}\{\varepsilon_i- F^{-1}_{\varepsilon_i}(\tau)\}$,
\end{enumerate}
where $\bx_i$ and $\varepsilon_i$ are generated the same way as in Section~\ref{subsec:estimation}. Moreover, we set $\tau=0.8$, $p=500$, $n=400$ and $m\in \{20,40,60,80,100,120\}$.

\begin{figure}[!htp]
\begin{center}
   \subfigure[]{\includegraphics[scale=0.485]{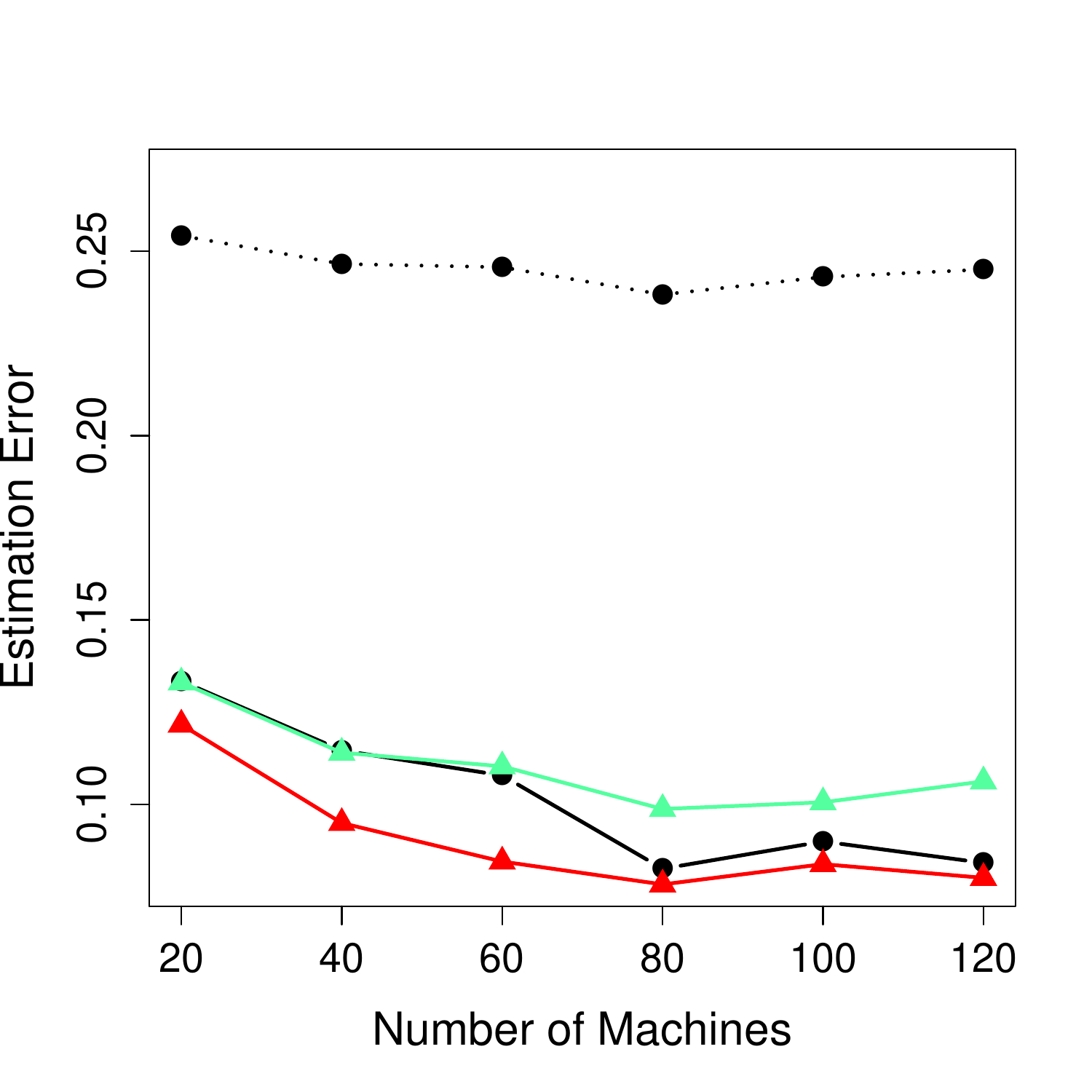}} 
   \subfigure[]{\includegraphics[scale=0.485]{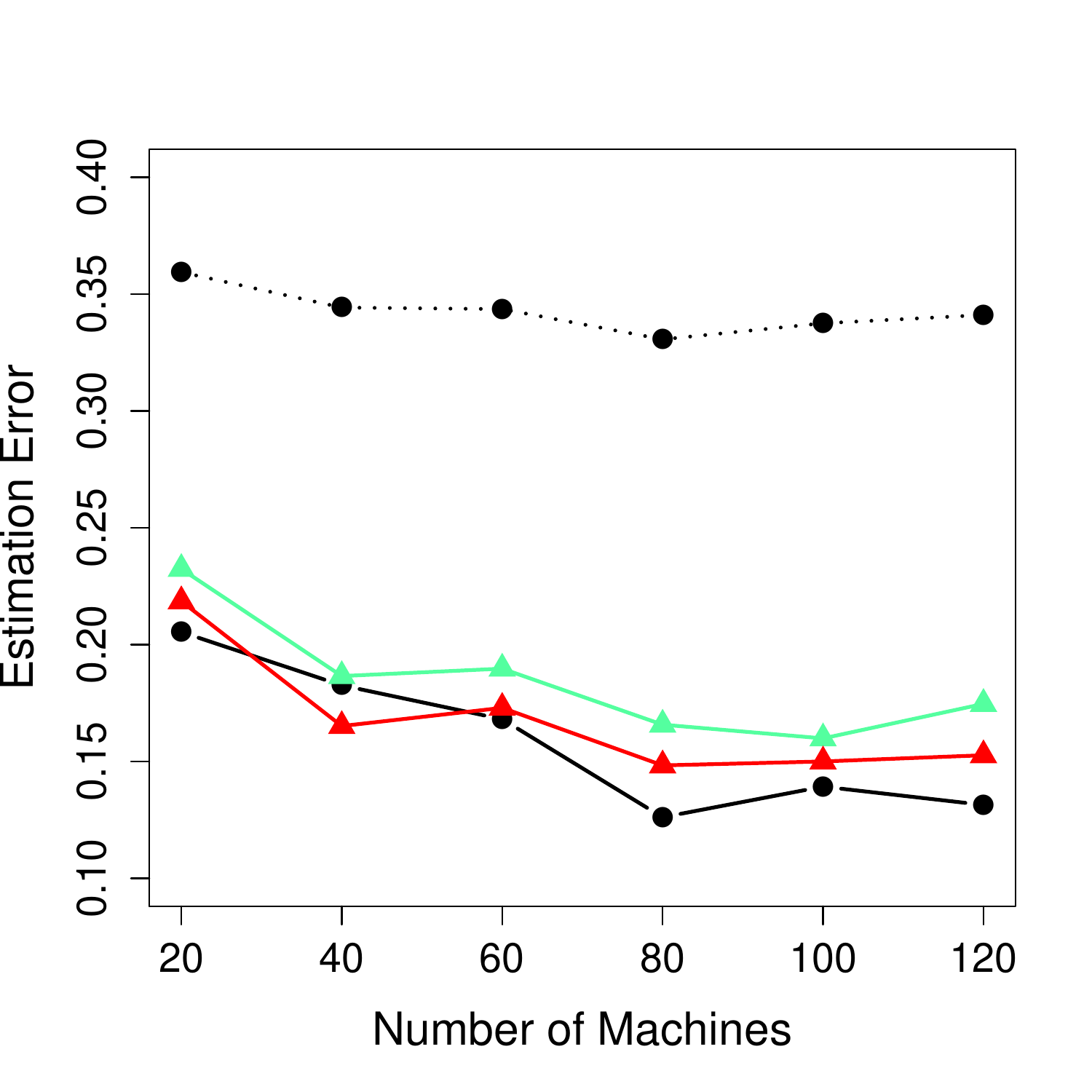}}
\end{center}
\caption{\label{fig:simhighd}  Estimation error as a function of $m$ under linear (panel (a)) and quadratic (panel (b)) heteroscedastic models with $t_{1.5}$ noise.  Each point corresponds to the average of 100 Monte Carlo replications for $(n,p)= (400,500)$. Three methods are implemented: (i) the multi-round method with $T=1$ (\protect\includegraphics[height=0.7em]{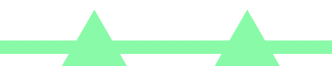}) and with $T=10$ as   in Algorithm~\ref{algo:dqr} (\protect\includegraphics[height=0.7em]{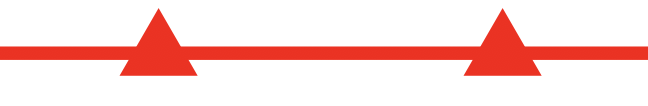}); (ii) the global $\ell_1$-conquer estimator  (\protect\includegraphics[height=0.7em]{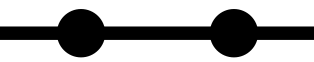}); (iii) the simple averaging estimator (\protect\includegraphics[height=0.7em]{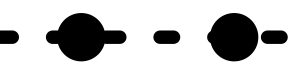}).  
}
\end{figure}

Guided by the theoretical results in Section~\ref{sec:hd}, we set the bandwidths $(b, h)$ as $b = 0.75 s^{1/2} \{\log (p)/n\}^{1/4}$ and $h = 0.75 \{s \log(p)/N\}^{1/4}$.  The regularization parameter $\lambda>0$ is selected using a validation set of size $n=200$  for easier illustration and comparison. 
Note that \cite{WKSZ2017} use 60\% of data for training, 20\% as held-out validation set for tuning the parameters, and the remaining 20\% for testing.
Figure~\ref{fig:simhighd} provides plots of the statistical error versus number of machines, averaged over 100 Monte Carlo replications, for the proposed distributed estimator and the simple averaging estimator. The latter performs poorly under both heteroscedastic models. This is not surprising because $\ell_1$-penalization induces visible finite-sample bias into the estimates, which is unaffected by aggregation no matter how many machines are available. 
Using this estimate as an initial value, the multi-round procedure considerably reduces the estimation error after one round of communication ($T=1$), and eventually performs almost as well as the global $\ell_1$-conquer when $T$ is automatically determined by the stopping criterion. 
Since $\lambda$ is tuned the same way for all the three methods, the global $\ell_1$-conquer estimator does not necessarily have the best performance but still provides a yardstick for distributed estimators.

\acks

\vspace{.1cm}
\noindent
We sincerely thank the Action Editor and two anonymous reviewers for their constructive comments that help improve the previous version of the manuscript.
K.\,M. Tan was supported by NSF Grant DMS-1949730 and DMS-2113356. H. Battey was supported by the EPSRC Fellowship EP/T01864X/1. W.-X. Zhou acknowledges the support of the NSF Grant DMS-2113409.

\newpage
\appendix

\section{Optimization Algorithms}
\subsection{First-order algorithm for solving~\eqref{one.step.conquer}}
\label{sec:algorithm}

Given an initial estimator $\tilde{\bbeta}^{(0)}$ of $\bbeta^*$, and the global and local bandwidths $b ,h>0$, recall from~\eqref{surrogate.loss} that the shifted conquer loss function takes the form 
$\tilde{\cQ}(\bbeta) = \hat{\cQ}_{1,b}(\bbeta)  - \langle\nabla \hat{\cQ}_{1,b}(\tilde{\bbeta}^{(0)}) - \nabla \hat{\cQ}_{h}(\tilde{\bbeta}^{(0)})  ,\bbeta  \rangle$. 
The communication-efficient procedure involves repeatedly minimizing $\bbeta \mapsto \tilde{\cQ}(\bbeta)$.
Since $\tilde{\cQ}(\bbeta)$ is smooth and convex, we employ the gradient descent (GD) method 
which, at the $k$th iteration,  computes 
\begin{equation}
\label{eq:gradientdescent}
\hat{\bbeta}^{k+1} = \hat{\bbeta}^{k} - \eta_k \cdot \nabla \tilde{\cQ}(\hat{\bbeta}^{k}),
\end{equation} 
where $\eta_k>0$ is the stepsize. The choice of stepsize is an important aspect of GD for achieving fast convergence, and has been thoroughly studied  in the optimization literature.
Alternatively, one can set the stepsize to be the inverse Hessian, namely, \{$\nabla^2 \tilde{\cQ}(\hat{\bbeta}^k)\}^{-1}$, which leads to the Newton-Raphson method.  The Newton step is computationally expensive at each iteration when $p$ is large. Moreover, when the quantile level $\tau$ is close to 0 or 1, $\nabla^2 \tilde{\cQ}(\hat{\bbeta}^k)$ may have a large condition number, thus causing unstableness in computing its inverse.

Motivated by the gradient-based method proposed by \cite{HPTZ2020} for solving \eqref{conquer}, we consider the use of Barzilai-Borwein stepsize in \eqref{eq:gradientdescent} \citep{BB1988}.
The main idea of the Barzilai-Borwein method is to seek a simple approximation of the inverse Hessian without having to compute it explicitly.  
In particular, for $k=1,2,\ldots$, the Barzilai-Borwein stepsizes are defined as
\#
\left\{\begin{array}{ll}
\eta_{1,k} = \frac{\langle \hat \bbeta^k - \hat \bbeta^{k-1}  , \hat \bbeta^k - \hat \bbeta^{k-1}   \rangle}{\langle \hat \bbeta^k - \hat \bbeta^{k-1}  ,  \nabla \tilde Q  (\hat \bbeta^k) - \nabla \tilde Q (\hat \bbeta^{k-1})\rangle} ,  \vspace{0.2cm}\\
\eta_{2,k} = \frac{\langle \hat \bbeta^k - \hat \bbeta^{k-1}  , \nabla \tilde Q  (\hat \bbeta^k) - \nabla \tilde Q (\hat \bbeta^{k-1})\rangle}{\langle \nabla \tilde Q  (\hat \bbeta^k) - \nabla \tilde Q (\hat \bbeta^{k-1})  ,  \nabla \tilde Q  (\hat \bbeta^k) - \nabla \tilde Q (\hat \bbeta^{k-1})\rangle}.
\end{array}\right. \label{bbstepsize}
\#
When the quantile level $\tau$ approaches 0 or 1, the objective function is flat in some directions and hence the Hessian matrix becomes more ill-conditioned.   To stablize the  algorithm, we set the stepsize to be $\eta_k = \min (\eta_{1,k},\eta_{2,k},C)$ ($k=1,2,\ldots$) for some constant $C>0$, say $C=20$.
The pseudo-code for the above Barzilai-Borwein GD method for solving \eqref{surrogate.loss} is given in Algorithm~\ref{algo:gdbb}.

\begin{algorithm}[!htp]
    \caption{ {\small  Gradient descent with Barzilai-Borwein stepsize (GD-BB) for solving \eqref{surrogate.loss}.}}
    \label{algo:gdbb}
    \textbf{Input:} Local data vectors $\{(y_i, \bx_i)\}_{i\in \cI_1}$, $\tau\in (0,1)$, bandwidth $b\in (0,1)$, initialization $\hat{\bbeta}^{0}=\wt{\bbeta}^{(0)}$, gradient vectors $\nabla \hat{\cQ}_{1,b}(\tilde{\bbeta}^{(0)})$ and $\nabla \hat{\cQ}_{h}(\tilde{\bbeta}^{(0)})$, and convergence tolerance $\delta$.
    
    \begin{algorithmic}[1]
    \STATE  Compute  $\hat\bbeta^1 \leftarrow \hat \bbeta^0 -  \nabla \tilde Q( \hat \bbeta^0)$.
      \FOR{$k=1,2 \ldots $}
          \STATE Compute step sizes $\eta_{1,k}$ and $\eta_{2,k}$ defined in~\eqref{bbstepsize}. 
          \STATE Set $\eta_k \leftarrow  \min\{\eta_{1,k},\eta_{2,k}, 20\}$ if $\eta_{1,k},\eta_{2,k} >0$, and $\eta_k \leftarrow 1$ otherwise.
          \STATE Update $ \hat \bbeta^{k+1} \leftarrow \hat \bbeta^k -  \eta_k \nabla \tilde \cQ (\hat \bbeta^{k})$.
      \ENDFOR~when $\| \nabla \tilde \cQ (\hat \bbeta^{k}) \|_2\le \delta$.
    \end{algorithmic}
\end{algorithm}

\subsection{Local adaptive majorize-minimize algorithm for solving~\eqref{one.step.l1.conquer}}
\label{sec:algorithmhighd}
In this section, we provide an algorithm to solve the $\ell_1$-penalized shifted conquer loss minimization.  In particular, given an initial estimator $\tilde{\bbeta}^{(0)}$, the $\ell_1$-penalized shifted conquer loss takes the form
\begin{equation}
\label{eq:l1drloss}
\tilde{\cQ}(\bbeta)+ \lambda   \|\bbeta_-\|_1, ~\mathrm{where}~ \tilde{\cQ}(\bbeta)=\hat{\cQ}_{1,b}(\bbeta)  - \langle\nabla \hat{\cQ}_{1,b}(\tilde{\bbeta}^{(0)}) - \nabla \hat{\cQ}_{h}(\tilde{\bbeta}^{(0)})  ,\bbeta  \rangle.
\end{equation}
Here $\bbeta_- \in \RR^{p-1}$ denotes the subvector of $\bbeta\in \RR^p$ with its first coordinate removed.
Due to the non-differentiability of the $\ell_1$-norm, the GD-BB method in Algorithm~\ref{algo:gdbb} is no longer applicable.
By extending the majorize-minimize (MM) algorithm \citep{HL2000} for standard quantile regressions, we employ a local adaptive majorize-minimize (LAMM)  principle \citep{FLSF2018} to minimize the penalized conquer loss $\bbeta \mapsto \tilde{\cQ}(\bbeta)+ \lambda    \|\bbeta_- \|_1$.

At the $k$th iteration with a previous estimate $\hat{\bbeta}^{k-1}$, the main idea of the LAMM algorithm is to construct an isotropic quadratic function that locally majorizes the shifted conquer loss function $\tilde{\cQ}(\cdot)$. 
Specifically, for some quadratic parameter $\phi_k>0$, we define the quadratic function
\[
F(\bbeta;\phi^{k},\hat{\bbeta}^{k-1}) = \tilde{\cQ}(\hat{\bbeta}^{k-1}) + \langle \nabla \tilde{\cQ}(\hat{\bbeta}^{k-1}) , \bbeta-\hat{\bbeta}^{k-1}   \rangle + \frac{\phi_{k}}{2} \|\bbeta-\hat{\bbeta}^{k-1}\|_2^2 ,
\]
and then compute the update $\hat \bbeta^k$ by solving
\#
\underset{\bbeta \in \RR^p}{\mathrm{minimize}}~  \big\{  F(\bbeta;\phi_{k},\hat{\bbeta}^{k-1}) + \lambda \| \bbeta_-  \|_1 \big\}. \label{iso.quad.min}
\#
The isotropic form of $F(\bbeta;\phi_{k},\hat{\bbeta}^{k-1}) $, as a function of $\bbeta$, permits a simple analytic solution $\hat \bbeta^k = (\hat \beta^k_1, \ldots , \hat \beta^k_p)^\T$ that takes the form  
\#
\left\{\begin{array}{ll}
 \hat{\beta}_1^{k}  = \hat{\beta}_1^{k-1} -  \phi_{k}^{-1}  \nabla_{\beta_1} \tilde{\cQ}(\hat{\bbeta}^{k-1}) ,  \vspace{0.2cm}\\
\hat{\beta}^{k}_j  = S(\hat{\beta}^{k-1}_j- \phi_{k}^{-1}   \nabla_{\beta_j} \tilde{\cQ}(\hat{\bbeta}^{k-1}) , \, \phi_{k}^{-1}   \lambda )~~\mathrm{ for~} j =2,\ldots,p ,
\end{array}\right.  \nn
\#
where $S(a,b) = \mathrm{sign}(a) \cdot \max (|a|-b,0)$ is the soft-thresholding operator.
To enforce overall descent of the function value, we need $\phi_k>0$ to be sufficiently large so that $F(\hat{\bbeta}^{k};\phi_{k},\hat{\bbeta}^{k-1}) \ge \tilde{\cQ}(\hat{\bbeta}^{k})$, and hence 
\$
\tilde{\cQ}(\hat{\bbeta}^{k}) + \lambda \| \hat \bbeta^k_- \|_1  &  \leq F(\hat{\bbeta}^{k};\phi_{k},\hat{\bbeta}^{k-1}) +   \lambda \| \hat \bbeta^k_- \|_1  \\
&  \leq F(\hat{\bbeta}^{k-1};\phi_{k},\hat{\bbeta}^{k-1})  + \lambda \| \hat \bbeta^{k-1}_- \|_1  = \tilde{\cQ}(\hat{\bbeta}^{k-1}) + \lambda \| \hat \bbeta^{k-1}_- \|_1 .
\$
To choose a proper $\phi_k$ in practice, we start from a relatively small  value $\phi_{k,0} = 0.001$, and successively inflate it by a factor $\gamma=1.1$---$\phi_{k,\ell} = \gamma \phi_{k,\ell-1}$ for $\ell=1,2,\ldots$---until the majorization requirement is met. See Algorithm~\ref{algo:ilamm} for the pseudo-code of the LAMM algorithm for solving \eqref{one.step.l1.conquer}.

\begin{algorithm}[!htp]
    \caption{ {\small  Local adaptive majorize-minimize (LAMM) algorithm for solving \eqref{one.step.l1.conquer}.}}
    \label{algo:ilamm}
    \textbf{Input:} Local data vectors $\{(y_i, \bx_i)\}_{i\in \cI_1}$, $\tau\in (0,1)$, bandwidth $b\in (0,1)$, initialization  $\hat{\bbeta}^{0} = \wt \bbeta^{(0)}$, gradient vectors $\nabla \hat{\cQ}_{1,b}(\tilde{\bbeta}^{(0)})$ and $\nabla \hat{\cQ}_{h}(\tilde{\bbeta}^{(0)})$,  regularization parameter $\lambda>0$, isotropic parameter $\phi_0$ and convergence tolerance $\delta$.
    
    \begin{algorithmic}[1]
          \FOR{$k=1,2 \ldots $}
          \STATE Set $\phi_k \leftarrow \max \{\phi_0,  \phi_{k-1}/1.1\}.$
          \REPEAT
          \STATE Update $\hat{\beta}_1^{k} \leftarrow \hat{\beta}_1^{k-1} - \phi_{k}^{-1} \nabla_{\beta_1} \tilde{\cQ}(\hat{\bbeta}^{k-1})$.
          \STATE Update $\hat{\beta}^{k}_j \leftarrow S(\hat{\beta}^{k-1}_j- \phi_{k}^{-1}  \nabla_{\beta_j} \tilde{\cQ}(\hat{\bbeta}^{k-1}) , \, \phi_{k}^{-1}  \lambda )~ \mathrm{for~} j =2,\ldots,p$.
                    \STATE \textbf{If} $F(\hat{\bbeta}^{k};\phi_{k},\hat{\bbeta}^{k-1}) < \tilde{\cQ}(\hat{\bbeta}^{k})$, set $\phi_{k} \leftarrow 1.1 \phi_{k}$.
          \UNTIL $F(\hat{\bbeta}^{k};\phi_{k},\hat{\bbeta}^{k-1}) \ge \tilde{\cQ}(\hat{\bbeta}^{k})$.
      \ENDFOR~when $\| \hat{\bbeta}^{k}-\hat{\bbeta}^{k-1} \|_2\le \delta$.
    \end{algorithmic}
\end{algorithm}

\section{Proof of Main Results}

\subsection{Supporting lemmas}

Recall that the data vector $( y, \bx) \in \RR \times \RR^p$ satisfies the conditional quantile model $Q_\tau(y|\bx) = F_{y|\bx}^{-1}(\tau) = \bx^\T \bbeta^*$. Equivalently, $y= \bx^\T \bbeta^* + \varepsilon $ with $Q_\tau(\varepsilon |\bx) = 0$.
Under Condition~(C2), $\bz = \Sigma^{-1/2} \bx \in \RR^p$ denotes the standardized  vector of covariates such that $\EE(\bz \bz^\T ) = \Ib_p$. For every $\delta \in (0, 1]$, define
\#
	\gamma_\delta = \inf\big\{ \gamma>0 :  \sup\nolimits_{\bu \in \mathbb{S}^{p-1}} \EE\big\{ (\bz^\T \bu )^2 \mathbbm{1}(|\bz^\T \bu |> \gamma_\delta )  \big\} \leq \delta  \big\}. \label{def:gamma}
\#
By the sub-Gaussian assumption on $\bx$, $\gamma_\delta$ depends only on $\delta$ and $\upsilon_1$, and the map $\delta \mapsto \gamma_\delta$ is non-increasing with $\gamma_\delta \downarrow 0$ as $\delta \to 1$. 
Under a   weaker condition that $\bz$ has {\it uniformly bounded fourth moments}, namely, $\mu_4 = \sup_{\bu \in  \mathbb{S}^{p-1}} \EE (\bz^\T \bu)^4 <\infty$, we have $\gamma_\delta \leq  (\mu_4 / \delta)^{1/2}$.

For the conditional density $f_{\varepsilon | \bx}(\cdot)$, Condition~(C1) ensures that as long as $b$ is sufficiently small, 
$$
	   \underbar{$f$}_b \leq \min_{|u|\leq b/2} f_{\varepsilon | \bx} (u) \leq  \max_{|u|\leq b/2} f_{\varepsilon | \bx} (u) \leq \bar f_b ~\mbox{ almost surely (over } \bx  )
$$
for some constants  $\bar f_b  \geq \underbar{$f$}_b  >0$. For example, we may take $\underbar{$f$}_b =  \underbar{$f$} -l_0 b/2$ and $\bar f_b  =  \bar{f} +l_0 b/2$.
Given a kernel $K(\cdot)$ and bandwidth $b>0$, recall that $\hat \cQ_{1,b}(\bbeta) = (1/n)\sum_{i\in \cI_1 } \ell_b( y_i - \bx_i^\T \bbeta)$ denotes the local smoothed loss function, where $\ell_b(\cdot) = (\rho_\tau * K_b)(\cdot)$. 
The proof of Theorem~\ref{thm:one-step} depends heavily on the local strong convexity of $\hat \cQ_{1,b}(\cdot)$ in a neighborhood of $\bbeta^*$. To this end, we first introduce the notion of symmetrized Bregman divergence. For any differentiable convex function $\psi : \RR^k \to \RR$ ($k\geq 1$), the corresponding Bregman divergence is given by $D_{\psi}(\bw', \bw) = \psi(\bw') - \psi(\bw) - \langle \nabla \psi(\bw), \bw' - \bw \rangle$. Define its symmetrized version as
\#
	\overbar D_{\psi}(\bw, \bw') = D_{\psi}(\bw , \bw') + D_{\psi}(\bw', \bw)  = \big\langle \nabla \psi(\bw) - \nabla \psi(\bw'), \bw - \bw' \big\rangle, \ \ \bw, \bw' \in \RR^k. \label{bregman.def}
\#
The first two lemmas, Lemma~\ref{lem:local.RSC} and \ref{lem:global.score}, provide a lower bound on the symmetrized Bregman divergence of the shifted conquer loss  $\wt \cQ(\cdot)$ given in \eqref{surrogate.loss} and an upper bound on the global gradient, respectively. These two results coincide with Lemmas~A.1 and A.2 in the supplement of \cite{HPTZ2020}.  We reproduce them here for the sake of readability. In particular, the former implies the restricted strong convexity property of $\wt \cQ(\cdot)$.

\begin{lemma}   \label{lem:local.RSC}
For any  $x>0$ and $0< r\leq b/(4\gamma_{0.25}) $, 
\#   \label{rsc.lbd}
	\inf_{\bbeta \in \Theta(r)} \frac{\overbar D_{\wt \cQ}(\bbeta, \bbeta^* ) }{\kappa_l \| \bbeta - \bbeta^* \|_{\Sigma}^2 } \geq \frac{3}{4} \underbar{$f$}_b - \bar{f}_b^{1/2} \Biggl( \frac{5}{4} \sqrt{\frac{b p}{r^2 n}} + \sqrt{\frac{b x}{8 r^2 n}} \, \Bigg) - \frac{b x}{3 r^2 n}
\#
with probability at least $1-e^{-x}$, where $\kappa_l = \min_{|u|\leq 1} K(u) >0$.
\end{lemma}

Consider the gradient $\nabla \hat \cQ_h(\cdot)$ evaluated at $\bbeta^*$, namely,
$$
	\nabla \hat \cQ_h(\bbeta^*) = \frac{1}{N} \sum_{i=1}^N \bigl\{ \overbar K(-\varepsilon_i / h ) - \tau \bigr\} \bx_i, 
$$
where $\varepsilon_i = y_i - \bx_i^\T \bbeta^*$. The following lemma provides an upper bound on the  $\ell_2$-norm of $\nabla \hat \cQ_h(\bbeta^*) $. Recall that $\Omega = \Sigma^{-1}$, and we write $\| \bu \|_{\Omega} = \|\Sigma^{-1/2} \bu \|_2$ for $\bu \in \RR^p$.

\begin{lemma} \label{lem:global.score}
Conditions~(C1)--(C3) ensure that, for any $x>0$, 
\#
	 \|   \nabla \hat  \cQ_h(\bbeta^* ) \|_{\Omega} \leq C_0 \bigg(   \sqrt{\frac{p+x}{N}} + h^2 \bigg) \label{global.score.bound}
\#
with probability at least $1-e^{-x}$ as long as $N \gtrsim p+x$, where  $C_0>0$ is a constant depending only on $( \tau,l_0, \upsilon_1,  \kappa_2)$.
\end{lemma}

Next, we extend the above results to the high-dimensional setting in which $p\gg n$. Recall that $\Theta(r)$ ($r>0$) denotes the local $\ell_2$ neighborhood of $\bbeta^*$ under $\|\cdot\|_{\Sigma}$-norm. Furthermore, define the $\ell_1$-cone $\Lambda$ as in \eqref{l1.cone}, that is,
\#
	\Lambda  = \bigl\{  \bbeta \in \RR^p: \|  \bbeta - \bbeta^* \|_1 \leq 4 s^{1/2} \|   \bbeta - \bbeta^*  \|_{\Sigma}   \bigr\}   .  \label{def.cone}
\#
 

\begin{lemma} \label{lem:hd.RSC}
Assume Conditions~(C1), (C2) and (C4) hold. Then, for any $x>0$, $0< r \leq b/(4\gamma_{0.25})$ and $L>0$,
\# \label{hd.rsc.lbd}
\inf_{\bbeta \in \Theta(r) \cap \Lambda } \frac{\overbar D_{\wt \cQ}(\bbeta, \bbeta^* ) }{\kappa_l \| \bbeta - \bbeta^* \|_{\Sigma}^2 } 
\geq   \frac{3}{4} \underbar{$f$}_b   - \bar f_b^{1/2}  \left\{  5 B  \sqrt{\frac{2 b s \log(2p)}{ r^2 n}} + \sqrt{\frac{b x}{8 r^2 n}}  \right\}  -  \frac{b x }{3 r^2 n}
\#
with probability at least $1-e^{-x}$.
\end{lemma}

\noindent \emph{Proof of Lemma~\ref{lem:hd.RSC}}:
Following the proof of Lemma~4.1 in \cite{TWZ2020}, it suffices to bound
\#
	\EE \bigg\| \frac{1}{n} \sn e_i \psi_{b/2}(\varepsilon_i) \bx_i \bigg\|_\infty = \EE \, \EE_e \bigg\| \frac{1}{n} \sn e_i \psi_{b/2}(\varepsilon_i) \bx_i \bigg\|_\infty  ,  \label{max.rad.average1}
\#
where $e_1,\ldots, e_n$ are i.i.d. Rademacher random variables,   $\psi_{b/2}(\varepsilon_i) = \mathbbm{1}(|\varepsilon_i |\leq b/2)$, and $\EE_e$ denotes the (conditional) expectation over $e_1, \ldots, e_n$ given all the remaining random variables.
Applying Hoeffding's moment inequality yields
\$
	& \EE_e  \bigg\| \frac{1}{n} \sn e_i \psi_{b/2}(\varepsilon_i) \bx_i \bigg\|_\infty \\
	&  \leq \max_{1\leq j\leq p} \bigg\{ \frac{1}{n} \sn  x_{ij}^2 \psi_{b/2}^2(\varepsilon_i)  \bigg\}^{1/2} \sqrt{\frac{2\log(2p)}{n}} \leq  \bigg\{ \frac{1}{n} \sn   \psi_{b/2} (\varepsilon_i)  \bigg\}^{1/2} B  \sqrt{\frac{2\log(2p)}{n}} .
\$
Moreover, note that $\EE \{ \psi_{b/2} (\varepsilon_i) | \bx_i  \} \leq \bar f_b \cdot b$. Substituting these bounds into \eqref{max.rad.average1}, we obtain that
\#
\EE \bigg\| \frac{1}{n} \sn e_i \psi_{b/2}(\varepsilon_i) \bx_i \bigg\|_\infty  \leq  (2\bar f_b )^{1/2} B\sqrt{\frac{ b\log(2p)}{n}} . \nn
\#
Keep the rest of the proof the same, we obtain the claimed bound \eqref{hd.rsc.lbd}.

For the empirical loss $\hat \cQ_h(\bbeta) = (1/N) \sum_{i=1}^N \ell_h(y_i - \bx_i^\T \bbeta)$, define its population counterpart $\cQ_h(\bbeta) = \EE \hat \cQ_h(\bbeta) = \EE_{(y,\bx) \sim P} \{ \ell_h( y - \bx^\T \bbeta)\}$.

\begin{lemma} \label{lem:hd.score}
Conditions~(C1), (C2) and (C4) ensure that, for any $x>0$, 
\#
  \|  \nabla \hat  \cQ_h(\bbeta^*) -   \nabla  \cQ_h(\bbeta^*) \|_\infty \leq C(\tau, h) \sqrt{  \frac{\log(2p) +x}{N}} + B\max(\tau, 1-\tau)  \frac{\log(2p) + x}{ 3N}  \nn
\#
with probability at least $1-e^{-x}$, where $C(\tau, h)  = \sigma_u \sqrt{2 \{\tau(1-\tau ) + (1+\tau)l_0 \kappa_2  h^2   \} }$ and $\sigma_u = \max_{1\leq j\leq p} \sigma_{jj}^{1/2}$.
\end{lemma}

\noindent \emph{Proof of Lemma~\ref{lem:hd.score}}:
To begin with, write
\$
 \|  \nabla \cQ_h(\bbeta^*) - \nabla  \cQ_h(\bbeta^*)  \|_\infty =\max_{1\leq j\leq p} \bigg| \frac{1}{N} \sum_{i=1}^N (1-\EE)  w_i  x_{ij} \bigg| ,
\$
where $w_i = \overbar K(-\varepsilon_i/ h) - \tau$ for $i=1,\ldots, n$, and note that $|w_i x_{ij}|\leq  B \bar \tau$ with  $\bar \tau := \max(\tau, 1-\tau)$. Using Taylor series expansion and integration by parts, we obtain that
\$
 | \EE \bigl( w_i  | \bx_i \bigr)  | \leq  0.5 l_0 \kappa_2 h^2 ~\mbox{ and }~\EE \bigl\{ \overbar K^2(-\varepsilon_i/h) | \bx_i \bigr\} \leq \tau +l_0 \kappa^2 h^2 ,
\$
which in turn implies $\EE (w_i^2 | \bx_i) \leq \tau(1-\tau ) + (1+\tau)l_0  \kappa_2 h^2 =\tau(1-\tau ) + C h^2 $. Hence, applying Bernstein's inequality yields that, for any $1\leq j\leq p$ and $z\geq 0$,
\#
 \bigg| \frac{1}{N} \sum_{i=1}^N (1-\EE ) w_i x_{ij}  \bigg|  \leq \sigma_{jj} ^{1/2}  \sqrt{ 2\bigl\{\tau(1-\tau ) + C h^2 \bigr\} \frac{  z}{N}} +  \frac{B \bar \tau }{3} \frac{z}{ N} \nn
\#
with probability at least $1-2e^{-z}$. Taking $z=  \log(2p) + x$, the claimed bound follows immediately from the union bound.

With the above preparations, we are ready to prove the main results in the paper.

\subsection{Proof of Theorem~\ref{thm:one-step}}

\noindent
{\sc Proof of \eqref{one-step.error}}.
The proof is carried out conditioning on the ``good" event that $\wt \bbeta^{(0)} \in \Theta(r_0)$. Let $\wt \bbeta = \wt \bbeta^{(1)}$ be the one-step estimator that minimizes $\wt \cQ(\cdot)$. Set $r_{\loc} = b/(4\gamma_{0.25})$ with $\gamma_{0.25}$ given in \eqref{def:gamma},  and define an intermediate estimator $\wt \bbeta_\eta  = \bbeta^* + \eta(\wt \bbeta - \bbeta^*)$ with 
\$
\eta  = \sup\big\{ u \in [0,1]: \bbeta^* + u(\wt \bbeta - \bbeta^*) \in \Theta(r_{\loc}) \big\} \begin{cases}
 = 1   & \mbox{ if } \wt \bbeta \in \Theta(r_{\loc}) ,  \\
 \in (0, 1) & \mbox{ if } \wt \bbeta \notin \Theta(r_{\loc}) .
\end{cases}
\$
In other words, $\eta$ is the largest value of $u\in (0,1]$ such that  the corresponding convex combination of $\bbeta^*$ and $\wt \bbeta$---namely, $(1-u)\bbeta^* + u \wt \bbeta$---falls into the region $\Theta(r_0)$. Hence, if $\wt \bbeta \notin \Theta(r_{\loc}) $,  we must have $\wt \bbeta_\eta \in \partial  \Theta(r_{\loc}) = \{ \bbeta \in \RR^p: \| \bbeta - \bbeta^* \|_{\Sigma}=r_{\loc} \}$.

By Lemma~C.1 in \cite{SZF2020}, the three  points $\wt \bbeta, \wt \bbeta_\eta$ and $\bbeta^*$ satisfy
$\overbar D_{\wt \cQ}(\wt \bbeta_\eta , \bbeta^*) \leq \eta \overbar D_{\wt \cQ}(\wt \bbeta,\bbeta^*)$, where by \eqref{bregman.def}, 
$$
	\overbar D_{\wt \cQ}( \bbeta , \bbeta^*) = \bigl\langle \nabla \wt \cQ(\bbeta) - \nabla  \wt \cQ(\bbeta^*),    \bbeta -\bbeta^* \bigr\rangle 
	=  \bigl\langle  \nabla \hat \cQ_{1,b}(\bbeta) - \nabla \hat  \cQ_{1,b}(\bbeta^*) ,  \bbeta - \bbeta^* \bigr\rangle  = \overbar D_{\hat \cQ_{1,b}}( \bbeta, \bbeta^*)
$$ 
for $\bbeta \in \RR^p$.
Taking into account the first-order optimality condition $\nabla \wt \cQ(\wt \bbeta) = \textbf{0}$, it follows that
\#
	\overbar D_{\wt \cQ }(\wt \bbeta_\eta, \bbeta^* ) \leq - \eta \big\langle \nabla \wt \cQ( \bbeta^* ) , \wt \bbeta - \bbeta^* \big\rangle \leq    \|   \nabla \wt \cQ(\bbeta^* )  \|_{\Omega} \cdot  \| \wt \bbeta_\eta - \bbeta^*  \|_{\Sigma} . \label{FOC}
\#
For the left-hand side of \eqref{FOC}, applying Lemma~\ref{lem:local.RSC} yields that with probability at least $1-e^{-x}$,
\#
\overbar D_{\wt \cQ}(\bbeta, \bbeta^* ) \geq 0.5\underbar{$f$}  \kappa_l \cdot  \| \bbeta - \bbeta^* \|_{\Sigma}^2 \label{local.rsc.bound}
\#
holds uniformly over all $\bbeta \in \Theta(r_{\loc})$ as long as $ (p+x)/n \lesssim b  \lesssim 1$.

To bound the right-hand side of \eqref{FOC}, we  define vector-valued random processes
\#
\left\{\begin{array}{ll}
	\Delta_1(\bbeta)   = \Sigma^{-1/2}  \big\{ \nabla \hat \cQ_{1,b}(\bbeta) - \nabla \hat \cQ_{1,b}( \bbeta^*)  - \Hb ( \bbeta - \bbeta^* )   \big\} ,  \vspace{0.2cm}\\
\Delta(\bbeta)   = \Sigma^{-1/2}  \big\{ \nabla \hat \cQ_h(\bbeta) - \nabla \hat \cQ_h( \bbeta^*) - \Hb  ( \bbeta - \bbeta^* ) \big\} ,
\end{array}\right. \label{def:Delta}
\#
where $\Sigma = \EE(\bx \bx^\T)$ and $\Hb = \EE \{ f_{\varepsilon | \bx}(0) \bx \bx^\T \}$. Following the proof of Theorem~4.2 in \cite{HPTZ2020}, it can be  shown that, with probability at least $1-2e^{-x}$,
\#
 \sup_{\bbeta \in \Theta(r)}  \| \Delta_1(\bbeta)  \|_2  & \leq C_1  r\bigg(   \sqrt{\frac{p+x}{n b}}  + r   + b  \bigg) ~\mbox{ and }~ \sup_{\bbeta \in \Theta(r)}  \| \Delta(\bbeta)  \|_2    \leq     C_1 r  \bigg(    \sqrt{\frac{p+x}{N h}} + r  + h \bigg) \label{diff.grad.unif.bound}
\#
as long as $b \gtrsim \sqrt{(p+x)/n}$, $h\gtrsim \sqrt{(p+x)/N}$ and $n\gtrsim p+x$, where $C_1>0$ is a constant independent of $(N,n,p,h,b)$.

Recall that $\nabla \wt \cQ(\bbeta) = \nabla \wt \cQ_{1,b}(\bbeta) - \nabla \wt \cQ_{1,b}(\wt \bbeta^{(0)})+ \nabla \wt \cQ_h(\wt \bbeta^{(0)})$ and $b\geq h$. Hence, applying \eqref{diff.grad.unif.bound} yields that, conditioned on the event $\cE_0(r_0) \cap  \cE_*(r_*)$,
\#
 \|   \nabla \wt \cQ (\bbeta^* )  \|_{\Omega}  &  =  \|    \Delta(\wt \bbeta^{(0)}) - \Delta_1(\wt \bbeta^{(0)}) + \Sigma^{-1/2} \nabla \hat \cQ_h(\bbeta^* )  \|_2 \nn \\
& \leq \|   \Delta(\wt \bbeta^{(0)})  \|_2 + \|   \Delta_1(\wt \bbeta^{(0)})  \|_2 +   \|  \nabla  \hat \cQ_h(\bbeta^* )   \|_{\Omega} \nn\\
& \leq C_1   \Bigg(    \sqrt{\frac{p+x}{n b}} +   \sqrt{\frac{p+x}{N h}}  + 2 r_0 + 2b \Bigg) \cdot r_0     +   r_* . \label{surrogate.score.bound}
\#

Together, the bounds \eqref{FOC}, \eqref{local.rsc.bound} and \eqref{surrogate.score.bound} imply that, conditioned on $\cE_0(r_0) \cap \cE_*(r_*)$,
\#
	&  \| \wt \bbeta_\eta - \bbeta^*   \|_{\Sigma}   \leq 2(\underbar{$f$} \kappa_l)^{-1}    \|   \nabla \wt \cQ (\bbeta^* )  \|_{\Omega} \nn \\
	& \leq 2(\underbar{$f$} \kappa_l)^{-1}  \Biggl\{ C_1  \Bigg(    \sqrt{\frac{p+x}{n b}} +   \sqrt{\frac{p+x}{N h}}  +2r_0 +2 b \Bigg) \cdot r_0 +   r_*\Bigg\}   \label{one.step.ubd}
\#
with probability at least $1-3e^{-x}$. 
We further let the bandwidths $b\geq h>0$ satisfy $b\gtrsim \max(r_0, r_*)$ and $\sqrt{(p+x)/(nb)} +b+ \sqrt{(p+x)/(Nh) } \lesssim 1$, so that the right-hand of \eqref{one.step.ubd} is strictly less than $r_{\loc}$. Then, the intermediate point $\wt \bbeta_\eta$---a convex combination of $\bbeta^*$ and $\wt \bbeta$---falls into the interior of the local region $\Theta(r_0)$ with high probability conditioned on $\cE_0(r_0) \cap \cE_*(r_*)$.  Via proof by contradiction, we must have $\wt \bbeta\in \Theta(r_{\loc})$ and hence $\wt \bbeta_\eta = \wt \bbeta$; otherwise if $\wt \bbeta \notin \Theta(r_{\loc})$, by construction $\wt \bbeta_\eta$ lies on the boundary of $\Theta(r_{\loc})$, which is a contradiction. As a result, the bound \eqref{one.step.ubd} also applies to $\wt \bbeta$, as desired. 

\medskip
\noindent
{\sc Proof of \eqref{one-step.bahadur}}. To establish the Bahadur representation, note that the random process $\Delta_1(\cdot)$ defined in \eqref{def:Delta} can be written as $\Delta_1(\bbeta)   =\Sigma^{-1/2}  \{ \nabla \wt \cQ (\bbeta) - \nabla \wt \cQ ( \bbeta^*)   - \Hb  ( \bbeta - \bbeta^* ) \}$. Moreover, note that
\#
 \nabla \wt \cQ ( \bbeta^*)   - \nabla  \hat \cQ_h( \bbeta^*)  = \nabla \hat \cQ_{1,b}(\bbeta^*) - \nabla \hat \cQ_{1,b}(\wt \bbeta^{(0)}) + \nabla  \hat \cQ_h(\wt \bbeta^{(0)}) -  \nabla  \hat \cQ_h(\bbeta^*)      , \nn
\#
which in turn implies
\#
 \|  \nabla \wt \cQ ( \bbeta^*) - \nabla \hat \cQ_h (\bbeta^* )   \|_{\Omega} \leq  \| \Delta_1(\wt \bbeta^{(0)})   \|_2 +   \| \Delta (\wt \bbeta^{(0)})   \|_2 ,	\nn
\#
where $\Delta(\cdot)$ is given in \eqref{def:Delta}.
Recall that $\nabla \wt \cQ(\wt \bbeta)=\textbf{0}$, and  conditioned on $\cE_0(r_0) \cap \cE_*(r_*)$, 
$$
	\| \wt \bbeta - \bbeta^* \|_{\Sigma} \leq r_1 \asymp    \left( \sqrt{\frac{p+x}{nb}} + \sqrt{\frac{p+x}{N h}} + b \right) \cdot r_0 + r_* 
$$  
with high probability. Under the assumed constraints on $b,h$ and $r_0 \gtrsim  r_*$, we may assume $r_1 \leq  r_0$.
Consequently,
\#
	&  \|    \nabla \hat \cQ_h(\bbeta^*) + \Hb  ( \wt \bbeta -\bbeta^* )     \|_{\Omega}  \nn \\
& =   \|      \Sigma^{-1/2}  \nabla \hat \cQ_h(\bbeta^* )  - \Sigma^{-1/2} \nabla \wt \cQ(\bbeta^*)  - \Delta_1(\wt \bbeta)  \|_2 \nn \\
& \leq      \| \Delta_1(\wt \bbeta^{(0)})   \|_2+   \| \Delta (\wt \bbeta^{(0)})   \|_2 +  \| \Delta_1(\wt \bbeta)   \|_2  \leq 2 \sup_{\bbeta \in \Theta(r_0 )}    \| \Delta_1(\bbeta)    \|_2 + \sup_{\bbeta \in \Theta(r_0 )}    \| \Delta (\bbeta)   \|_2 \label{Bahadur1}
\#
 with the same probability conditioned on $\cE_0(r_0) \cap \cE_*(r_*)$. Combining \eqref{Bahadur1} with the bounds in \eqref{diff.grad.unif.bound} completes the proof of \eqref{one-step.bahadur}.

\subsection{Proof of Theorem~\ref{thm:multi-step}}

Recall that, for $ t =1, 2, \ldots$,  $\wt \cQ^{(t)}(\cdot)$ given in \eqref{surrogate.loss.l} denotes the shifted smoothed QR loss at iteration $t$, whose gradient and Hessian are 
$$
 \nabla \wt \cQ^{(t)}(\bbeta) = \nabla \hat \cQ_{1,b}(\bbeta) - \nabla \hat \cQ_{1,b}(\wt \bbeta^{(t-1)} )  +  \nabla \hat \cQ_h(\wt \bbeta^{(t-1) }  ) ~\mbox{ and }~  \nabla^2 \wt \cQ^{(t)}(\bbeta) =\nabla^2 \hat\cQ_{1,b}(\bbeta).
$$
Let $\Delta_1(\bbeta) $ and  $\Delta(\bbeta)$ be the stochastic processes defined in the proof of Theorem~\ref{thm:one-step}. The gradient   $\nabla \wt \cQ^{(t)}(\bbeta^*) $ can thus be written as $\Sigma^{-1/2}\nabla \wt \cQ^{(t)}(\bbeta^*)  =  \{ \Delta(\wt \bbeta^{(t-1)} ) - \Delta_1(\wt \bbeta^{(t-1)})  \} +  \Sigma^{-1/2} \nabla \hat \cQ_h(\bbeta^*)$, so that 
\#
   \|  \nabla \wt \cQ^{(t)}(\bbeta^*)    \|_{\Omega}   \leq     \| \Delta(\wt \bbeta^{(t-1)} )  \|_2 +    \| \Delta_1(\wt \bbeta^{(t-1)})  \|_2   +   \| \nabla \hat \cQ_h(\bbeta^*)   \|_{\Omega} . \label{shift.gradient.ubd}
\#

Given a sequence of iterates $\{\wt \bbeta^{(t)} \}_{t = 0, 1, \ldots , T}$, we define the ``good" events 
\#
	\cE_t( r_t ) = \big\{  \wt \bbeta^{(t)} \in \Theta(r_t) \big\} , \ \  t= 0 ,\ldots, T,  \nn
\#
for some sequence of radii $r_0 \geq   r_1 \geq   \cdots \geq  r_T >0$ to be determined. Moreover, note that all the shifted loss functions $\wt \cQ^{(t)}(\cdot)$ have the same symmetrized Bregman divergence, denoted by
\$
\overbar D(\bbeta_1, \bbeta_2 ) = \langle \nabla \wt \cQ^{(t)}(\bbeta_1) - \nabla \wt \cQ^{(t)}(\bbeta_2), \bbeta_1 - \bbeta_2 \rangle =\langle \nabla \hat \cQ_{1,b}(\bbeta_1) -\nabla  \hat \cQ_{1,b}(\bbeta_2), \bbeta_1 - \bbeta_2 \rangle .
\$
In other words,  the curvature of shifted loss functions  is only determined by the local loss $\hat \cQ_{1,b}(\cdot)$. Define the local radius $r_{\loc}= b/(4\gamma_{0.25})$ the same way as in the proof of Theorem~\ref{thm:one-step}. As long as $ (p+x)/n \lesssim b  \lesssim 1$, Lemma~\ref{lem:local.RSC} ensures that with probability at least $1-e^{-x}$,
\#
\overbar D(\bbeta, \bbeta^* ) \geq 0.5\underbar{$f$}  \kappa_l \cdot  \| \bbeta - \bbeta^* \|_{\Sigma}^2 =  \kappa \cdot \| \bbeta - \bbeta^* \|_{\Sigma}^2 \label{local.curvature}
\#
holds uniformly over all $\bbeta \in \Theta(r_{\loc})$, where $\kappa :=0.5\underbar{$f$}  \kappa_l$ is the curvature parameter. Let $\cE_{\loc}$ be the event that the local strong convexity \eqref{local.curvature} holds.

Proceeding via proof by contradiction, at each iteration $t\geq 1$, we may construct an intermediate estimator $\wt \bbeta^{(t)}_{\imd}$---as a convex combination of $\wt \bbeta^{(t)}$ and $\bbeta^*$---which falls in $\Theta(r_{\loc})$. If event $\cE_*(r_*) \cap \cE_{\loc}$ occurs, then the bounds \eqref{FOC}, \eqref{shift.gradient.ubd} and \eqref{local.curvature} guarantee 
\#
	& \|  \wt \bbeta^{(t)}_{\imd} - \bbeta^* \|_{\Sigma} \leq \kappa^{-1}  \|  \nabla \cQ^{(t)} (\bbeta^* ) \|_{\Omega}  \leq   \kappa^{-1}  \big\{ \|   \Delta(\wt \bbeta^{(t-1)})  \|_2 + \|   \Delta_1(\wt \bbeta^{(t-1)})  \|_2 +  r_* \big\} , \label{general.bound1}
\#
where $\Delta(\cdot)$ and $\Delta_1(\cdot)$ are the random processes defined in \eqref{def:Delta}. For $\wt \bbeta^{(t)}$ which minimizes the shifted loss $\wt \cQ^{(t)}(\cdot)$, the first-order condition $\nabla \wt \cQ^{(t)}( \wt \bbeta^{(t)} ) = \textbf{0}$ holds, and hence
\#
	&    \|     \Hb (   \wt \bbeta^{(t)} - \bbeta^* )   + \nabla \hat \cQ_h(\bbeta^* )    \|_{\Omega}    \nn \\
&  = \| \Sigma^{-1/2}    \{   \nabla \wt \cQ^{(t)} (  \wt \bbeta^{(t)}  ) -  \nabla \wt \cQ^{(t)}  (\bbeta^*) -      \Hb (  \wt \bbeta^{(t)} - \bbeta^* )   \}  + \Sigma^{-1/2}  \{  \nabla \wt \cQ^{(t)}  (\bbeta^*) -   \nabla \hat \cQ_h(\bbeta^* )  \} \|_2    \nn \\  
& \leq  \| \Delta_1 ( \wt \bbeta^{(t)} ) \|_2 + \|  \Delta_1(\wt \bbeta^{(t-1)}) \|_2 + \| \Delta(\wt \bbeta^{(t-1)} ) \|_2 . \label{general.bound2}
\#
In what follows we deal with $\{ ( \wt \bbeta^{(t)}_{\imd} , \wt \bbeta^{(t)} ) \}_{t=1,2,\ldots}$ sequentially, conditioning on $\cE_0(r_0)\cap\cE_*(r_*) \cap \cE_{\loc}$.
In view of the basic inequalities \eqref{general.bound1} and \eqref{general.bound2}, the key is to control the random processes  $\Delta(\cdot)$ and $\Delta_1(\cdot)$ as we have done in \eqref{diff.grad.unif.bound}. Define the event 
\$
 \cF(r)  = \left\{     \sup_{\bbeta \in \Theta(r)}  \big\{  \| \Delta_1(\bbeta)  \|_2  + \| \Delta(\bbeta) \|_2  \big\} \leq \delta(x) \cdot r \right\}  
\$ 
with $\delta(x) = C \{ \sqrt{(p+x)/(nb)} + \sqrt{(p+x)/(Nh)} +b \}$ for some $C>0$, so that $\PP\{ \cF(r) \}\geq 1- 2e^{-x}$ for every $0<r\lesssim b$.   

At iteration 1, the bound \eqref{general.bound1} yields that, conditioned on $\cE_0(r_0)\cap \cE_*(r_*) \cap \cE_{\loc}  \cap \cF(r_0)$,
\#
\|  \wt \bbeta^{(1)}_{\imd} - \bbeta^* \|_{\Sigma} \leq r_1 :=   \kappa^{-1} \delta(x) \cdot  r_0 + \kappa^{-1} r_*      . \nn
\#
The imposed constraints on $(b,h,r_0,r_*)$ ensure that $\kappa^{-1}\delta(x)<1$,  $r_1 < r_{\loc} \asymp b$ and $r_1 \leq r_0$. Via proof by contradiction, we must have $\wt \bbeta^{(1)}  = \wt \bbeta^{(1)}_{\imd}\in \Theta(r_{\loc})$, which in turn certifies the event $\cE_1(r_1)=\{ \wt \bbeta^{(1)}  \in \Theta(r_1)\}$. This, combined with \eqref{general.bound2} implies that, conditioned on $\cE_0(r_0)\cap \cE_*(r_*) \cap \cE_{\loc}  \cap \cF(r_0)$,
\#
\left\{\begin{array}{ll}
	\| \wt \bbeta^{(1)} - \bbeta^*   \|_{\Sigma} \leq      \kappa^{-1}  \delta(x) \cdot r_0 + \kappa^{-1}    r_*  =  r_1  \leq r_0 ,  \vspace{0.2cm}\\
\|   \Hb (  \wt \bbeta^{(1)}- \bbeta^*  ) +   \nabla \hat \cQ_h(\bbeta^* )    \|_{\Omega} \leq 2 \delta(x) \cdot r_0.
\end{array}\right.  \label{iterate1.bound}
\#



Now assume that for some $t\geq 1$, $\wt \bbeta^{(t)} \in \Theta(r_t)$ with $r_t  := \kappa^{-1} \delta(x)  \cdot  r_{t-1} + \kappa^{-1} r_* \leq  r_{t-1}$, and $r_\ell < r_{\loc}$ for all $\ell=1,\ldots, t$.
At iteration $t+1$, applying the general bound \eqref{general.bound1} again we see that, if  event $\cE_{t}(r_t) \cap \cE_*(r_*)  \cap \cE_{\loc}\cap \cF(r_t)$ occurs, 
\#
\|  \wt \bbeta^{(t+1)}_{\imd} - \bbeta^* \|_{\Sigma} \leq     \kappa^{-1} \delta(x)  \cdot  r_t+ \kappa^{-1} r_*    . \nn
\#
Set $ r_{t+1} =  \kappa^{-1} \delta(x)  \cdot  r_t+ \kappa^{-1} r_*$, and note that $r_{t+1}  \leq   \kappa^{-1} \delta(x)  \cdot r_{t-1}  + \kappa^{-1}   r_* = r_t < r_{\loc}$. This means that $ \wt \bbeta^{(t+1)}_{\imd}$ falls into the interior of $\Theta(r_{\loc})$, which in turn implies   $\wt \bbeta^{(t+1)} =  \wt \bbeta^{(t+1)}_{\imd} \in \Theta(r_{\loc})$ and certifies event $\cE_{t+1}(r_{t+1})$. Conditioned on $\cE_{t}(r_t) \cap \cE_*(r_*)  \cap \cE_{\loc}\cap \cF(r_t)$, we combine the consequence that $\wt \bbeta^{(t+1) } \in \Theta(r_{t+1}) \subseteq \Theta(r_t)$ with the general bound \eqref{general.bound2}, thereby obtaining
\#
\left\{\begin{array}{ll}
	\| \wt \bbeta^{(t+1)} - \bbeta^*   \|_{\Sigma} \leq      \kappa^{-1}  \delta(x) \cdot r_t + \kappa^{-1}    r_*  =  r_{t+1}  \leq r_t ,  \vspace{0.2cm}\\
 \|    \Hb  (  \wt \bbeta^{(t+1)}- \bbeta^*  ) +  \nabla \hat \cQ_h(\bbeta^* )  \|_{\Omega}  \leq 2 \delta(x) \cdot r_t.
\end{array}\right.  \label{iterate2.bound}
\#

Repeat the above argument until we obtain $ \wt \bbeta^{(T)}$ for some $T\geq 1$. For every $1\leq t\leq T$, note that conditioned on $\cE_{t-1}(r_{t-1}) \cap \cE_*(r_*) \cap \cE_{\loc} \cap \cF(r_{t-1})$, the event $\cE_t(r_t)$ must happen. Therefore, conditioned on $\cE_0(r_0) \cap  \cE_*(r_*) \cap \cE_{\loc} \cap \{  \cap_{t=0}^{T-1} \cF(r_{t-1}) \}$, $ \wt \bbeta^{(T)}$ satisfies the bounds
\#
\left\{\begin{array}{ll}
	\| \wt \bbeta^{(T)} - \bbeta^*   \|_{\Sigma} \leq      \kappa^{-1}  \delta(x) \cdot r_{T-1} + \kappa^{-1}    r_*  =: r_T  \leq r_{T-1} ,  \vspace{0.2cm}\\
\|    \Hb (  \wt \bbeta^{(T)}- \bbeta^*  ) +   \nabla \hat \cQ_h(\bbeta^* )    \|_{\Omega} \leq 2 \delta(x) \cdot r_{T-1}.
\end{array}\right.     \label{iterateT.bound}
\#
It is easy to see that  $r_t = \{ \kappa^{-1} \delta(x) \}^t r_0 +   \frac{1- \{ \kappa^{-1} \delta(x)\}^t}{1 -  \kappa^{-1}\delta(x)}  \kappa^{-1} r_* $ for $t=1,\ldots, T$. We thus take $T = \lceil \log(  r_0/r_*) / \log (\kappa/\delta(x)) \rceil +1$, the smallest integer such that $\{ \kappa^{-1} \delta(x) \}^{T-1} r_0 \leq   r_*$.

Finally, conditioned on $\cE_0(r_0) \cap \cE_*(r_*)$,  we combine \eqref{iterate1.bound}--\eqref{iterateT.bound} with  \eqref{diff.grad.unif.bound}, \eqref{local.curvature} and the union bound to conclude that, with probability at least $1- (2T+1)e^{-x}$,
\#
\left\{\begin{array}{ll}
\|  \wt \bbeta^{(T)} - \bbeta^*  \|_{\Sigma} \leq  \kappa^{-1}  \delta (x) \cdot r_*  + \tfrac{1}{\kappa -  \delta(x)}  r_*  \lesssim r_*    ,  \vspace{0.2cm}\\
\|  \Hb (  \wt \bbeta^{(T)}- \bbeta^*  ) +   \nabla \hat \cQ_h(\bbeta^* )    \|_{\Omega}  \leq 2 \delta(x) \big\{  r_*  + \tfrac{1}{\kappa -   \delta(x)}   r_*  \big\} \lesssim \delta(x) \cdot r_*  \nn
\end{array}\right.\nn
\#
Under the constraints $\sqrt{(p+x)/n}\lesssim b \lesssim 1$ and $\sqrt{(p+x)/N} \lesssim h \leq b$, we have $\delta(x) \lesssim b^{1/2}$ and hence $\log(\kappa/\delta(x)) \gtrsim \log(1/b)$. This completes the proof of \eqref{multi-step.bound}. 

\subsection{Proof of Theorem~\ref{thm:final.rate}}

Let $\wt \bbeta^{(0)}$ be the initial estimator given in \eqref{initial.estimator}. For $x>0$, we can apply either Theorem~2.1 in \cite{PZ2020} if $\wt \bbeta^{(0)}$ is a local standard QR estimator, or Theorem~3.1 in \cite{HPTZ2020} with a bandwidth $b \asymp \{ (p+x)/n \}^{1/3}$ if $\wt \bbeta^{(0)}$ is a  local  conquer---convolution smoothed quantile regression---estimator. In either case,  $\wt \bbeta^{(0)}$ satisfies the bound $\| \wt \bbeta^{(0)} - \bbeta^*   \|_{\Sigma} \leq r_0 \asymp \sqrt{(p+x)/n}$ with probability at least $1-2e^{-x}$ as long as $n\gtrsim p + x$.
For the second event $\cE_*(r_*)$ in \eqref{event0}, it follows from Lemma~\ref{lem:global.score} with $r_*   \asymp \sqrt{(p+x)/N} + h^2$ that $\PP\{ \cE_*(r_*) \} \geq 1-e^{-x}$. Putting together the pieces, we conclude that the event $\cE_0(r_0) \cap \cE_*(r_*)$ occurs with probability at least $1-3e^{-x}$.

Set  $x = \log(n \log m)$. Given the specified choice of the bandwidths $b, h>0$, we have 
$$
	r_0 \asymp \sqrt{\frac{p +\log(n \log m) }{n}}, \quad  r_* \asymp \sqrt{\frac{p+   \log(n \log m)  }{N}}
$$ 
and 
$$
\sqrt{\frac{p+x }{n b}} + \sqrt{\frac{p+x}{N h }} + b \asymp  \left( \frac{p +  \log(n \log m)}{n} \right)^{1/3} + \sqrt{\frac{p+  \log(n\log m)}{N h}} .
$$
Finally, applying the high-level result in Theorem~\ref{thm:multi-step} yields \eqref{final.rate} and \eqref{final.bahadur}.

\subsection{Proof of Theorem~\ref{corollary:clt}}

To simplify the presentation, we set  $q=p +  \log(n \log m)$ throughout the proof.
For an arbitrary vector $\ba \in \RR^p$, define the partial sums $S_N = N^{-1/2} \SN w_i  v_i$ and $S_N^0= S_N - \EE S_N$, where $w_i = \overbar K(-\varepsilon_i/h) - \tau $ and $v_i = (\Hb^{-1}\ba)^\T \bx_i$. Recall that $|\EE(w_i | \bx_i) | \leq  0.5 l_1 \kappa_2 h^2$ and hence $|\EE(w_i v_i)| \leq  0.5 l_1 \kappa_2  \| \Hb^{-1} \ba \|_{\Sigma} \cdot  h^2$. Now we are ready to prove the normal approximation for $\wt \bbeta$. To begin with, we have
\#
	&  | N^{1/2} \ba^\T (\wt \bbeta - \bbeta^* ) + S_N^0 |  \nn \\
	 & \leq    N^{1/2}  \Bigg|  \Bigg\langle  \Sigma^{1/2} \Hb^{-1} \ba , \Sigma^{-1/2} \Hb (\wt \bbeta - \bbeta^* ) + \Sigma^{-1/2} \frac{1}{N} \SN \{ \overbar K(-\varepsilon_i/h) - \tau \} \bx_i \Bigg\rangle  \Bigg| + | \EE S_N |  \nn \\
	 & \leq N^{1/2} \| \Hb^{-1} \ba \|_{\Sigma} \cdot \Bigg\|  \Hb (\wt \bbeta - \bbeta^* ) + \frac{1}{N} \SN \{ \overbar K(-\varepsilon_i/h) - \tau \} \bx_i   \Bigg\|_{\Omega} + 0.5 l_1 \kappa_2  \| \Hb^{-1} \ba \|_{\Sigma} \cdot N^{1/2}  h^2  . \nn
\#
By \eqref{final.bahadur}, it follows that with probability at least $1-C n^{-1}$,
\#
 | N^{1/2} \ba^\T (\wt \bbeta - \bbeta^* ) + S_N^0 |  \leq C_1   \|    \Hb^{-1} \ba \|_{\Sigma}  \cdot    \big\{ q^{5/6} n^{-1/3} + q (N h)^{-1/2} + N^{1/2} h^2\big\} .  \label{GAR1}
\#

For the centered partial sum $S_N^0$, applying the Berry-Esseen inequality (see, e.g. \cite{S2013}) yields
\#
 \sup_{x\in \RR}  \big|  \PP\big\{  S_N^0     \leq \var (S_N)^{1/2} x \big\} - \Phi(x) \big| \leq 0.5 N^{-1/2}     \var(wv)^{-3/2}  \EE | wv - \EE (wv) |^3 , \label{GAR2}
\#
where $w=\overbar K(-\varepsilon/h)-\tau$ and $v= (\Hb^{-1} \ba)^\T \bx$. Following the proof of Lemma~\ref{lem:hd.score}, it can be shown that $\EE(w^2 | \bx) \leq  \tau (1-\tau) + (1+\tau) l_0 \kappa_2 h^2$ and  $| \EE (w^2 | \bx) -  \tau(1-\tau) | \lesssim  h$. Consequently, $\var(wv) = \{ \tau(1-\tau) + O(h) \} \| \Hb^{-1} \ba \|_{\Sigma}^2$ and $\EE |wv|^3 \leq \max(\tau, 1-\tau ) \EE( w^2 |v|^3 ) \leq \mu_3 \{ \tau(1-\tau) + O(h^2) \} \| \Hb^{-1}\ba \|_{\Sigma}^3$, where $\mu_3 = \sup_{\bu \in \mathbb{S}^{p-1}} \EE |\bz^\T \bu|^3$. Substituting these bounds into \eqref{GAR2} gives
\#
 \sup_{x\in \RR}  \big|  \PP\big\{  S_N^0     \leq  \var(S_N)^{1/2} x \big\} - \Phi(x) \big| \leq C_2 N^{-1/2} .  \label{GAR3}
\#
Write $\sigma_{\tau, h}^2 = \EE \{ \overbar K(-\varepsilon/h)-\tau\}^2 \langle \Hb^{-1} \ba, \bx \rangle^2$, and note that $| \var(S_N) - \sigma_{\tau, h}^2 | = (\EE  w v )^2   \leq (0.5 l_0 \kappa_2 h^2)^2 \cdot  \| \Hb^{-1} \ba \|_{\Sigma}^2$. Comparing the distribution functions of two Gaussian random variables shows that
\#
 \sup_{x\in \RR}  \big|  \Phi\big(x/ \var(S_N^0)^{1/2} \big) - \Phi\big(x/ \sigma_{\tau, h}\big)   \big|  \leq C_3 h^4 . \label{GAR4}
\#

Let $G\sim \cN(0,1)$.  Applying the bounds \eqref{GAR1}, \eqref{GAR3} and \eqref{GAR4}, we conclude that for any $x\in \RR$ and $\ba \in \RR^p$, 
\#
	&	\PP\big\{ N^{1/2} \ba^\T (\wt \bbeta - \bbeta^* ) \leq x \big\} \nn \\
	& \leq \PP\left[   S_N^0 \leq x + C_1 \| \Hb^{-1} \ba \|_{\Sigma} \cdot \big\{ q^{5/6} n^{-1/3} + q (N h)^{-1/2} + N^{1/2} h^2\big\} \right] + C n^{-1} \nn \\
	& \leq  \PP\left[  \var(S_N^0)^{1/2} G \leq  x + C_1 \| \Hb^{-1} \ba \|_{\Sigma} \cdot \big\{ q^{5/6} n^{-1/3} + q (N h)^{-1/2} + N^{1/2} h^2\big\} \right] \nn  \\
	&~~~~~ + C n^{-1} + C_2 N^{-1/2} \nn \\
	& \leq  \PP\left[  \sigma_{\tau, h} G \leq  x + C_1 \| \Hb^{-1} \ba \|_{\Sigma} \cdot \big\{ q^{5/6} n^{-1/3} + q (N h)^{-1/2} + N^{1/2} h^2\big\} \right]  \nn \\
	&~~~~~ + C n^{-1} + C_2 N^{-1/2}+ C_3 h^4 \nn \\
	& \leq \PP\big(\sigma_{\tau, h} G \leq x \big) +    C n^{-1} +  \frac{C_1  \| \Hb^{-1} \ba \|_{\Sigma}}{(2\pi)^{1/2} \sigma_{\tau, h}}    \big\{ q^{5/6} n^{-1/3} + q (N h)^{-1/2} + N^{1/2} h^2\big\} \nn \\
	&~~~~~ + C_2 N^{-1/2} + C_3 h^4 . \nn
\#
A similar argument leads to a series of reverse inequalities. The claimed bound then follows by noting that $| \sigma_{\tau, h}^2 - \tau(1-\tau ) \| \Hb^{-1} \ba \|_{\Sigma}^2 | \lesssim h$.

\subsection{Proof of Proposition~\ref{prop:covariance.estimation}}

Without loss of generality, assume $\cI_1 =\{1,\ldots, n\}$, and write $\Hb_1(\bbeta) =  \EE \hat \Hb_1(\bbeta)$.
Consider the change of variable $\bdelta = \Sigma^{1/2} (\bbeta - \bbeta^*)$, so that $\bbeta \in \Theta(r)$ is equivalent to $\bdelta \in \BB^p(r)$.
Recall that $\bz_i = \Sigma^{-1/2} \bx_i \in \RR^p$ are isotropic random vectors. Define
\#
	\hat \Hb(\bdelta ) =   \frac{1}{n} \sn \phi_b(\varepsilon_i - \bz_i^\T \bdelta ) \bz_i \bz_i^\T  ~~\mbox{ and }~~ \Hb(\bdelta ) = \EE \bigl\{ \hat \Hb(\bdelta) \bigr\} ,
\#
so that $\hat \Hb(\bdelta ) = \Sigma^{-1/2} \hat \Hb_1(\bbeta) \Sigma^{-1/2}$ and $ \Hb(\bdelta ) = \Sigma^{-1/2}  \Hb_1(\bbeta) \Sigma^{-1/2}$, where $\phi_b(u) = (1/b) \phi(u/b)$.
For any $\epsilon\in (0,r)$, there exists an $\epsilon$-net $\{ \bdelta_1, \ldots, \bdelta_{d_\epsilon}\}$ with $d_\epsilon \leq (1+ 2r/\epsilon)^p$ satisfying that, for each $\bdelta \in \BB^p(r)$, there exists some $1\leq j\leq d_\epsilon$ such that $\| \bdelta - \bdelta_j \|_2 \leq \epsilon$. Hence,
 \#
	& \| \hat \Hb(\bdelta) - \Hb(\bdelta)    \|_2  \nn \\
	& \leq \| \hat \Hb(\bdelta) - \hat \Hb(\bdelta_j)   \|_2 +    \| \hat \Hb(\bdelta_j) -  \Hb(\bdelta_j)    \|_2 +    \| \Hb(\bdelta_j) - \Hb(\bdelta)    \|_2 \nn \\
	& =: I_1(\bdelta) + I_2 (\bdelta_j)+ I_3(\bdelta). \nn
\# 
Starting with $I_1(\bdelta)$, note that $|\phi_b(u) -\phi_b(v)| \leq  \sup_{t} |\phi'(t)|  \cdot  b^{-2} |u-v| \leq (2b)^{-2} |u-v|$ for all $u,v \in \RR$.
It follows that
\#
  I_1(\bdelta)  & \leq \sup_{\bu, \bv  \in \mathbb{S}^{p-1}} \frac{1}{n} \sn  | \phi_b(\varepsilon_i - \bz_i^\T \bdelta ) -  \phi_b(\varepsilon_i -\bz_i^\T \bdelta_j )   | \cdot   | \bz_i^\T \bu \cdot \bz_i^\T \bv  |  \nn \\
  & \leq  (2 b)^{-2}\sup_{\bu , \bv \in \mathbb{S}^{p-1}} \frac{1}{n} \sn   | \bz_i^\T ( \bdelta-\bdelta_j ) \cdot   \bz_i^\T \bu  \cdot  \bz_i^\T \bv   | \nn \\
  & \leq (2b)^{-2} \epsilon \cdot  \max_{1\leq i\leq n} \| \bz_i \|_2  \cdot \bigg\| \frac{1}{n} \sn \bz_i \bz_i^\T \bigg\|_2 .  \label{I1.ubd1}
\#
To bound $ \max_{1\leq i\leq n} \| \bz_i \|_2$, using a standard covering argument we have, for any $\epsilon_1 \in (0,1)$,  an $\epsilon_1$-net $\cN_{\epsilon_1} \subseteq \mathbb{S}^{p-1}$ with $|\cN_{\epsilon_1} | \leq (1+2/\epsilon_1)^p$ such that $\max_{1\leq i\leq n} \| \bz_i \|_2 \leq (1-\epsilon_1)^{-1} \max_{1\leq i\leq n} \max_{\bu \in \cN_{\epsilon_1}}  \bz_i^\T \bu$.
Given $1\leq i\leq n$ and $\bu \in \cN_{\epsilon_1}$,  recall that $\PP( |   \bz_i^\T \bu | \geq \upsilon_1 u) \leq 2 e^{-u^2/2}$ for any $u \geq 0$. Taking the union bound over $i$ and $\bu$, and setting $u=\sqrt{2x + 2\log (2n)+2p\log(1+2/\epsilon_1)}$, we obtain that with probability at least $1-2 n (1+2/\epsilon_1)^p e^{-u^2/2}= 1-e^{-x}$, $\max_{1\leq i\leq n} \| \bz_i \|_2 \leq (1-\epsilon_1)^{-1} \upsilon_1 \sqrt{2x + 2\log (2n)+2p\log(1+2/ \epsilon_1 )}$.
By minimizing this upper bound with respect to $ \epsilon_1 \in (0,1)$, we obtain that with probability at least $1-e^{-x}$,
\#
	 \max_{1\leq i\leq n} \| \bz_i \|_2  \lesssim (p+\log n + x)^{1/2}  . \nn
\#
For $\| (1/n) \sn \bz_i \bz_i^\T \|_2$, it follows from the covering argument along with Bernstein's inequality that, with probability at least $1-e^{-x}/3$,
\#
	\biggl\| \frac{1}{n} \sn \bz_i \bz_i^\T  - \Ib_p \biggr\|_2  \lesssim    \sqrt{\frac{p+ x}{n}} \bigvee  \frac{p+ x}{n}. \nn 
\#
Plugging the above bounds into \eqref{I1.ubd1} yields 
\#
  \sup_{\bdelta \in \BB^p(r)} I_1(\bdelta) \lesssim (p + \log n+x)^{1/2}  b^{-2} \epsilon \label{I1.ubd2}
\#
with probability at least $1- 2 e^{-x} $ as long as $n\gtrsim p+x$. For $I_3(\bdelta)$, it can be similarly obtained that
\#
  I_3(\bdelta)  \leq (2 b)^{-2} \sup_{\bu, \bv \in \mathbb{S}^{p-1}} \EE    |  \bz^\T (  \bdelta - \bdelta_j ) \cdot  \bz^\T \bu \cdot   \bz^\T \bv   | \leq   \mu_3 (2b )^{-2}\epsilon  \label{I3.ubd}
\#
uniformly over all $\bdelta \in \BB^p(r)$.

Turning to $I_2(\bdelta_j)$, note that $\hat \Hb(\bdelta_j) - \Hb(\bdelta_j )  = (1/n) \sn (1-\EE) \phi_{ij} \bz_i \bz_i^\T$, where $\phi_{ij} =\phi_b(\varepsilon_i - \bz_i^\T  \bdelta_j )$ satisfy $|\phi_{ij}| \leq  (2\pi)^{-1/2} b^{-1}$ and
\# 
\EE \bigl(  \phi_{ij}^2 | \bx_i \bigr)     = \frac{1}{ b^2} \int_{-\infty}^\infty \phi^2 \biggl( \frac{   \langle \bz_i ,  \bdelta \rangle- t}{b} \biggr) f_{\varepsilon_i  | \bx_i  } (t) \, {\rm d} t   = \frac{1}{b} \int_{-\infty}^\infty \phi^2(u) f_{\varepsilon_i  | \bx_i  } (\bz_i^\T \bdelta - b u  ) \, {\rm d} u  \leq     \frac{  \bar f   }{ 2\pi^{1/2} b}    \nn
\#
almost surely.
Given $\epsilon_2\in (0,1/2)$, there exits an $\epsilon_2$-net $\cM$ of the sphere $\mathbb{S}^{p-1}$ with $|\cM|\leq (1+2/\epsilon_2)^p$ such that $\| \hat \Hb(\bdelta_j) - \Hb(\bdelta_j )\|_2 \leq (1-2\epsilon_2)^{-1} \max_{\bu \in \cM} |  \bu^\T \{ \hat \Hb(\bdelta_j) - \Hb(\bdelta_j ) \}\bu   |$. Given $\bu \in \cM$ and $k=2,3,\ldots$, we bound the higher order moments of $\phi_{ij}   (\bz_i^\T  \bu )^2$ by
\#
	\EE |\phi_{ij}   (\bz_i^\T  \bu )^2 |^k &  \leq  \bar f  (2\pi^{1/2} b )^{-1} \cdot   \{  (2\pi)^{-1/2} b^{-1} \}^{k-2}   \upsilon_1^{2k}   \cdot 2k \int_0^\infty \PP\bigl( |\bz_i^\T \bu  | \geq\upsilon_1 u \bigr) u^{2k-1} {\rm d} u  \nn \\
	& \leq \bar f  (2\pi^{1/2} b)^{-1} \cdot   \{  (2\pi)^{-1/2} b^{-1} \}^{k-2}   \upsilon_1^{2k} \cdot 4k  \int_0^\infty  u^{2k-1} e^{-u^2/2} {\rm d} u   \nn \\
	& \leq  \bar f  (2\pi^{1/2} b)^{-1} \cdot   \{  (2\pi)^{-1/2} b^{-1} \}^{k-2}   \upsilon_1^{2k}  \cdot  2^{k+1} k! .\nn
\#
In particular, $\EE  \phi_{ij}^2 (\bz_i^\T \bu)^4 \leq  8\pi^{-1/2}\upsilon_1^4  \bar f    b^{-1} $, and for each $k\geq 3$, $\EE |\phi_{ij} (\bz_i^\T \bu)^2 |^k\leq \frac{k!}{2} \cdot  8\pi^{-1/2}  \upsilon_1^4  \bar f    b^{-1} \cdot (  \sqrt{2/\pi}  \,\upsilon_1^2  b^{-1} )^{k-2}$. Applying  Bernstein's inequality and the union bound, we find that for any $u\geq 0$,
\#
 & \| \hat \Hb(\bdelta_j) - \Hb(\bdelta_j )  \|_2 \nn \\
 & \leq \frac{1}{1-2\epsilon_2} \max_{\bu \in \cM} \bigg|   \frac{1}{n} \sn (1-\EE) \phi_{ij}  (\bz_i^\T \bu)^2 \bigg| \leq \frac{  \upsilon_1^2}{1- 2\epsilon_2} \bigg(   4 \pi^{-1/4} \bar f^{1/2} \sqrt{    \frac{u}{n b }} + \sqrt{\frac{2}{\pi}}  \frac{u}{n b} \bigg) \nn 
\#
with probability at least $1-2 (1+2/\epsilon_2)^p e^{-u} = 1 - e^{\log(2) + p\log(1+2/\epsilon_2) - u}$. Setting $\epsilon_2=2/(e^3-1)$ and $u= \log(2) + 3p + v$, it follows that with probability at least $1 -e^{-v}$,
\#
	I_2(\bdelta_j)  \lesssim    \sqrt{  \frac{ p + v  }{n b } }  + \frac{ p + v }{n b}   . \nn
\#
Once again,  taking the union bound over $j=1,\ldots, d_{\epsilon}$ and setting $v= p \log(1 + 2r/\epsilon)+ x$, we obtain that with probability at least $1-d_\epsilon e^{-v} \geq 1-e^{-x}$, 
\#
\max_{1\leq j\leq N_{\epsilon}} I_2(\bdelta_j) \lesssim \sqrt{\frac{p\log(3 e r/\epsilon)+x }{n b }} +   \frac{p\log(3e r/\epsilon)+x }{n b }  . \label{I2.ubd}
\#

Combining \eqref{I1.ubd2}, \eqref{I3.ubd} and \eqref{I2.ubd}, and taking $\epsilon = r/n^2 \in (0, r)$ in the beginning of the proof, we conclude that with probability at least $1-  3 e^{-x}$,
\#
 & \sup_{\bbeta \in \Theta(r)}  \|   \hat \Hb_1(\bbeta) -    \Hb_1(\bbeta)    \|_{\Omega}  \lesssim \sqrt{\frac{p \log n + x}{n b }} +  \frac{p\log n+x}{n b} + \frac{(p + \log n+  x)^{1/2}  r }{ (nb)^2}  \nn
\#
as long as $n\gtrsim p+x $. 
Moreover, note that for every $\bbeta \in \Theta(r)$,
\#
	 &    \|   \Hb_1(\bbeta)  - \Hb  \|_{\Omega} \nn \\
& =   \bigg\|  \EE  \int_0^1 \int_{-\infty}^\infty \phi(u)  \bigl\{ f_{\varepsilon |\bx} ( t \langle \bx, \bbeta - \bbeta^* \rangle   - b u) - f_{\varepsilon | \bx} (0) \bigr\}  \, {\rm d} u \,{\rm d} t \cdot   \bz\bz^\T  \bigg\|_2 . \nn
\#
By the Lipschitz continuity of $f_{\varepsilon| \bx}(\cdot)$ and $f'_{\varepsilon | \bx}(\cdot)$, we have $| f_{\varepsilon |\bx} ( t \langle \bx, \bbeta - \bbeta^* \rangle   - b u) -f_{\varepsilon |\bx} (    - b u) | \leq l_0  \cdot t \cdot |\bx^\T (\bbeta -\bbeta^* ) |$ and 
$ | f_{\varepsilon |\bx} (    - b u) -  f_{\varepsilon | \bx} (0 ) + b   f'_{\varepsilon | \bx} (0)  \cdot u   | \leq   | \int_0^{-bu}     \{ f_{\varepsilon | \bx}'(v)  -  f'_{\varepsilon | \bx} (0)    \} \, {\rm d}v   | \leq  0.5 l_1 b^2 u^2$.
Plugging these into the above inequality yields
\#
& 	  \|   \Hb_1(\bbeta)  - \Hb   \|_{\Omega} \nn \\
&\leq   0.5 l_0 \sup_{\bu \in \mathbb{S}^{p-1}} \EE  \bigl\{ | \bx^\T(\bbeta -\bbeta^*) | (\bz^\T \bu)^2 \bigr\}  + 0.5 l_1 b^2 
\leq 0.5 \bigl(l_0 \mu_3  r +  l_1 b^2 \bigr)  . \nn
\#
Putting together the pieces proves the claimed bound, provided that $b\gtrsim (p\log n + x)/n$. 

\subsection{Proof of Theorem~\ref{thm:hd.one-step}}

Without loss of generality, assume $\cI_1 = \{ 1,\ldots, n\}$.
Let $\cS = {\rm supp}(\bbeta^*) \subseteq \{ 1,\ldots, p\}$ be the true active set with cardinality $| \cS| \leq s$, and write $\wt \bdelta = \wt \bbeta - \bbeta^*$ with $\wt \bbeta = \wt \bbeta^{(1)}$ for simplicity. By the first-order optimality condition, there exits  a subgradient $\wt \bxi \in \partial \| \wt  \bbeta \|_1$ such that $\nabla \wt \cQ( \wt \bbeta ) + \lambda  \cdot  \wt \bxi = \textbf{0}$ and $\wt \bxi^\T \wt \bbeta = \| \wt \bbeta \|_1$. Hence,
\#
	\bigl\langle \wt  \bxi , \bbeta^* - \wt \bbeta \bigr\rangle \leq \| \bbeta^* \|_1 - \| \wt \bbeta \|_1 = \| \bbeta_{\cS}^* \|_1 - 
	\| \wt \bdelta_{\cS^{{\rm c}}} \|_1  - \| \wt \bdelta_{\cS} + \bbeta^*_{\cS} \|_1 \leq \| \wt \bdelta_{\cS} \|_1 - \| \wt \bdelta_{\cS^{{\rm c}}} \|_1 . \nn
\#
This, together with the convexity of $ \wt \cQ(\cdot )$, implies
\#
	 0  & \leq \overbar D_{\wt \cQ}(  \wt \bbeta , \bbeta^*) = \bigl\langle \nabla \wt \cQ(\wt \bbeta)  - \nabla \wt \cQ(\bbeta^* ), \wt \bbeta - \bbeta^* \bigr\rangle  \nn \\
	  & =  \lambda \bigl\langle \wt \bxi , \bbeta^* - \wt \bbeta  \,\bigr\rangle   -  \bigl\langle \nabla \wt \cQ(\bbeta^*) , \wt \bdelta \, \bigr\rangle \nn \\
	 & \leq  \lambda \bigl(  \| \wt \bdelta_{\cS} \|_1 - \| \wt \bdelta_{\cS^{{\rm c}}} \|_1 \bigr)  -  \bigl\langle \nabla \wt \cQ(\bbeta^*) , \wt \bdelta \, \bigr\rangle ,  \label{hd.foc}
\#
where $\nabla \wt \cQ(\bbeta^*) =  \nabla  \hat \cQ_{1,b}(\bbeta^*) -  \nabla  \hat \cQ_{1,b}(\wt \bbeta^{(0)}) + \nabla  \hat \cQ_h(\wt \bbeta^{(0)})$. Define gradient-based random processes
\#
	D_1 (\bbeta) =   \nabla \hat \cQ_{1,b}( \bbeta) -   \nabla  \hat \cQ_{1,b}(\bbeta^*)  , \quad D(\bbeta) =  \nabla  \hat  \cQ_h(\bbeta) -  \nabla  \hat  \cQ_h(\bbeta^*) , \nn
\#
and their means $E_1(\bbeta) = \EE D_1(\bbeta)$ and $E(\bbeta) = \EE D(\bbeta)$. Moreover,  let $\cQ_h(\bbeta) = \EE (\rho_\tau * K_h)(y-\bx^\T \bbeta)$ be the population smoothed loss function. It is easy to see that
$ \EE \hat \cQ_h(\bbeta) = \cQ_h(\bbeta)$ and $\EE \hat \cQ_{1,b}(\bbeta) = \cQ_b(\bbeta)$.
Then, the gradient $\nabla \wt \cQ(\bbeta^*)$ can be decomposed as
\$
    \bigl\{  D(\bbeta) - E(\bbeta) \bigr\} \Big|_{\bbeta=\wt \bbeta^{(0)}}   & +    \bigl\{ E_1(\bbeta) -  D_1(\bbeta)   \bigr\} \Big|_{\bbeta= \wt \bbeta^{(0)}}+  \nabla \hat \cQ_h(\bbeta^*)   - \nabla \cQ_h(\bbeta^*) \\
  &~~~~~~~~~~~~~  +  \bigl\{  E(\bbeta) - E_1(\bbeta) \bigr\} \Big|_{\bbeta = \wt \bbeta^{(0)}}  +    \nabla \cQ_h(\bbeta^*)    .
\$
For $r>0$, define the suprema of random processes over the local $\ell_1$/$\ell_2$ region $\Theta(r) \cap \Lambda$
\#
	\Pi_1(r) = \sup_{\bbeta \in \Theta(r) \cap \Lambda }  \|  D_1(\bbeta ) - E_1(\bbeta )    \|_\infty, ~~ \Pi(r) = \sup_{\bbeta \in \Theta(r) \cap \Lambda }   \|  D(\bbeta ) - E(\bbeta )    \|_\infty , \label{def.Pi}
\#
and the deterministic quantities
\#
	  \omega(r) =  \sup_{\bbeta \in \Theta(r)   }    \|    E(\bbeta) - E_1(\bbeta)      \|_{\Omega}   
, ~~ \omega^* =   \|  \nabla \cQ_h(\bbeta^*)   \|_{\Omega} . \label{def.omega}
\#
If event $\cE_*(\lambda_*) \cap \cE_0(r_0)$ occurs, then using H\"older's inequality gives
\#
 |  \langle \nabla \wt \cQ(\bbeta^* ) , \wt \bdelta \rangle |    \leq  \bigl\{ \Pi(r_0) + \Pi_1(r_0) + \lambda_* \bigr\} \cdot \| \wt \bdelta \|_1 +  \{  \omega(r_0 )  + \omega^* \} \cdot \| \wt \bdelta \|_{\Sigma} . \nn
\#
Let  $\lambda = 2.5(\lambda^* + \varrho)$ with $\varrho $ satisfying
\#
 \varrho \geq   \max\left\{     \Pi(r_0) + \Pi_1(r_0), \,   \frac{ \omega(r_0) + \omega^* }{s^{1/2}}   \right\} , \label{varrho.constraint}
\#
so that $\Pi(r_0) + \Pi_1(r_0) + \lambda_*  \leq  0.4 \lambda  ~~\mbox{ and }~~\omega(r_0 )  + \omega^* \leq   0.4 s^{1/2} \lambda$.
Substituting the above bounds into \eqref{hd.foc} yields
$ 0\leq  1.4 \| \wt \bdelta_{\cS} \|_1 -   0.6  \| \wt \bdelta_{\cS^{{\rm c}}} \|_1   +  0.4  s^{1/2}   \|  \wt \bdelta \|_{\Sigma} $. Consequently, $ \| \wt \bdelta \|_1 \leq   (10/3) \| \wt \bdelta_{\cS} \|_1 +  (2/3) s^{1/2}    \|  \wt \bdelta \|_{\Sigma}   \leq 4s^{1/2}   \|  \wt \bdelta \|_{\Sigma}$, showing that $\{ \wt \bbeta \in \Lambda \}$  occurs.


Throughout the rest of the proof, we assume event $\cE_*(\lambda_*) \cap \cE_0(r_0)$ occurs.
Turning to the left-hand side of \eqref{hd.foc}, for $r_{\loc} :=  b/(4\gamma_{0.25})$, we define $\wt \bbeta_\eta = \bbeta^* + \eta(\wt \bbeta - \bbeta^*)$ with $0<\eta \leq 1$ the same way as in the first paragraph in the proof of Theorem~\ref{thm:one-step}.  Under the requirement \eqref{varrho.constraint} on $\varrho$, we have $\wt \bbeta_\eta \in  \Theta(r_{\loc} ) \cap \Lambda$ and hence by \eqref{hd.foc}, 
\#
	\overbar D_{\wt \cQ}( \wt \bbeta_\eta  , \bbeta^*)  & \leq \eta \cdot  \overbar D_{\wt \cQ}(  \wt \bbeta , \bbeta^*) 
\leq   \eta  \cdot   \bigl(  1.4  \lambda    \|  \wt \bdelta_{  \cS} \|_1  +  0.4 s^{1/2} \lambda   \|    \wt \bdelta \|_{\Sigma} \bigr)   
 \leq   1.8 s^{1/2}\lambda  \cdot \|  \wt \bbeta_\eta - \bbeta^*  \|_{\Sigma} . \nn
\#
For the lower bound,  Lemma~\ref{lem:hd.RSC} implies
\#
\overbar D_{\wt \cQ}( \wt \bbeta_\eta  , \bbeta^*) \geq  0.5 \underbar{$f$} \kappa_l   \cdot  \| \wt \bbeta_\eta - \bbeta^*   \|_{\Sigma}^2 \nn
\#
with probability at least $1-e^{-x}$ as long as $(s \log p + x)/n \lesssim b \lesssim 1$. We thus conclude that
\#
	  \| \wt \bbeta_\eta - \bbeta^*     \|_{\Sigma} \leq   3.6  (\underbar{$f$} \kappa_l  )^{-1}   s^{1/2} \lambda  . 
	\label{intermediate.bound1}
\#

It remains to choose a sufficiently large $\lambda$, or equivalently $\varrho$, so that   \eqref{varrho.constraint}  is satisfied.
The following two lemmas provide upper bounds on the suprema $\Pi(r_0)$, $\Pi_1(r_0)$ and $\omega(r_0)$, defined in \eqref{def.Pi} and \eqref{def.omega}.

\begin{lemma} \label{lem:diff.ubd1}
Assume Conditions~(C1), (C2) and (C4) hold.
For any $r>0$ and $x>0$,
\# \label{glob.grad.uniform}
& \Pi(r) = \sup_{\bbeta \in \Theta(r ) \cap \Lambda }   \| D (  \bbeta) -  E  ( \bbeta)  \|_\infty  \nn \\
&   \leq  \left[  c_1   h^{-1} \sqrt{  2 s\log(2p)/ N } + c_2 \sqrt{ \{ \log (2p) + x\}/(Nh )} + c_3   s^{1/2} \{ \log(2p) + x  \}/(Nh) \right]  \cdot  r
\#
with probability at least $1- e^{-x}$, where $c_1 = 20   \kappa_u B^2$, $c_2 =  (2\kappa_u \bar f   \mu_4  )^{1/2} \sigma_u$ and $c_3 = (16+4/3) \kappa_u B^2 $.
The same high probability bound, with $(N,h)$ replaced by $(n,b)$, holds for $\Pi_1(r)$.
\end{lemma}

\begin{lemma} \label{lem:diff.ubd2}
Conditions~(C1), (C2) and (C4) ensure that $\omega(r)\leq l_0 \kappa_1 | b-h| r$ for any $r>0$ and $\omega^* \leq l_0 \kappa_2 h^2/2$, where $\kappa_1 = \int_{-\infty}^\infty |u|  K(u) \, {\rm d} u $ and $\kappa_2 = \int_{-\infty}^\infty u^2  K(u) \, {\rm d} u$. 
\end{lemma}

Given the bandwidth $0< h\leq b\lesssim 1$, applying Lemmas~\ref{lem:diff.ubd1} and \ref{lem:diff.ubd2} yields that
\$
  \Pi(r_0) + \Pi_1(r_0) \lesssim  s^{1/2} r_0 \cdot   \Bigg( \frac{1}{b} \sqrt{\frac{ \log p+ x  }{n}} +  \frac{1}{h} \sqrt{\frac{ \log p+ x  }{N}} \,\Bigg)  
\$
with probability at least $1-2e^{-x}$, and $\omega(r_0) +\omega^* \leq l_0 ( \kappa_1 b   r_0 + \kappa_2 h^2/2)$. Hence,  a sufficiently large $\varrho$, which is of order 
$$
 \varrho \asymp   \max \left[  \left\{ b^{-1}  \sqrt{ s \cdot   (\log p + x)  / n}   +  h^{-1} \sqrt{ s \cdot    (\log p + x)   / N }   \right\}  r_0  ,  ~ s^{-1/2} \bigl( b  r_0 + h^2 \bigr)  \right]   ,
$$
guarantees that \eqref{varrho.constraint} holds with high probability.  Consequently,  conditioning on $\cE_*(\lambda_*)\cap \cE_0(r_0)$, the intermediate ``estimator" $\wt \bbeta_\eta$ satisfies the error bound \eqref{intermediate.bound1} with probability at least $1-3e^{-x}$. We then set $\delta=3e^{-x} \in (0,1)$, so that $\log p +x = \log(3p/\delta) \asymp \log(p/\delta)$ and 
$$
 \varrho \asymp   \max \left[  \left\{ b^{-1}  \sqrt{ s\log(p/\delta)  / n}   +  h^{-1} \sqrt{ s \log(p/\delta)  / N }   \right\}  r_0  ,  ~ s^{-1/2} \bigl( b  r_0 + h^2 \bigr)  \right]   .
$$

With the above choice of $\varrho$, let $b > 14.4 \gamma_{0.25}(\underbar{$f$} \kappa_l)^{-1}  s^{1/2} \lambda$ so that the right-hand side of \eqref{intermediate.bound1} is strictly less than $r_{\loc}$. Via proof by contradiction, we must have $\wt \bbeta = \wt \bbeta_\eta \in \Theta(r_{\loc})$ and hence the bound \eqref{intermediate.bound1} also applies to $\wt \bbeta$, as claimed.

\subsection{Proof of Theorem~\ref{hd.final.rate}}

The whole proof will be carried out conditioning on $\cE_*(\lambda_*)\cap \cE_0(r_0)$ for some prespecified $r_0, \lambda_* >0$, and write $r_* = s^{1/2} \lambda_*$. Examine the proof of Theorem~\ref{thm:hd.one-step}, we see that to obtain the desired error bound for the first iterate $\wt \bbeta^{(1)}$,
the regularization parameter $\lambda_1$ needs to be sufficiently large. Given $\delta \in (0,1)$, we set $\lambda_1 =2.5 ( \lambda_* + \varrho_1)$, where $\varrho_1>0$ is of order
$$
	\varrho_1 \asymp \max\left[    \left\{ b^{-1}  \sqrt{ s\log(p/\delta) / n   } + h^{-1} \sqrt{ s \log(p/\delta)  / N  }   \right\}     r_0 , \,  s^{-1/2} ( b r_0 + h^2 ) \right]
$$
Provided that the bandwidths $b\geq h>0$ satisfy
$$
	r_* +  \max \left[  \left\{ b^{-1} s \sqrt{\log(p/\delta)/n}     + h^{-1} s \sqrt{\log(p/\delta)/N} \right\} r_0 ,  \, b r_0 + h^2  \right] \lesssim b \lesssim 1 ,
$$
the first iterate $\wt \bbeta^{(1)}$ satisfies $\wt \bbeta^{(1)} \in \Lambda$ and 
\#
  \| \wt \bbeta^{(1)} - \bbeta^*   \|_{\Sigma} \leq  \underbrace{ C_1  \left\{  b^{-1} s \sqrt{\log(p/\delta)/n}  +  b+   h^{-1} s \sqrt{\log(p/\delta)/N}  \right\} }_{=: \, \gamma  }  \cdot r_0 + C_2 (  r_* + h^2)     =: r_1  \label{hd.ubd1}
\#
with probability at least $1-\delta$, where $\gamma = \gamma(s,p,n,N, h,b,\delta)>0$ is a contraction factor.  
With the stated choice of bandwidths $b\asymp  s^{1/2}\{ \log(p/\delta) / n \}^{1/4}$ and $h\asymp  \{ s \log(p/\delta) /N \}^{1/4}$, we have
\$
 \gamma \asymp \big\{  s^2 \log(p/\delta) /n  \big\}^{1/4} + \bigl\{ s^3 \log(p/\delta) /N \bigr\}^{1/4}  \\
  \mbox{ and }~   \varrho_1 \asymp  \max\big\{ \gamma s^{-1/2} r_0  , \,  \sqrt{  \log(p/\delta )/N} \big\} .
\$
A  sufficiently accurate initial estimator---say, $r_0\lesssim \min\{1 , (m/ s)^{1/4} \}$---ensures that $s^{1/2} \varrho_1 \lesssim b$. Moreover, we need the local and total sample sizes to be sufficiently large---namely, $n\gtrsim s^2 \log(p/\delta)$ and $N\gtrsim s^3 \log(p/\delta)$---so that the contraction factor $\gamma$ is strictly less than 1. As a result, the one-step procedure reduces the estimation error of $ \wt \bbeta^{(0)}$ by a factor of $\gamma$.


For $t=2, 3, \ldots, T$, define the events $\cE_t(r_t ) := \{ \wt \bbeta^{(t)} \in \Theta(r_t ) \cap \Lambda \}$ and  
$$
	r_t  := \gamma r_{t-1} + C_2 ( r_* + h^2)    =   \gamma^t  r_0 + C_2 \frac{1-\gamma^t}{1-\gamma}  ( r_* + h^2)     .
$$
At iteration $t\geq 2$, we set $\lambda_t = 3( \lambda_* + \varrho_t)$ with 
$$
	\varrho_t  \asymp    \max\left\{  \gamma s^{-1/2} r_{t-1}  , \,  \sqrt{  \log(p/\delta )/N} \right\} .
$$
Together, the last two displays imply
$$
 \varrho_t \asymp  \max\left\{   \gamma^t s^{-1/2} r_0 + \gamma s^{-1/2}  ( \lambda_* +h^2)\mathbbm{1}(t\geq 2)  , \, \sqrt{  \log(p/\delta )/N}  \right\}. 
$$
Under the stated conditions on $r_0$ and $(n,N)$, we have $s^{1/2} \varrho_t \lesssim b$ for every $t\geq 2$. Applying Theorem~\ref{thm:hd.one-step} repeatedly, we obtain that conditioned on the event $\cE_*(\lambda_*)  \cap \cE_{t-1} (r_{t-1} )$, the $t^{{\rm th}}$ iterate $\wt \bbeta^{(t)}$ satisfies $\wt \bbeta^{(t)} \in \Lambda$ and
\#
 \| \wt \bbeta^{(t)} - \bbeta^* \|_{\Sigma} \leq   \gamma r_{t-1} + C_2 ( r_* + h^2)    = r_t =   \gamma^t  r_0 + C_2 \frac{1-\gamma^t}{1-\gamma} ( r_* + h^2)     \label{hd.ubd2}
\#
with probability at least $1-\delta$. 

Note that $r_* = s^{1/2} \lambda^*$ corresponds to the optimal rate under $\ell_2$-norm.
We thus choose the number of iterations $T$ to be the smallest integer such that $ \gamma^T  r_0 \leq r_* $, that is, $T= \lceil  \log(r_0/r_*) / \log(1/\gamma) \rceil$.
Applying the union bound over $t=1,2,\ldots, T$ yields that conditioned on $\cE_*(\lambda_*) \cap \cE_0(r_0 )$, the $T^{{\rm th}}$ iterate $\wt \bbeta^{(T)}$ satisfies the error bounds
\#
	\|  \wt \bbeta^{(T)} - \bbeta^* \|_{\Sigma} \lesssim s^{1/2} \lambda^* +h^2  ~~\mbox{ and }~~ \| \wt \bbeta^{(T)}  - \bbeta^* \|_1 \lesssim s \lambda^* + s^{1/2} h^2  \nn
\#
with probability at least $1- T\delta$. This completes the proof of the theorem.

\subsection{Proof of Proposition~\ref{prop:hd.initial}}

The proof is similar in spirit to that of Theorem~\ref{thm:hd.one-step}, while certain modifications are required.  We thus provide a sketch proof for completeness. Note that the shifted  loss $\wt \cQ(\cdot)$ in the proof of Theorem~\ref{thm:hd.one-step} shares the Hessian as well as symmetrized Bregman divergence with the local loss $\hat \cQ_{1,b}(\cdot)$.
With slight abuse of notation, let $\wt \bdelta = \wt \bbeta - \bbeta^*$ with $\wt \bbeta = \wt \bbeta^{(0)}$. Inequality \eqref{hd.foc} implies
\#
	 0 \leq \overbar D_{\hat \cQ_{1,b}} ( \wt \bbeta , \bbeta^* ) \leq   \lambda_0 \bigl(  \| \wt \bdelta_{\cS} \|_1 - \| \wt \bdelta_{\cS^{{\rm c}}} \|_1 \bigr)  -  \bigl\langle \nabla \hat \cQ_{1,b}(\bbeta^*) , \wt \bdelta \, \bigr\rangle .  \label{local.foc}
\#
By H\"older's inequality,
\$
&   |  \langle \nabla \hat \cQ_{1,b}(\bbeta^*) , \wt \bdelta  \, \rangle   |    \leq    \|   \nabla \hat \cQ_{1,b}(\bbeta^*) - \nabla \cQ_{1,b}(\bbeta^*)   \|_\infty \cdot \| \wt \bdelta \|_1 +   \|   \nabla \cQ_{1,b}(\bbeta^*)   \|_{\Omega} \cdot \| \wt \bdelta \|_{\Sigma} ,  
\$
where $\cQ_{1,b}(\bbeta) = \EE \hat \cQ_{1,b}(\bbeta)$ is the population loss. By the Lipschitz continuity of $f_{\varepsilon |\bx}(\cdot)$, it can be shown that $\|  \nabla \cQ_{1,b}(\bbeta^*) \|_{\Omega} \leq 0.5 l_0 \kappa_2 b^2$. Moreover, let the regularization parameter $\lambda_0$ satisfy
\#
	\lambda_0 \geq    2.5    \|   \nabla \hat \cQ_{1,b}(\bbeta^*) - \nabla \cQ_{1,b}(\bbeta^*)    \|_\infty   .  \label{l0.constraint}
\#
Then, $|   \langle \nabla \hat \cQ_{1,b}(\bbeta^*) , \wt \bdelta   \rangle   | \leq 0.4 \lambda_0  \| \wt \bdelta \|_1 + 0.5 l_0 \kappa_2 b^2 \| \wt \bdelta \|_{\Sigma}$.
Substituting these into \eqref{local.foc} yields 
\#
	\overbar D_{\hat \cQ_{1,b}} ( \wt \bbeta , \bbeta^* )  \leq  \bigl(   1.4  s^{1/2} \lambda_0 +  0.5 l_0 \kappa_2 b^2 \bigr) \cdot \| \wt \bdelta \|_{\Sigma}  \nn  \\ 
 \mbox{and }~~  0 \leq   \lambda_0 \bigl(  \| \wt \bdelta_{\cS} \|_1 - \| \wt \bdelta_{\cS^{{\rm c}}} \|_1 \bigr) +  0.4 \lambda_0    \| \wt \bdelta  \|_1 +  0.5 l_0 \kappa_2 b^2  \| \wt \bdelta \|_{\Sigma} . \nn
\#
The latter implies 
$$
	\| \wt \bdelta  \|_1 \leq  (10/3) \| \wt \bdelta_{\cS} \|_1 +  (5/6)  l_0 \kappa_2 \lambda_0^{-1} b^2   \| \wt \bdelta \|_{\Sigma} \leq  L \| \wt \bdelta \|_{\Sigma} ~\mbox{ with }~ L:= (10/3) s^{1/2} + (5/6)  l_0 \kappa_2 \lambda_0^{-1} b^2 .
$$

Starting from here, we introduce an intermediate ``estimator" $\wt \bbeta_\eta = \bbeta^* + \eta(\wt \bbeta-\bbeta^*)$, for some $0<\eta\leq 1$, the same way as in the proof of Theorem~\ref{thm:hd.one-step}, so that $\wt \bbeta_\eta \in \Theta(r_{\loc})$ with $r_{\loc}=b/(4\gamma_{0.25})$. Since $\wt \bbeta_\eta - \bbeta^* = \eta(\wt \bbeta_\eta - \bbeta^*)$, we also have 
$\| \wt \bbeta_\eta - \bbeta^* \|_1 \leq L \| \wt \bbeta_\eta - \bbeta^* \|_{\Sigma}$ for the same $L>1$ given above.
Applying Lemma~4.1 in \cite{TWZ2020} and Lemma~\ref{lem:hd.RSC}, we obtain that with probability at least $1-e^{-x}$,
\#
	\overbar D_{\hat \cQ_{1,b}} ( \wt \bbeta_\eta , \bbeta^* ) \geq  0.5 \underbar{$f$}   \kappa_l\cdot  \| \wt \bbeta_\eta - \bbeta^*   \|_{\Sigma}^2, ~\mbox{ provided that }  n \gtrsim  b^{-1} \{ L^2 \log (p) + x\}. \label{local.breg.lbd}
\#
Consequently, 
\#
	0.5   \underbar{$f$}   \kappa_l   \cdot   \| \wt \bbeta_\eta - \bbeta^*   \|_{\Sigma}^2  &  \leq \overbar D_{\hat \cQ_{1,b}} ( \wt \bbeta_\eta , \bbeta^* )   \leq \eta \overbar D_{\hat \cQ_{1,b}} ( \wt \bbeta , \bbeta^* )  \nn \\
 &\leq  \bigl(  1.4 s^{1/2} \lambda_0 + 0.5 l_0 \kappa_2 b^2 \bigr) \cdot   \| \wt \bbeta_\eta - \bbeta^*  \|_{\Sigma}  \nn
\#
with probability at least $1-e^{-x}$. Canceling  $  \| \wt \bbeta_\eta - \bbeta^*  \|_{\Sigma}$ on both sides yields
$\| \wt \bbeta_\eta - \bbeta^*   \|_{\Sigma}  \leq   (  \underbar{$f$}   \kappa_l  )^{-1}  ( 2.8 s^{1/2} \lambda_0  + l_0 \kappa_2 b^2)$.
Provided that 
\#
	b > 4\gamma_{0.25} (\underbar{$f$} \kappa_l   )^{-1}    \bigl( 2.8 s^{1/2} \lambda_0 + l_0 \kappa_2 b^2  \big)  ,  \label{loc.band.constraint}
\#
$\wt \bbeta_\eta$ falls in the interior of $\Theta(r_{\loc})$ (with high probability). Thus we must have $\wt \bbeta_\eta = \wt \bbeta$, and the same error bound holds for $\wt \bbeta$, that is, $\| \wt \bbeta - \bbeta^* \|_{\Sigma} \lesssim  s^{1/2} \lambda_0 + b^2$.

It remains to tune the regularization parameter $\lambda_0$ and  bandwidth $b$ so that  \eqref{l0.constraint} and \eqref{loc.band.constraint} hold, and to determine the scaling of the sample size required to ensure the lower bound \eqref{local.breg.lbd}.
Applying a local version of Lemma~\ref{lem:hd.score} yields that with probability at least $1-e^{-x}$,
\#
	   \|   \nabla \hat \cQ_{1,b}(\bbeta^*) - \nabla \cQ_{1,b}(\bbeta^*)   \|_\infty  \leq  C(\tau, b) \sqrt{\frac{\log(2p)+ x}{n}} + B \max(\tau, 1-\tau) \frac{\log(2p)+ x}{3n}  \label{local.grad.ubd}
\#
for the same constants therein. Given $\delta\in (0,1)$, we take $x= \log(2/\delta)$ in \eqref{local.breg.lbd} and \eqref{local.grad.ubd}, and set 
$$
  \lambda_0 \asymp \sqrt{\tau(1-\tau) \log(p/\delta) / n}  ,
$$
so that \eqref{l0.constraint} holds with high probability.  Furthermore, as long as the bandwidth $b$ and sample size $n$ are such that $\sqrt{s\log(p/\delta) / n} \lesssim b \lesssim 1$, both requirements in \eqref{local.breg.lbd} and \eqref{loc.band.constraint} are satisfied.
This completes the proof of \eqref{initial.hd.rate}.

If in addition $\lambda_0 \geq  1.25 l_0 \kappa_2 s^{-1/2}b^2$, then $\| \wt \bdelta \|_1 \leq 4s^{1/2} \| \wt \bdelta \|_{\Sigma}$ and hence $\wt \bbeta \in \Lambda$.

\section{Proof of Auxiliary Lemmas}
 
 \subsection{Proof of Lemma~\ref{lem:diff.ubd1}}
 
For $r_1 , r_2>0$, define the parameter set $\Theta_0(r_1,r_2) = \{ \bdelta \in \RR^p: \| \bdelta \|_1 \leq r_1, \| \bdelta \|_{\Sigma}\leq r_2\}$.
Consider the change of variable $\bv = \bbeta -\bbeta^*$, so that  $\bv \in \Theta_0(4s^{1/2} r ,r)$ for $\bbeta \in \Theta(r) \cap \Lambda$.
Consequently,
 \#
 &  \sup_{\bbeta \in \Theta(r_1, r_2)}   \| D  (  \bbeta) - E  ( \bbeta)     \|_\infty \nn \\
  &  = \max_{1\leq j\leq p} \sup_{ \bv \in \Theta_0(r_1 , r_2)}  \Biggl|  
   \frac{1}{N} \sum_{i=1}^N (1- \EE)  \underbrace{\bigg\{   \overbar K\bigg( \frac{  \bx_i^\T \bv - \varepsilon_i }{h} \bigg) - \overbar K\bigg( \frac{   - \varepsilon_i }{h} \bigg) \bigg\} x_{ij}  }_{=: \psi_{ij}(\bv)}\Biggr|    =:  \max_{1\leq j\leq p} \Psi_j , \label{diff.grad.bound1}
 \#
 where $\Psi_j = \sup_{ \bv \in \Theta_0(r_1 , r_2) } |(1/N)\SN (1-\EE) \psi_{ij}(\bv)|$. 
Since $K(\cdot) = \overbar K'(\cdot)$ is uniformly bounded,
 $$
 	\sup_{\bv \in \Theta_0(r_1 , r_2)} | \psi_{ij}(\bv) | \leq   \kappa_u B^2 \frac{r_1}{h}  .
 $$ 
By Bousquet's version of Talagrand's inequality \citep{B2003}, we obtain that for any $z>0$,
 \#
 	\Psi_j \leq  \frac{5}{4} \EE \Psi_j + \sup_{  \bv \in \Theta_0(r_1 , r_2) } \big\{ \EE \psi_{ij}^2(\bv)  \bigr\}^{1/2} \sqrt{ \frac{2 z }{N}  }  + (4 +1/3)\kappa_u B^2   \frac{  r_1 z }{ N h }    \label{diff.grad.concentration}
 \#
 with probability at least $1-2e^{-z}$. For   $\bv \in \Theta_0(r_1 , r_2)$, 
\#
 	\EE \psi_{ij}^2(\bv) & = \EE \Biggl[  x_{ij}^2     \int_{-\infty}^\infty  \bigl\{ \overbar K\bigl(    \bx^\T \bv /h - u/h \bigr) - \overbar K(-u/h) \bigr\}^2 f_{\varepsilon | \bx} (u) \, {\rm d} u\Biggr] \nn \\
 & =  h \, \EE \Biggl(   x_{ij}^2    \int_{-\infty}^\infty  \bigl\{ \overbar K\bigl(  \bx^\T \bv /h  + v \bigr) - \overbar K(v) \bigr\}^2 f_{\varepsilon | \bx} (-v h )  \, {\rm d} v\nn\biggr) \\ 
 & \leq  \bar f  h^{-1}  \EE \Biggl[ x_{ij}^2\, (\bx^\T \bv )^2 \int_{-\infty}^\infty \bigg\{  \int_0^1 K \bigl( v+ w  \bx^\T  \bv /h   \bigr) \, {\rm d} w \bigg\}^2  \,  {\rm d} v\Biggr] \nn \\
 & \leq  \bar f  h^{-1}  \EE \Biggl( x_{ij}^2\,  (\bx^\T \bv )^2 
 \Biggl[ \int_0^1  \bigg\{ \int_{-\infty}^\infty K^2 \bigl(  v+ w  \bx^\T  \bv /h   \bigr) \, {\rm d} v  \bigg\}^{1/2} {\rm d}w \Biggr]^2 \Biggr)  \nn\\ 
 & ~~~~~~~~~~~~~~~~~~~~~~~~~~~~~~~~~~~~~~~~~~~~~~~~~~~~~~~~~~~~\mbox{(by Minkowski's integral inequality)} \nn\\
   & \leq  \kappa_u \bar f \cdot   h^{-1}  \EE \bigl( x_{ij}  \cdot \bx^\T \bv  \bigr)^2 \leq   \kappa_u \bar f  \cdot   h^{-1}  \bigl( \EE x_{ij}^4 \bigr)^{1/2} \bigl\{ \EE(\bx^\T \bv )^4 \bigr\}^{1/2}  \leq   \kappa_u \bar f  \sigma_{jj} \mu_4 \cdot  h^{-1} r_2^2 , \nn
 \#
 where the last inequality uses the bound $\EE x_{ij}^4 = \EE \langle \Sigma^{-1/2} \bx, \Sigma^{1/2} \be_j \rangle^4 \leq \mu_4 \| \Sigma^{1/2} \be_j \|_2^4  = \sigma_{jj}^2 \mu_4$.

We next bound the mean $\EE \Psi_j$. By Rademacher symmetrization,
\$
\EE \Psi_j \leq 2 \EE \sup_{\bv \in \Theta_0(r_1 , r_2) }  \bigg| \frac{1}{N} \SN e_i \psi_{ij}(\bv ) \bigg|  = 2 \EE \Biggl\{  \EE_e\sup_{\bv \in \Theta_0(r_1 , r_2)  }  \bigg| \frac{1}{N} \SN e_i \psi_{ij}(\bv ) \bigg| \Biggr\},
\$
where $\EE_e$ denotes the conditional expectation over $e_1,\ldots , e_n$ given the remaining variables, and $e_1,\ldots, e_n$ are i.i.d. Rademacher random variables. For each $i$, write $\psi_{ij}(\bv) = \varphi_{i}(\bx_i^\T \bv)$, where $\varphi_i(\cdot)$ is such that $\varphi_i(0)=0$ and  $| \varphi_i(u) - \varphi_i(v) | \leq \kappa_u  |x_{ij}| \cdot h^{-1} |u-v|$. Then, by Talagrand's contraction principle (see, e.g., Theorem~4.12 in \cite{LT1991}), 
\$
 \EE_e\sup_{\bv \in \Theta_0(r_1 , r_2) }  \bigg| \frac{1}{N} \SN e_i \psi_{ij}(\bv ) \bigg|  & \leq 2 \kappa_u  \max_{1\leq i\leq N} | x_{ij} |   \cdot    \EE_e \sup_{  \bv \in \Theta_0(r_1 , r_2) } \bigg| \frac{1}{N h } \SN  e_i  \bx_i^\T \bv \bigg| \\ 
& \leq  2  \kappa_u   B \,  \frac{r_1}{h}   \EE_e \bigg\| \frac{1}{N} \SN e_i  \bx_i \bigg\|_\infty  . 
\$
By Hoeffding's moment inequality,
\$
\EE_e \bigg\| \frac{1}{N} \SN e_i  \bx_i \bigg\|_\infty \leq  \max_{1\leq k\leq p}  \bigg( \frac{1}{N} \SN x_{i k}^2 \bigg)^{1/2}\sqrt{\frac{2\log(2p)}{N}} . 
\$
Putting together the above three inequalities yields
\#
  \EE \Psi_j \leq 4  \kappa_u B^2 \frac{r_1}{h} \sqrt{\frac{2\log(2p)}{N}}  ~\mbox{ for any } j =1,\ldots, p. \label{max.mean.bound}
\#
To sum up, we take $z=\log(2p) +x$, $r_1 = 4s^{1/2}r $ and $r_2=r$ in \eqref{diff.grad.concentration}, which combined with \eqref{diff.grad.bound1}, \eqref{max.mean.bound} and the union bound, completes the proof of \eqref{glob.grad.uniform}.

\subsection{Proof of Lemma~\ref{lem:diff.ubd2}}

Under the conditional quantile model \eqref{qr.model}, note that $E(\bbeta)  = \nabla \cQ_h(\bbeta) - \nabla \cQ_h(\bbeta^*)$, where $\cQ_h(\bbeta) = \EE \hat \cQ_h(\bbeta)$ is the population smoothed loss which is twice-differentiable with Hessian $\nabla^2 \cQ_h(\bbeta) = \EE \{ K_h( (\bx^\T \bbeta - y)/h) \bx \bx^\T \}$.
Moreover, define $\Hb_0= \EE\{ f_{\varepsilon |\bx}(0) \bz \bz^\T\} = \Sigma^{-1/2} \Hb \Sigma^{-1/2}$, where $\Hb = \EE\{ f_{\varepsilon |\bx}(0) \bx \bx^\T\}$ and $\bz = \Sigma^{-1/2}\bx$.  By the mean value theorem for vector-valued functions,
\#
  \Sigma^{-1/2}  E(\bbeta)   & =  \Sigma^{-1/2} \EE \int_0^1 \nabla^2 \cQ_{h}\bigl(   (1-t) \bbeta^* + t \bbeta \bigr)  \, {\rm d}t \, \Sigma^{-1/2}  \cdot   \Sigma^{1/2} (\bbeta - \bbeta^* )   \nn  \\
& = \EE \int_0^1 \int_{-\infty}^\infty K(u)    f_{\varepsilon |\bx} (t \cdot \bz^\T \bdelta - h u)  \, {\rm d} u \, {\rm d} t \cdot \bz \bz^\T   \bdelta  , \nn
\#
where $\bdelta = \Sigma^{1/2}( \bbeta -\bbeta^*)$.
Similarly,
\#
\Sigma^{-1/2}  E_1(\bbeta)  = \EE \int_0^1 \int_{-\infty}^\infty K(u)    f_{\varepsilon |\bx} (t \cdot \bz^\T \bdelta - b u)  \, {\rm d} u \, {\rm d} t \cdot \bz \bz^\T   \bdelta. \nn
\#
where $\cQ_{1,b}(\bbeta) = \EE \hat \cQ_{1,b}(\bbeta)$. This, together with the Lipschitz continuity of $f_{\varepsilon |\bx}(\cdot)$ (implied by Condition~(C1)), implies that for any $\bbeta \in \Theta(r)$,
\#
 &   \|    E(\bbeta) - E_1(\bbeta)    \|_{\Omega} \nn \\
 &  \leq  \sup_{\bu \in \mathbb{S}^{p-1}}    \EE \int_0^1 \int_{-\infty}^\infty K(u)   |   f_{\varepsilon |\bx} (t \cdot \bz^\T \bdelta - h u)  -  f_{\varepsilon |\bx} (t \cdot \bz^\T \bdelta - b u)  | \, {\rm d} u \, {\rm d} t \cdot   | \bz^\T   \bdelta \cdot \bz^\T \bu |  \nn \\
 & \leq   l_0   \int_{-\infty}^\infty |u| K(u)    \, {\rm d} u \cdot  |b-h| \sup_{\bu \in \mathbb{S}^{p-1}}   \bigl\{ \EE  ( \bz^\T \bu )^2 \bigr\}^{1/2} \| \bdelta \|_{\Sigma}  =  l_0   \kappa_1  |b-h| r , \nn
\#
as claimed.

Turning to $\Sigma^{-1/2}\nabla \cQ_h(\bbeta^*) = \EE \{ \overbar K(-\varepsilon /h )   -\tau \} \bz $, by integration by parts we get
\$
&  \EE \big\{ \overbar K(-\varepsilon /h ) | \bx \big\}  
 =  \int_{-\infty}^\infty \overbar  K(-t/h)\, {\rm d} F_{\varepsilon | \bx}(t) \\
 &  = -\frac{1}{h} \int_{-\infty}^\infty  K(-t/h) F_{\varepsilon |\bx}(t) \, {\rm d} t = \int_{-\infty}^\infty K(u) F_{\varepsilon |\bx} (  - hu ) \, {\rm d} t \\
 & = \tau + \int_{-\infty}^\infty   K(u) \int_0^{-hu} \big\{ f_{\varepsilon |\bx}(t) - f_{\varepsilon |\bx} (0)  \big\} \, {\rm d} t  \, {\rm d}u .
\$
Combined with the Lipschitz continuity of $f_{\varepsilon |\bx}(\cdot)$, this implies 
\$
&     \|  \nabla \cQ_h(\bbeta^*)   \|_{\Omega}  = \sup_{\bu \in \mathbb{S}^{p-1}}  \EE \int_{-\infty}^\infty   K(u) \int_0^{-hu}  \big\{ f_{\varepsilon |\bx}(t) - f_{\varepsilon |\bx} (0)  \big\} \, {\rm d} t  \, {\rm d}u   \cdot \bz^\T \bu  \leq   \frac{1}{2} l_0 \kappa_2 h^2 ,
\$
thus completing the proof.

\section{Estimation at Extreme Quantile Levels}
\label{sec:extreme}

As highlighted in the introduction, the causal mechanisms underpinning extreme behavior are of high relevance in numerous fields, and QR at extreme quantile levels operationalizes attempts to understand these. 
Rather than treating variables on an equal footing as \cite{EH2020}, QR singles out a particular variable for which understanding is sought. Just as the statistical aspects of extreme value theory are challenged by the limitation of data beyond extreme thresholds, QR coefficients at extreme quantiles are notoriously hard to estimate. The following minor adaptation of our procedure improves its performance at extreme quantile levels.

Recall from Section~\ref{sec:smoothing} that the conquer method is evolved from a smoothed estimating equation approach \citep{KS2017}. The latter constructs a smoothed sample analog of the moment condition.  As observed by both \cite{FGH2019} and \cite{HPTZ2020}, the smoothing bias primarily affects the intercept estimation especially in the random design setting.
Now let us take a closer look at the moment condition. For every $p$-vector $\bu=(u_1,\ldots, u_p)^\T$, we use $\bu_-\in \RR^{p-1}$ to denote its $(p-1)$-subvector  with the first coordinate removed, i.e. $\bu_- = (u_2, \ldots, u_p)^\T $. Then, the first-order moment condition can be written as
\$
\left\{\begin{array}{ll}
	\EE  \{ \mathbbm{1}( y < \bx_-^\T \bbeta_- + \beta_1) - \tau \} = 0 ,  \vspace{0.2cm}\\
	\EE  \{ \mathbbm{1}( y < \bx_-^\T \bbeta_- + \beta_1) - \tau \} x_j = 0 , \ \  j = 2, \ldots, p ,
\end{array}\right.
\$
whose sample counterpart is 
\#
\left\{\begin{array}{ll}
	\sum_{i=1}^N  \{ \mathbbm{1}( y_i< \bx_{i,-}^\T \bbeta_- + \beta_1) - \tau \} =0,   \vspace{0.2cm}\\
	\sum_{i=1}^N   \{ \mathbbm{1}( y_i< \bx_{i,-}^\T \bbeta_- + \beta_1) - \tau   \}  x_{ij}  =0 ,  \ \  j=2,\dots, p. \\
\end{array}\right. \label{sample.moment.condition}
\#
To find the solution of the above system of equations, note that given $\bbeta_-\in \RR^{p-1}$, the first equation can be (approximately) solved by taking $\beta_1$ to be the sample $\tau$-quantile of $\{ y_i - \bx_{i,-}^\T \bbeta_- \}_{i=1}^N$, which only allows for an error $1/N$. The main difficulty then arises from solving the remaining $p-1$ equations, for which analytical solutions do not exist. To mitigate the smoothing bias of conquer for intercept estimation, we consider a hybrid estimating equation approach that solves
\begin{align}\label{hybrid.method}
	\left\{\begin{array}{ll}
		\sum_{i=1}^N  \{ \mathbbm{1}( y_i< \bx_{i,-}^\T \bbeta_- + \beta_1) - \tau \} =0,     \vspace{0.2cm} \\ 
		\sum_{i=1}^N   \{  \overbar K   ( - ( y_i - \beta_1 - \bx_{i,-}^\T \bbeta_- ) /h   ) - \tau   \}  \bx_{i,-} = \textbf{0}_{p-1}, \\
	\end{array}\right .
\end{align}
where $\overbar K(u ) = \int_{-\infty}^u K(v) \, {\rm d} v$ for some kernel function $K(\cdot)$. Note that, given $\bbeta_- \in \RR^{p-1}$, the first equation in \eqref{hybrid.method} can be solved by taking $\beta_1$ to be the sample $\tau$-quantile of $\{ y_i - \bx_{i,-}^\T \bbeta_- \}_{i=1}^N$, and given $\beta_1$, solving the second vector equation is equivalent to minimizing the conquer loss $\bbeta_- \mapsto \hat \cQ_h(\beta_1, \bbeta_-)$. This motivates the following iterative procedure,  starting at iteration 0 with initial estimates $\beta_1^{(0)}$ and $\bbeta_-^{(0)}$ of the intercept and slope coefficients, respectively. The procedure involves two steps. At iteration $t=1,2,\ldots$:

\noindent
{\it Refitted intercept}. Using the current slope coefficients estimate $\bbeta^{(t-1)}_-$, we compute the residuals $r_i^{(t-1)} = y_i - \bx_{i,-}^\T \bbeta^{(t-1)}_-$, and then update the intercept $\beta_1^{(t)}$ as the sample $\tau$-quantile of $\{ r_i^{(t-1)} \}_{i=1}^N$.

\noindent
{\it Adjusted conquer}. With a refitted intercept $\beta_1^{(t)}$, we take any solution to the optimization problem
\$
\min_{\bbeta_- \in \RR^{p-1}} \hat \cQ_h(\beta_1^{(t )} , \bbeta_-) = \min_{\bbeta_- \in \RR^{p-1} } \frac{1}{N} \sum_{i=1}^N (\rho_\tau * K_h) ( y_i - \beta_1^{(t )} - \bx_{i,-}^\T \bbeta_- ) 
\$
as the updated slope estimator, denoted by $\hat \bbeta^{(t)}_-$.
We refer to the above method as   {\it two-step conquer}.

Using two-step conquer, in Algorithm~\ref{algo2} we present a modified multi-round distributed algorithm, which is particularly suited for extreme quantile regressions with $\tau$ close to either 0 or 1.

\begin{algorithm}[!t]
	\caption{ {\small  Efficient Distributed Quantile Regression via Two-Step Conquer.}}
	\label{algo2}
	\textbf{Input}: data batches $\{(y_i, \bx_i)\}_{i\in \cI_j}$, $j=1,\ldots, m$, stored at $m$ sites, quantile level $\tau\in (0,1)$, bandwidths $b , h>0$, initialization $\wt{\bbeta}^{(0)} \in \RR^p$, maximum number of iterations $T$, $g_0 = 1$.
	
	\begin{algorithmic}[1]
		\FOR{$t = 1, 2 \ldots, T$}
		\STATE Broadcast $\wt \bbeta^{(t-1)}_- \in \RR^{p-1}$ to all local machines.
		\FOR{$j=1,\ldots, m$}
		\STATE At the $j$th site, compute the sample $\tau$-quantile of $\{ \hat r_i^{(t-1)} := y_i - \langle \bx_{i,-}, \wt \bbeta_-^{(t-1)}\rangle\}_{i\in \cI_j}$, denoted by $\hat q^{(t-1)}_j$, and send it the master (first) machine.
		\ENDFOR
		\STATE Calculate $\hat q^{(t)} = (1/m) \sum_{j=1}^m  \hat q^{(t-1)}_j$ on the master, and send it to every local machine.
		\FOR{$j=1,\ldots, m$}
		\STATE  On the $j$th machine, compute the gradient vector
		$$
		\hat \bg_{j,h}^{(t-1)} = -\frac{1}{n} \sum_{i\in \cI_j } \ell'_h(\hat r_i^{(t-1)} -\hat q^{(t)}   ) \bx_{i,-} \in \RR^{p-1} ,
		$$
		and send it to the master.
		\ENDFOR
		\STATE On the master machine, calculate 
		$$
		\hat  \bg_h^{(t-1)} = \frac{1}{m}\sum_{j=1}^m  \hat \bg_{j,h}^{(t-1)} ~~\mbox{and}~~ g_t = \| 	\hat  \bg_h^{(t-1)}\|_\infty .
		$$
		\vspace{-0.3cm}
		\STATE {\bf if} $g_t > g_{t-1}$ or $g_t < 10^{-5}$ {\bf break}
		\STATE {\bf otherwise} Calculate $ \hat \bg_{1,b}^{(t-1)} = (-1/n)\sum_{i\in \cI_1 } \ell'_b(\hat r_i^{(t-1)} -\hat q^{(t)}  ) \bx_{i,-} $, and solve the shifted conquer loss minimization
		$$
		\hat  \btheta^{(t)}  \in \argmin_{\btheta \in \RR^{p-1}}  ~ \hat \cQ_{1,b} ( \hat  q^{(t)} ,  \btheta )  - \big\langle  \hat \bg_{1,b}^{(t-1)}  - \hat \bg_{h}^{(t-1)}  , \btheta \big\rangle 
		$$
		on the master machine. Define $\wt \bbeta^{(t)} =( \hat q^{(t)} , (\hat \btheta^{(t)})^\T)^\T$ as the $t^{{\rm th}}$ iterate.
		\ENDFOR 
	\end{algorithmic}
	\textbf{Output}: $\wt \bbeta^{(T)}$.
\end{algorithm}

\section{Additional Simulation Studies}
\label{sec:extraSim}
In this section, we provide numerical studies for score-based confidence sets to complement those in Section \ref{subsec:inferencesim} of the paper.  Specifically, in each of 200 Monte Carlo replications, $p=10$ covariates are generated at random from a uniform distribution on $[-1,1]$ and a response variable is generated according to the linear heteroscedastic model
\[
y_{i} = \bx_i^\T \bbeta^* + (0.25x_{i1}+0.25x_{i2}+0.75)\{\varepsilon_i- F_{\varepsilon_i}^{-1}(\tau)\}, \quad i=1,\ldots, N,
\]
where $\tau=0.9$ and $\varepsilon_i$ is drawn from the $t$-distribution with 1.5 degrees of freedom. The intercept $\beta_0^*$ is taken as 2 and all other elements of $\bbeta^*$ as unity. 

For $n=200$ and $m=100$, Tables \ref{tab:n200m100coverage} and \ref{tab:n200m100width} report the simulated coverage probabilities and mean width for each confidence set construction described in Section \ref{subsec:inference}. In addition to the aforementioned methods, the construction in equation \eqref{eq:scoreset} is referred to as CE-Score.


\begin{table}[!htp]
	\begin{tabular}{ccccccccccc}
		& $\beta^*_1$ & $\beta^*_2$ & $\beta^*_3$ & $\beta^*_4$ & $\beta^*_5$ & $\beta^*_6$ & $\beta^*_7$ & $\beta^*_8$ & $\beta^*_9$ & $\beta^*_{10}$ \\
		\hline
		DC-Normal        & 0.810       & 0.805       & 0.995       & 1.000       & 0.985       & 0.995       & 0.995       & 0.990       & 1.000       & 0.995        \\
		CE-Normal        & 0.935       & 0.865       & 0.925       & 0.930       & 0.895       & 0.940       & 0.895       & 0.925       & 0.910       & 0.910        \\
		CE-Boot (a) & 0.970       & 0.940       & 0.960       & 0.945       & 0.940       & 0.975       & 0.950       & 0.955       & 0.950       & 0.955        \\
		CE-Boot (b) & 0.965       & 0.910       & 0.945       & 0.950       & 0.935       & 0.975       & 0.935       & 0.955       & 0.930       & 0.950        \\
		CE-Score         & 0.960       & 0.905       & 0.960       & 0.970       & 0.915       & 0.980       & 0.920       & 0.955       & 0.945       & 0.950      \\ 
		\hline
	\end{tabular}
\caption{Monte Carlo coverage probabilities for the case when $p=10$, $n=200$, and $m=100$.}\label{tab:n200m100coverage}
\end{table}

\begin{table}[!htp]
	\begin{tabular}{ccccccccccc}
		& $\beta^*_1$ & $\beta^*_2$ & $\beta^*_3$ & $\beta^*_4$ & $\beta^*_5$ & $\beta^*_6$ & $\beta^*_7$ & $\beta^*_8$ & $\beta^*_9$ & $\beta^*_{10}$ \\
		\hline
		DC-Normal   & 0.359       & 0.344       & 0.338       & 0.336       & 0.348       & 0.347       & 0.344       & 0.336       & 0.337       & 0.340        \\
		CE-Normal   & 0.207       & 0.197       & 0.191       & 0.187       & 0.192       & 0.192       & 0.191       & 0.186       & 0.189       & 0.189        \\
		CE-Boot (a) & 0.238       & 0.234       & 0.216       & 0.217       & 0.228       & 0.225       & 0.222       & 0.216       & 0.216       & 0.219        \\
		CE-Boot (b) & 0.222       & 0.215       & 0.206       & 0.203       & 0.210       & 0.209       & 0.208       & 0.202       & 0.204       & 0.204        \\
		CE-Score    & 0.162       & 0.162       & 0.162       & 0.162       & 0.162       & 0.161       & 0.161       & 0.161       & 0.161       & 0.161    \\ 
		\hline   
	\end{tabular}
	\caption{Monte Carlo mean width of the constructed confidence intervals for the case when $p=10$, $n=200$, and $m=100$.}\label{tab:n200m100width}
\end{table}

Apart from those based on the averaging estimator $\hat{\bbeta}^{\text{dc}}$, which has appreciable undercoverage for the fist two coefficients, all constructions are broadly comparable in terms of coverage probability. Confidence intervals based on the score statistic are considerably narrower than the others, although at higher computational cost due to the need to calculate $\hat{T}_k(c_k)$ from equation \eqref{eq:scoreset} over a grid of $c_k$ values.

Tables \ref{tab:n200m200coverage} and \ref{tab:n200m200width} for $n=200$ and $m=200$ are qualitatively similar. For roughly similar coverage, the higher total sample size $N=nm$ reduces the widths of the confidence intervals. The larger value of $m$ also has the effect of reducing the DC-Normal coverage probability due to the bias in $\hat{\bbeta}^{\text{dc}}$. This bias does not disappear asymptotically in $m$ but rather the variation of $\hat{\bbeta}^{\text{dc}}$ around the wrong point is diminished, leading to poor coverage.


\begin{table}[!htp]
	\begin{tabular}{ccccccccccc}
		& $\beta^*_1$ & $\beta^*_2$ & $\beta^*_3$ & $\beta^*_4$ & $\beta^*_5$ & $\beta^*_6$ & $\beta^*_7$ & $\beta^*_8$ & $\beta^*_9$ & $\beta^*_{10}$ \\
		\hline
		DC-Normal   & 0.595       & 0.615       & 0.990       & 0.995       & 1.000       & 0.990       & 0.995       & 1.000       & 0.990       & 0.995        \\
		CE-Normal   & 0.895       & 0.925       & 0.935       & 0.950       & 0.915       & 0.910       & 0.940       & 0.950       & 0.920       & 0.900        \\
		CE-Boot (a) & 0.955       & 0.945       & 0.960       & 0.985       & 0.960       & 0.960       & 0.985       & 0.980       & 0.955       & 0.945        \\
		CE-Boot (b) & 0.945       & 0.935       & 0.950       & 0.975       & 0.955       & 0.935       & 0.985       & 0.975       & 0.950       & 0.930        \\
		CE-Score    & 0.955       & 0.955       & 0.950       & 0.970       & 0.955       & 0.935       & 0.970       & 0.980       & 0.940       & 0.965    \\ 
		\hline   
	\end{tabular}
	\caption{Monte Carlo coverage probabilities for the case when $p=10$, $n=200$,  and $m=200$.}\label{tab:n200m200coverage}
\end{table}

\begin{table}[!htp]
	\begin{tabular}{ccccccccccc}
		& $\beta^*_1$ & $\beta^*_2$ & $\beta^*_3$ & $\beta^*_4$ & $\beta^*_5$ & $\beta^*_6$ & $\beta^*_7$ & $\beta^*_8$ & $\beta^*_9$ & $\beta^*_{10}$ \\
		\hline
		DC-Normal   & 0.259       & 0.257       & 0.257       & 0.247       & 0.251       & 0.251       & 0.254       & 0.244       & 0.252       & 0.245        \\
		CE-Normal   & 0.149       & 0.145       & 0.138       & 0.137       & 0.135       & 0.138       & 0.138       & 0.134       & 0.135       & 0.134        \\
		CE-Boot (a) & 0.173       & 0.173       & 0.165       & 0.162       & 0.160       & 0.164       & 0.164       & 0.158       & 0.162       & 0.160        \\
		CE-Boot (b) & 0.165       & 0.163       & 0.155       & 0.154       & 0.150       & 0.154       & 0.154       & 0.150       & 0.152       & 0.151        \\
		CE-Score    & 0.114       & 0.114       & 0.113       & 0.113       & 0.113       & 0.113       & 0.113       & 0.112       & 0.113       & 0.112      \\ 
		\hline 
	\end{tabular}
	\caption{Monte Carlo mean width of the constructed confidence intervals for the case when $p=10$, $n=200$, and $m=200$.}\label{tab:n200m200width}
\end{table}

Further simulation results for $n=400$ with $m=100$ and $m=200$ are reported in Tables~\ref{appendix:table1}--\ref{appendix:table4}. Similar conclusions can be drawn.


\begin{table}[H]
	\begin{tabular}{ccccccccccc}
		& $\beta^*_1$ & $\beta^*_2$ & $\beta^*_3$ & $\beta^*_4$ & $\beta^*_5$ & $\beta^*_6$ & $\beta^*_7$ & $\beta^*_8$ & $\beta^*_9$ & $\beta^*_{10}$ \\
		\hline
		DC-Normal   & 0.765       & 0.715       & 0.995       & 1.000       & 0.995       & 0.980       & 0.990       & 1.000       & 0.985       & 1.000        \\
		CE-Normal   & 0.965       & 0.950       & 0.935       & 0.960       & 0.940       & 0.920       & 0.960       & 0.960       & 0.960       & 0.960        \\
		CE-Boot (a) & 0.990       & 0.970       & 0.970       & 0.980       & 0.950       & 0.940       & 0.965       & 0.975       & 0.960       & 0.955        \\
		CE-Boot (b) & 0.975       & 0.965       & 0.945       & 0.975       & 0.960       & 0.930       & 0.975       & 0.965       & 0.960       & 0.975        \\
		CE-Score    & 0.950       & 0.950       & 0.945       & 0.970       & 0.950       & 0.935       & 0.980       & 0.980       & 0.930       & 0.965       \\ 
		\hline
	\end{tabular}

	\caption{	\label{appendix:table1}Monte Carlo coverage probabilities ($p=10$, $n=400$, $m=100$).}
\end{table}

\begin{table}[H]
	\begin{tabular}{ccccccccccc}
		& $\beta^*_1$ & $\beta^*_2$ & $\beta^*_3$ & $\beta^*_4$ & $\beta^*_5$ & $\beta^*_6$ & $\beta^*_7$ & $\beta^*_8$ & $\beta^*_9$ & $\beta^*_{10}$ \\
		\hline
		DC-Normal   & 0.191       & 0.189       & 0.187       & 0.184       & 0.184       & 0.184       & 0.185       & 0.180       & 0.185       & 0.182        \\
		CE-Normal   & 0.141       & 0.139       & 0.132       & 0.130       & 0.130       & 0.130       & 0.131       & 0.130       & 0.133       & 0.131        \\
		CE-Boot (a) & 0.153       & 0.154       & 0.147       & 0.144       & 0.144       & 0.144       & 0.146       & 0.143       & 0.146       & 0.145        \\
		CE-Boot (b) & 0.147       & 0.147       & 0.139       & 0.138       & 0.137       & 0.137       & 0.139       & 0.137       & 0.140       & 0.138        \\
		CE-Score    & 0.114       & 0.114       & 0.112       & 0.113       & 0.112       & 0.113       & 0.113       & 0.112       & 0.112       & 0.113       \\ 
		\hline
	\end{tabular}

	\caption{		\label{appendix:table2}Monte Carlo mean width ($p=10$, $n=400$, $m=100$).}
\end{table}

\begin{table}[H]
	\begin{tabular}{ccccccccccc}
		& $\beta^*_1$ & $\beta^*_2$ & $\beta^*_3$ & $\beta^*_4$ & $\beta^*_5$ & $\beta^*_6$ & $\beta^*_7$ & $\beta^*_8$ & $\beta^*_9$ & $\beta^*_{10}$ \\
		\hline
		DC-Normal   & 0.495       & 0.450       & 0.985       & 0.995       & 0.985       & 0.990       & 0.985       & 0.980       & 0.980       & 0.990        \\
		CE-Normal   & 0.945       & 0.915       & 0.940       & 0.960       & 0.950       & 0.935       & 0.925       & 0.950       & 0.930       & 0.930        \\
		CE-Boot (a) & 0.965       & 0.940       & 0.960       & 0.980       & 0.975       & 0.950       & 0.960       & 0.980       & 0.945       & 0.970        \\
		CE-Boot (b) & 0.955       & 0.935       & 0.955       & 0.975       & 0.970       & 0.950       & 0.945       & 0.970       & 0.930       & 0.960        \\
		CE-Score    & 0.930       & 0.930       & 0.965       & 0.965       & 0.955       & 0.930       & 0.965       & 0.950       & 0.955       & 0.960       \\ 
		\hline
	\end{tabular}

	\caption{\label{appendix:table3}Monte Carlo coverage probabilities ($p=10$, $n=400$, $m=200$).}
\end{table}

\begin{table}[H]
	\begin{tabular}{ccccccccccc}
		& $\beta^*_1$ & $\beta^*_2$ & $\beta^*_3$ & $\beta^*_4$ & $\beta^*_5$ & $\beta^*_6$ & $\beta^*_7$ & $\beta^*_8$ & $\beta^*_9$ & $\beta^*_{10}$ \\
		\hline
		DC-Normal   & 0.131       & 0.129       & 0.128       & 0.127       & 0.127       & 0.126       & 0.129       & 0.124       & 0.124       & 0.125        \\
		CE-Normal   & 0.097       & 0.096       & 0.091       & 0.091       & 0.092       & 0.091       & 0.092       & 0.089       & 0.089       & 0.090        \\
		CE-Boot (a) & 0.106       & 0.105       & 0.100       & 0.101       & 0.101       & 0.100       & 0.103       & 0.099       & 0.097       & 0.099        \\
		CE-Boot (b) & 0.103       & 0.102       & 0.097       & 0.097       & 0.097       & 0.097       & 0.098       & 0.095       & 0.094       & 0.095        \\
		CE-Score    & 0.078       & 0.078       & 0.077       & 0.077       & 0.077       & 0.077       & 0.077       & 0.078       & 0.077       & 0.077       \\ 
		\hline
	\end{tabular}

	\caption{		\label{appendix:table4}Monte Carlo mean width ($p=10$, $n=400$, $m=200$).}
\end{table}



\begin{thebibliography}{9}

\bibitem[Barzilai and Borwein(1988)]{BB1988}	
	{\sc Barzilai, J.} and {\sc Borwein, J.\,M.} (1988).
 	Two-point step size gradient methods.
 	{\it IMA Journal of Numerical Analysis} {\bf 8} 141--148.
 	  
\bibitem[Battey {\it et~al.}(2018)]{BFLLZ2018}
	{\sc Battey, H., Fan, J., Liu, H., Lu, J.} and {\sc Zhu, Z.} (2018).
	Distributed testing and estimation under sparse high-dimensional models.
	{\it Annals of  Statistics} {\bf 46} 1352--1382.

\bibitem[Belloni and Chernozhukov(2011)]{BC2011}
{\sc Belloni, A.} and {\sc Chernozhukov, V.} (2011).
$\ell_1$-regularized quantile regression in high-dimensional sparse models.
{\it Annals of  Statistics} {\bf 39} 82--130.


\bibitem[Belloni et al.(2017)]{BCFH2017}
	{\sc Belloni, A., Chernozhukov, V., Fern\'andez-Val, I.} and {\sc Hansen, C.} (2017).
	Program evaluation and causal inference with high-dimensional data.
	{\it Econometrica} {\bf 85} 233-298.	
	
\bibitem[Bickel(1975)]{B1975}
	{\sc Bickel, P.\,J.} (1975).
	One-step Huber estimates in the linear model.
	{\it Journal of the American Statistical Association} {\bf 70} 428--434.

 	
\bibitem[{Bousquet(2003)}]{B2003}
{\sc Bousquet, O.} (2003).
Concentration inequalities for sub-additive functions using the entropy method.
{\it In Stochastic Inequalities and Applications. Progress in Probability} {\bf 56} 213--247. Birkh\"auser, Basel.
  
 
\bibitem[Boyd {\it et al.}(2011)]{Boyd2011}	
	{\sc Boyd, S., Parikh, N., Chu, E., Peleato, B.} and {\sc Eckstein, J.} (2011).
 	Distributed optimization and statistical learning via the alternating direction method of multipliers.
 	{\it Foundations and Trends in Machine Learning} {\bf 3} 1--122. 
 
\bibitem[Bradic and Kolar(2017)]{BK2017}
	{\sc Bradic, J.} and {\sc Kolar, M.} (2017).
	Uniform inference for high-dimensional quantile regression: Linear functionals and regression rank scores.
	{\it arXiv preprint arXiv:1702.06209.}
	
\bibitem[Chen, Liu and Zhang(2021)]{CLZ2018}
	{\sc Chen, X., Liu, W.} and {\sc Zhang, Y.} (2021).
	First-order Newton-type estimator for distributed estimation and inference.
	{\it Journal of American Statistical Association}  \href{https://doi.org/10.1080/01621459.2021.1891925}{DOI:
10.1080/01621459.2021.1891925}.	 
	
	\bibitem[Chen, Liu and Zhang(2019)]{CLZ2019}
	{\sc Chen, X., Liu, W.} and {\sc Zhang, Y.} (2019).
	Quantile regression under memory constraint.
	{\it Annals of Statistics} {\bf 47} 3244--3273.
	
	\bibitem[Cochran(1938)]{Cochran1938}	
	{\sc Cochran, W.~G.} (1938).
 	The omission or addition of an independent variable in multiple linear regression.
 	{\it Supplement to the Journal of the  Royal Statistical Society} {\bf 5} 171--176.	
	
	\bibitem[Cox(2007)]{Cox2007}	
	{\sc Cox, D.~R.} (2007).
 	On a generalization of a result of W.~G.~Cochran.
 	{\it Biometrika} {\bf 94} 755--759.
 	
 	
\bibitem[de la Pa\~na, Lai and Shao(2009)]{DLS2009}	
	{\sc de la Pe\~na, V.\,H., Lai, T.\,L.} and {\sc Shao, Q.-M.} (2009).
 	{\it Self-Normalized Processes: Theory and Statistical Applications.}
	Springer, Berlin.
 
\bibitem[Douglas and Rachford(1956)]{DR1956}
	{\sc Douglas, J.} and {\sc Rachford, H.\,H.} (1956).
	On the numerical solution of heat conduction problems in two and three space variables.
 	{\it Transactions of the American Mathematical Society} {\bf 82} 421--439. 
 	
     \bibitem[Efron(1969)]{Efron1969}	
 {\sc Efron, B.} (1969).
 Student’s t-test under symmetry conditions.
 {\it Journal of the American Statistical Association} {\bf 64} 1278--1302.	
 
 	
    \bibitem[Engelke and Hitz(2020)]{EH2020}	
	{\sc Engelke, S.} and {\sc Hitz, A.\,S.} (2020).
 	Graphical models for extremes.
 	{\it Journal of the  Royal Statistical Society: Series B} {\bf 82} 869--931.	


\bibitem[Fan, Guo and Wang(2021)]{FGW2021}	
	{\sc Fan, J.}, {\sc Guo, Y.} and {\sc Wang, K.} (2021).
	Communication-efficient accurate statistical estimation.
	{\it Journal of the American Statistical Association} \href{https://doi.org/10.1080/01621459.2021.1969238}{DOI:
10.1080/01621459.2021.1969238}.
	  
   \bibitem[Fan {\it et al.}(2018)]{FLSF2018}
	{\sc Fan, J., Liu, H., Sun, Q.} and {\sc Zhang, T.} (2018).
	 I-LAMM for sparse learning: Simultaneous control of algorithmic complexity and statistical error.
 	{\it Annals of  Statistics} {\bf 46} 814--841.


\bibitem[Fernandes, Guerre and Horta(2021)]{FGH2019}
	{\sc Fernandes, M., Guerre, E.} and {\sc Horta, E.} (2021).
	Smoothing quantile regressions.
 	{\it Journal of Business and Economics Statistics} {\bf 39} 338--357.

    \bibitem[Fieller(1954)]{Fieller1954}	
	{\sc Fieller, E.\,C.} (1954).
	Some problems in interval estimation.
	{\it Journal of the  Royal Statistical Society: Series B} {\bf 16} 175--185.	

\bibitem[Gabay and Mercier(1976)]{GM1976}
	{\sc Gabay, D.} and {\sc Mercier, B.} (1976).
	A dual algorithm for the solution of nonlinear variational problems via finite element approximation.
 	{\it Computers \& Mathematics with Applications} {\bf 2} 17--40. 
 	
\bibitem[Galvao and Kato(2016)]{GK2016}
	{\sc Galvao, A.\,F.} and {\sc Kato, K.} (2016).
	Smoothed quantile regression for panel data.
 	{\it Journal of Econometrics} {\bf 193} 92--112.

\bibitem[Gu {\it et al.}(2018)]{Gu2018}	
	{\sc Gu, Y.}, {\sc Fan, J.}, {\sc Kong, L.}, {\sc Ma, S.} and {\sc Zou, H.}  (2018).
 	ADMM for high-dimensional sparse regularized quantile regression.
 	{\it Technometrics} {\bf 60} 319--331.
		 	
 
\bibitem[Hall and Sheather(1988)]{HS1988}	
	{\sc Hall, P.} and {\sc Sheather, S.\,J.} (1988).
 	On the distribution of a studentized quantile.
 	{\it Journal of the  Royal Statistical Society: Series B} {\bf 50} 381--391.		
	
\bibitem[He {\it et al.}(2021)]{HPTZ2020}
	{\sc He, X., Pan, X., Tan, K.\,M.} and {\sc Zhou, W.-X.} (2021).
	Smoothed quantile regression with large-scale inference.
	{\it Journal of Econometrics} \href{https://doi.org/10.1016/j.jeconom.2021.07.010}{DOI: 10.1016/j.jeconom.2021.07.010}.
	 
\bibitem[{Horowitz(1998)}]{H1998}
	{\sc Horowitz, J.\,L.} (1998).
	Bootstrap methods for median regression models.
	{\it Econometrica} {\bf 66} 1327--1351.
	
	\bibitem[Hunter and Lange(2000)]{HL2000}
	{\sc Hunter, D.\,R.} and {\sc Lange, K.} (2000).
	Quantile regression via an MM algorithm.
 	{\it Journal of Computational and Graphical Statistics} {\bf 9} 60--77.

\bibitem[Jiang and Yu(2021)]{JY2021}	
	{\sc Jiang, R.} and {\sc Yu, K.} (2021).
	Smoothing quantile regression for a distributed system.
 	{\it Neurocomputing} {\bf 466} 311--326.	

\bibitem[Jordan, Lee and Yang(2019)]{JLY2018}
	{\sc Jordan, M.\,I., Lee, J.\,D.} and {\sc Yang, Y.} (2019).
	Communication-efficient distributed statistical inference.
	{\it Journal of  the American Statistical  Association} {\bf 114} 668--681. 	

\bibitem[Kaplan and Sun(2017)]{KS2017}	
	{\sc Kaplan, D.\,M.} and {\sc Sun, Y.} (2017).
 	Smoothed estimating equations for instrumental variables quantile regression.
 	{\it Econometric Theory} {\bf 33} 105--157.
 

\bibitem[Koenker(2005)]{K2005}	
	{\sc Koenker, R.} (2005).
 	{\it Quantile Regression.}
	Cambridge University Press, Cambridge.
	

\bibitem[Koenker and Bassett(1978)]{KB1978}	
	{\sc Koenker, R.} and {\sc Bassett, G.} (1978).
 	Regression quantiles.
 	{\it Econometrica} {\bf 46} 33--50.
 	
 \bibitem[Koenker {\it et al.}(2017)]{KCHP2017}	
	{\sc Koenker, R.}, {\sc Chernozhukov, V.}, {\sc He, X.} and {\sc Peng, L.} (2017).
 	{\it Handbook of Quantile Regression.}
	CRC Press, New York.	 	

	\bibitem[Lan {\it et al.}(2020)]{LLZ2020}
		{\sc Lan, G., Lee, S.} and {\sc Zhou, Y.} (2020).
	Communication-efficient algorithms for decentralized and stochastic optimization.
	{\it Mathematical Programming} {\bf 180} 237--284.	
	
\bibitem[{Ledoux and Talagrand(1991)}]{LT1991}
{\sc Ledoux, M.} and {\sc Talagrand, M.} (1991).
{\it Probability in Banach Spaces: Isoperimetry and Processes}.
Springer-Verlag, Berlin.


	\bibitem[Lee {\it et al.}(2017)]{LSLT2017}
		{\sc Lee, J., Sun, Y., Liu, Q.} and {\sc Taylor, J.} (2017).
	Communication-efficient	sparse regression.
	{\it Journal of Machine Learning Research} {\bf 18}(5): 1--30.	
	
\bibitem[Li {\it et al.}(2020)]{LSTS2020}
	{\sc Li, T., Sahu, A.\,K., Talwalkar, A.} and {\sc Smith, V.} (2020).
	Federated learning: Challenges, methods, and future directions.
	{\it IEEE Signal Processing Magazine} {\bf 37} 50--60.
	
		
\bibitem[Li and Zhu(2008)]{LZ2008}	
	{\sc Li, Y.} and {\sc Zhu, J.} (2008).
	$L_1$-norm quantile regression.
 	{\it Journal of Computational and Graphical Statistics} {\bf 17} 163--185.	


\bibitem[Pan and Zhou(2021)]{PZ2020}	
	{\sc Pan, X.} and {\sc Zhou, W.-X.} (2021).
 	Multiplier bootstrap for quantile regression: Non-asymptotic theory under random design.
 	{\it Information and Inference: A Journal of the IMA} {\bf 10} 813--861.

\bibitem[Portnoy and Koenker(1997)]{PK1997}	
	{\sc Portnoy, S.} and {\sc Koenker, R.} (1997).
 	The Gaussian hare and the Laplacian tortoise: Computability of squared-error versus absolute-error estimators.
 	{\it Statistical Science} {\bf 12} 279--300.

 \bibitem[Powell(1991)]{P1991}
	{\sc Powell, J.\,L.} (1991).
	Estimation of monotonic regression models under quantile restrictions.
	In {\it Nonparametric and Semiparametric Methods in Econometrics} (eds W. Barnett, J. Powell and G. Tauchen). 
	Cambridge Univ. Press, Cambridge.

\bibitem[Shamir, Srebro and Zhang(2014)]{SSZ2014}
	{\sc Shamir, O., Srebro, N.} and {\sc Zhang, T.} (2014).
	Communication efficient distributed optimization using an approximate Newton-type method.
	In {\it Proceedings of the 31st International Conference on Machine Learning} {\bf 32} 1000--1008.

 	
\bibitem[Shevtsova(2013)]{S2013}
	{\sc Shevtsova, I.\,G.} (2013).
	On the absolute constants in the Berry--Esseen inequality and its structural and nonuniform improvements.
 	{\it Informatika i Ee Primeneniya} {\bf 7} 124--125. 	
 	
\bibitem[Sun, Zhou and Fan(2020)]{SZF2020}
	{\sc Sun, Q., Zhou, W.-X.} and {\sc Fan, J.} (2020).
	Adaptive Huber regression.
 	{\it Journal of  the American Statistical  Association} {\bf 115} 254--265.


\bibitem[Tan, Wang and Zhou(2022)]{TWZ2020}
	{\sc Tan, K.\,M., Wang, L.} and {\sc Zhou, W.-X.} (2022).
	High-dimensional quantile regression: Convolution smoothing and concave regularization.
	{\it Journal of the  Royal Statistical Society: Series B} {\bf 84} 205--233.

\bibitem[Tibshirani(1996)]{Tib1996}	
	{\sc Tibshirani, R.} (1996).
 	Regression shrinkage and selection via the lasso.
 	{\it Journal of the  Royal Statistical Society: Series B} {\bf 58} 267--288.

\bibitem[Volgushev, Chao and Cheng(2019)]{VCC2019}
	{\sc Volgushev, S., Chao, S.-K.} and {\sc Cheng, G.} (2019).
	Distributed inference for quantile regression models.
	{\it Annals of Statistics} {\bf 47} 1634--1662. 	

\bibitem[Wang(2013)]{W2013}	
	{\sc Wang, L.} (2013).
	The $L_1$ regularized LAD estimator for high-dimensional linear regression.
 	{\it Journal of Multivariate Analysis} {\bf 120} 135--151.

\bibitem[Wang and He(2021)]{W2019}	
	{\sc Wang, L.} and {\sc He, X.} (2021).
 	Analysis of global and local optima of regularized quantile regression in high dimension: A subgradient approach. \href{https://github.com/wangx346/myweb/blob/gh-pages/quantile_largeP_rev1.pdf}{{\it Preprint}}. 
	
\bibitem[Wang, Li and Jiang(2007)]{WLJ2007}	
	{\sc Wang, H.}, {\sc Li, G.} and {\sc Jiang, G.} (2007).
 	Robust regression shrinkage and consistent variable selection through the LAD-Lasso.
 	{\it Journal of Business and Economics Statistics} {\bf 25} 347--355.
	

\bibitem[Wang {\it et al.}(2017)]{WKSZ2017}
	{\sc  Wang, J., Kolar, M., Srebro, N.} and {\sc Zhang, T.} (2017).
	Efficient distributed learning with sparsity.
	In {\it Proceedings of the 34th International Conference on Machine Learning} {\bf 70} 3636--3645.

\bibitem[Wang, Stefanski and Zhu(2012)]{WSZ2012}
	{\sc Wang, H.\,J., Stefanski, L.\,A.} and {\sc Zhu, Z.} (2012).
	Corrected-loss estimation for quantile regression with covariate measurement errors.
	{\it Biometrika} {\bf 99} 405--421.
	

\bibitem[Wang, Wu and Li(2012)]{WWL2012}	
	{\sc Wang, L.}, {\sc Wu, Y.} and {\sc Li, R.} (2012).
	Quantile regression for analyzing heterogeneity in ultra-high dimension.
	{\it Journal of  the American Statistical  Association} {\bf 107} 214--222. 
	
	
\bibitem[Whang(2006)]{W2006}
	{\sc Whang, Y.-J.} (2006).
	Smoothed empirical likelihood methods for quantile regression models.
	{\it Econometric Theory} {\bf 22} 173--205.

 
\bibitem[Wu and Lange(2008)]{WL2008}	
	{\sc Wu. T.\,T.} and {\sc Lange, K.} (2008).
	 Coordinate descent algorithms for lasso penalized regression.
 	{\it Annals of Applied Statistics} {\bf 2} 224--244.
 
\bibitem[Wu, Ma and Yin(2015)]{WMY2015}
	{\sc Wu, Y., Ma, Y.} and {\sc Yin, G.} (2015).
	Smoothed and corrected score approach to censored quantile regression with measurement errors.
	{\it Journal of  the American Statistical  Association} {\bf 110} 1670--1683.

\bibitem[Yi and Huang(2017)]{YH2016}	
	{\sc Yi, C.} and {\sc Huang, J.} (2017).
	Semismooth Newton coordinate descent algorithm for elastic-net penalized Huber loss regression and quantile regression.
 	{\it Journal of Computational and Graphical Statistics} {\bf 26} 547--557.	 
 
\bibitem[Yu, Chao and Cheng(2020)]{YCC2020}
	{\sc Yu, Y., Chao, S.-K.} and {\sc Cheng, G.} (2020).
	Simultaneous inference for massive data: distributed bootstrap.
	In {\it Proceedings of the 37th International Conference on Machine Learning} {\bf 119} 10892--10901.	
\bibitem[Yu, Lin and Wang(2017)]{Yu2017}	
	{\sc Yu, L.}, {\sc Lin, N.} and {\sc Wang, L.} (2017).
	A parallel algorithm for large-scale nonconvex penalized quantile regression.
 	{\it Journal of Computational and Graphical Statistics} {\bf 26} 935--939.	 
 		
 		
\bibitem[Zhang, Duchi and Wainwright(2015)]{ZDW2015}	
	{\sc Zhang, Y., Duchi, J.} and {\sc Wainwright, M.} (2015).
 	Divide and conquer kernel ridge regression: A distributed algorithm with minimax optimal rates.
 	{\it Journal of Machine Learning Research} {\bf 16}(102): 3299--3340.
 	
\end{thebibliography}
\end{document}